\documentclass[default]{aastex631}

\shorttitle{SNRs in clumpy medium: Model of HD and radio synchrotron evolution}
\shortauthors{Short authors}
\graphicspath{{./}{figures/}}

\begin{document}

\title{Supernova remnants in clumpy medium: A model of hydrodynamic and radio synchrotron evolution}

\author[0000-0001-6986-9570]{Petar Kosti\'c}
\affiliation{Astronomical Observatory of Belgrade \\
Volgina 7, P.O.Box 74, 11060 Belgrade, Serbia}

\author[0000-0002-8036-4132]{Bojan Arbutina}
\affiliation{Department of Astronomy, Mathematical Faculty, University of Belgrade \\ Studentski Trg 16, 11000 Belgrade, Serbia}

\author[0000-0001-9393-8863]{Branislav Vukoti\'c}
\affiliation{Astronomical Observatory of Belgrade \\
Volgina 7, P.O.Box 74, 11060 Belgrade, Serbia}

\author[0000-0003-0665-0939]{Dejan Uro\v{s}evi\'c}
\affiliation{Department of Astronomy, Mathematical Faculty, University of Belgrade \\ Studentski Trg 16, 11000 Belgrade, Serbia}
\affiliation{Isaac Newton Institute of Chile, Yugoslavia Branch}

\begin{abstract}

   {We present an analytical model of $\Sigma$--$D$ relation for supernova remnants (SNRs) evolving in a clumpy medium. The model and its approximations were developed using the hydrodynamic simulations of SNRs in environments of low-density bubbles and clumpy media with different densities and volume-filling factors. For calculation of SNR luminosities we developed the synchrotron emission model, implying the test-particle approximation. The goal of this work is to explain the flattened part of $\Sigma$--$D$ relation for Galactic SNRs at $D\approx14$--50 pc. Our model shows that the shock collision with the clumpy medium initially enhances the brightness of individual SNRs, which is followed by a steeper fall of their $\Sigma$--$D$ curve. We used the analytical model to generate large SNR samples on $\Sigma$--$D$ plane, within a span of different densities and distances to clumpy medium, keeping the observed distribution of diameters. After comparison with the Galactic sample, we conclude that the observed $\Sigma$--$D$ flattening and scatter originates in sporadic emission jumps of individual SNRs while colliding with the dense clumps. Statistically, the significant impact of the clumps starts at diameters of $\approx14$~pc, up to $\sim70$~pc, with the average density jump at clumpy medium of $\sim2$–20 times, roughly depending on the low density of circumstellar region. However, additional analysis considering the selection effects is needed, as well as the improvement of the model, considering radiation losses and thermal conduction.}

\end{abstract}

\keywords{Supernova remnants(1667) --- Interstellar clouds(834) --- Hydrodynamical simulations(767)}


\section{Introduction} \label{sec:intro}

Supernova remnants (SNR) have been studied significantly over the past several decades and many aspects of their dynamical evolution and radiation mechanisms are now known or placed in a good frame of research. The three phases of SNR evolution: the ejecta-dominated, the Sedov-Taylor (ST) phase, and the radiative phase, are the standard theoretical concept of their dynamical evolution, but it is known that the phases cannot be sharply distinct from one another. Also, there are observed remnants whose parts are in different phases of evolution due to a complex structure of the ambient medium or the anisotropies in supernova (SN) explosions. The extensive observations in X-ray, optical or radio spectrum had lead to several SNR classifications (shell-type, plerionic/composite, mixed-morphology, etc.).

Here, we investigate the hydrodynamic (HD) and radio synchrotron evolution of shell-type SNRs expanding in the clumpy medium, using the 3D simulations. We do not include the radiative cooling nor thermal conduction to the simulations and modeling. As the conservation of energy throughout the evolution is implied, we are able to analytically model the shock hydrodynamics. Such HD model is convenient for generating large samples of artificial SNRs in different ambient media to test some hypotheses against the observations. At this stage, we focus on the basic HD physics and geometry of the shock evolution (using the test-particle radiation), so these could be distinguished from more complex effects of radiative cooling, and more sophisticated emission model, which will be investigated in the consequent stages of research. With these simplifications, our main goal is to investigate the impact of the ambient density on the dynamical and brightness evolution of SNRs. Also, this work aims to find whether this impact can be responsible for the scatter in $\Sigma$--$D$ evolution for Galactic supernova remnants.

\subsection{$\Sigma$--$D$ relation} \label{subsec:SDrel}

The radio-surface-brightness-to-diameter relation, $\Sigma$--$D$, is used as one of the few methods for estimating the distance to an SNR and its evolutionary status \citep{Urosevic2020}, because the surface brightness and angular diameter are the quantities obtained by observations. It was proposed by \citet{Shklovskii1960}:
\begin{equation} \label{eq:SD}
	\Sigma_\nu(D) = AD^{-\beta}.
\end{equation}
The parameter $A$ depends on supernova (SN) properties such as the explosion energy $E_0$, ejected mass $M_{\text{e}}$, SN type, but also on the average ambient density $\rho_0$ and magnetic field $B_0$ \citep{ArbutinaUrosevic2005}. The power-law index $\beta$ (i.e. the slope in log-log form of the relation) depends mainly on the particle acceleration process through the spectral index $\alpha$ of the radio emission, but it might as well depend on the radial distribution of ambient density \citep{Kosticetal2016}. Theoretical and empirical studies of $\Sigma$--$D$ relation give slopes within a large range of values, $\beta=2$--6 \citep[][and references therein]{Kosticetal2016}. It is clear that distribution of SNRs in Galactic $\Sigma$--$D$ plane shows an average decrease of $\Sigma$ with increasing $D$. However, the scatter is so large that uncertainties in finding distances using $\Sigma$--$D$ relation are 50$\%$ on average \citep[see][]{RanasingheLeahy2023}, which puts severe limitations in its usefulness for determination of diameters/distances of individual remnants \citep{Green2005}. Knowing that the large scatter is probably caused by different parameters such as explosion energy and ambient density, it is unlikely that a single relation could be valid for the Galactic SNR sample \citep{RanasingheLeahy2023}. Using the Galactic SNR sample from \citet{Vukoticetal2019} we found that the average $\Sigma$--$D$ relation has an apparent change in slope at some $\sim$20~pc in diameter, where also the scatter becomes larger. This might be caused by the shock transition from low-density wind-blown cavities to much denser regions of the surrounding ISM \citep{Green2005}.

\subsection{SNRs in presence of ISM inhomogeneities} \label{subsec:nonuniISM}

It is reasonable to assume that the medium of non-uniform density might affect the slope of $\Sigma$--$D$ relation of the remnants \citep{Kosticetal2016}. We will investigate the hydrodynamic and radio evolution in the presence of some typical, although simple, ISM models of density clumps.

The density profiles of the wind-blown bubbles (WBB) around the stars are usually inverse square functions of radius, $\rho(r) \propto r^{-2}$, and radii of the bubbles can vary from few to few tens of parsecs. This may significantly alter the expected slope of the $\Sigma$--$D$ curve in earlier stages of the evolution in comparison with the constant density distribution. Moreover, when the shock hits the dense shell, or cloudy medium, and continue to expand in the higher density, it may result in the break of the $\Sigma$--$D$ evolution at various radii for different environments. Considering this, maybe the concept of straight line $\Sigma$--$D$ relations (in a log-log form), should be replaced with the ones with the variable slope.

We will consider the impact of the dense clumps on the SNR shocks and the corresponding radio emission. When the shock wave hits the higher density its speed abruptly decrease. However, opposed to a situation where the shock hits the uniformly higher density environment and continue to propagate slower, e.g. leaving the WBB cavity, in a clumpy medium the shock finds the way around the clumps, partially avoiding to slow down. The luminosity of such shocks is investigated in this paper.

In Section~\ref{sec:model_emission} we present the model of radio synchrotron emission at an SNR shock, while the simulation details and setups are given in Section~\ref{sec:simulation_setups}. The semi-analytical model of HD evolution and emission is presented in Section~\ref{sec:semi-analytical_model}. The results, discussion, and conclusions are given in Sects.~\ref{sec:results}, \ref{sec:discussion}, and \ref{sec:conclusions}, respectively.

\section{Model of emission} \label{sec:model_emission}

In a non-linear diffusive shock acceleration process \citep[DSA mechanism, see][]{BlandfordOstriker1978, Axfordetal1977, Krymsky1977, Bell1978a} the accelerated particles give significant contribution to the pressure, affecting the shock structure, making it more compressed than a factor of 4 (for $\gamma=5/3$). As a crucial result the spectrum of the particles is changed, becoming more concave-up at the tail of distribution \citep{Blasietal2005}. Since these modifications cannot be obtained using the HD code, we work with the test-particle approximation (TPA) with inclusion of magnetic field amplification model. The total luminosity is calculated as a sum of test-particle radiation over the entire shock, and as such it is implemented in HD simulations. Although this seems like a rough approximation, most of the radio SNR spectra are in the plain power-law form, suggesting weak modification, especially in ST phase. For more details on the radio spectra of SNRs, see \citet{Urosevic2014}. Also, for recent summary of radio spectral indices in SNRs, see \citet{RanasingheLeahy2023}.

For a power-law distribution of accelerated electrons $N(E)dE = KE^{-\Gamma}dE$, which is implied in TPA, the emissivity of synchrotron radiation in cgs units is \citep{Pacholczyk1970}:
\begin{equation} \label{eq:emissivity}
	\varepsilon_\nu = c_5 KB^{\alpha+1}_2 \left( \frac{\nu}{2c_1} \right)^{-\alpha} \, [\textrm{erg s}^{-1}~\textrm{Hz}^{-1}~\textrm{cm}^{-3}~\textrm{sr}^{-1}],
\end{equation}
where $E$ is particle energy, $B_2$ is downstream magnetic field, $\alpha = (\Gamma-1)/2$ is the radio spectral index and $\nu$ is the radiation frequency ($\nu=1\,\text{GHz}$ in our model). The constants $c_1 = 6.27\cdot 10^{18}$ and $c_5=1.8786\cdot10^{-23}$ are defined in \citet{Pacholczyk1970}. The constant $K$ is \citep{Arbutina2017}:
\begin{equation} \label{eq:K}
	K =  N(\Gamma-1)(p^{\text{e}}_{\text{inj}}c)^{(\Gamma-1)},
\end{equation} 
where $N=\eta_{\text{e}} n_2$ is the total number of accelerated particles per unit volume, $p^{\text{e}}_{\text{inj}}$ is the injection momentum of electrons above which they accelerate and $c$ is the speed of light. The acceleration efficiency $\eta_{\text{e}}$, is the fraction of particles that crossed the shock (with downstream concentration $n_2$) taking the part in acceleration process.

We assume the equal injection momenta of electrons and protons, $p^{\text{e}}_{\text{inj}} = p^{\text{p}}_{\text{inj}}$, and the electron-to-proton number ratio $\text{K}_{ep}=10^{-2}$ based on the observation of Galactic cosmic rays (CRs) \citep{BeckKrause2005}, which implies $\eta_{\text{e}}=\eta_{\text{p}}10^{-2}$. Following the recipe for injection of particles from thermal pool from \citet{Blasietal2005}, $p^{\text{p}}_{\text{inj}}$ is obtained relative to the proton momentum at the thermal peak of the Maxwellian distribution in downstream region:
\begin{equation} \label{eq:p_inj}
	p^{\text{p}}_{\text{inj}} = \xi p^{\text{p}}_{\text{th}},~
	p^{\text{p}}_{\text{th}} = \sqrt{2m_{\text{p}}k_{\text{B}}T^{\text{p}}_2},
\end{equation}
where $\xi\sim3$--4 is the injection parameter that determines the momentum above which all the particles will be accelerated as CRs and $T^{\text{p}}_2$ is downstream proton temperature. Although the thermal pool injection is not very realistic, the same assumption was applied in \citet{ArbutinaZekovic2021} for shock-drift or micro-DSA preacceleration of protons and electrons. For a test-particle case $\xi \approx 4$ and $T^{\text{p}}_2$ is derived from Rankine-Hugoniot relations \citep{TidmanKrall1971}:
\begin{equation} \label{eq:T2}
	k_{\text{B}}T^{\text{p}}_2 = \frac{1}{X} \left( \frac{1}{\gamma M^2} + 1 - \frac{1}{X} \right) m_{\text{p}} v^2_{\text{s}},
\end{equation}
where $M$ is sonic Mach number and $X$ is the compression ratio:
\begin{equation} \label{eq:X}
	X = \frac{\rho_2}{\rho_1} = \frac{(\gamma+1)M^2}{2+(\gamma-1)M^2}.
\end{equation}
The proton injection efficiency highly depends on $\xi$ due to the exponential behavior of the Maxwellian distribution at high momenta \citet{Blasietal2005}:
\begin{equation} \label{eq:eta_p}
	\eta_{\text{p}} = \frac{4}{3\sqrt{\pi}}(X-1) \xi^3 \text{e}^{-\xi^2},
\end{equation}
but for a fixed injection parameter it depends only on shock compression. Now we know all the constituents of Eq.~(\ref{eq:K}).

For the magnetic field (MF) amplification model we adopt the approach of \citet{Sarbadhicaryetal2017} where the amplified MF in the upstream region incorporates two regimes: the resonant, which dominates the late stages of the SNR evolution when the shock weakens, and the non-resonant, which dominates the earlier stages at large Mach numbers.\footnote{We use the model of \citet{Sarbadhicaryetal2017} with the caveat, given that \citet{Leahyetal2022} have shown that it fails to correctly reproduce observed radio-luminosities for 54 analyzed Galactic SNRs.} The upstream MF $B_1$ depends on the energy density of CRs $\epsilon_{\text{CR}}$, the Alfv\'en Mach number of the shock $M_{\text{A}}$, and the shock speed $v_{\text{s}}$ as \citep{Sarbadhicaryetal2017}:
\begin{equation} \label{eq:B_1}
	\frac{B_1^2}{8\pi} = \frac{\epsilon_{\text{CR}}}{2} \left( \frac{v_{\text{s}}}{c} + \frac{1}{M_{\text{A}}} \right).
\end{equation}
The first and second term in the parentheses give the contribution from non-resonant and resonant modes, respectively. Instead of Alfv\'en Mach number, here we use sonic Mach number, $M_{\text{A}}=M$, approximating their equality. Only the transverse components of $B_1$ is compressed downstream, so if the isotropic MF is assumed, the amplified MF downstream is:
\begin{equation} \label{eq:B_2}
	B_2 = \sqrt{\frac{1+2X^2}{3}}B_1.
\end{equation}

\begin{figure*}
	\includegraphics[width=0.321\textwidth]{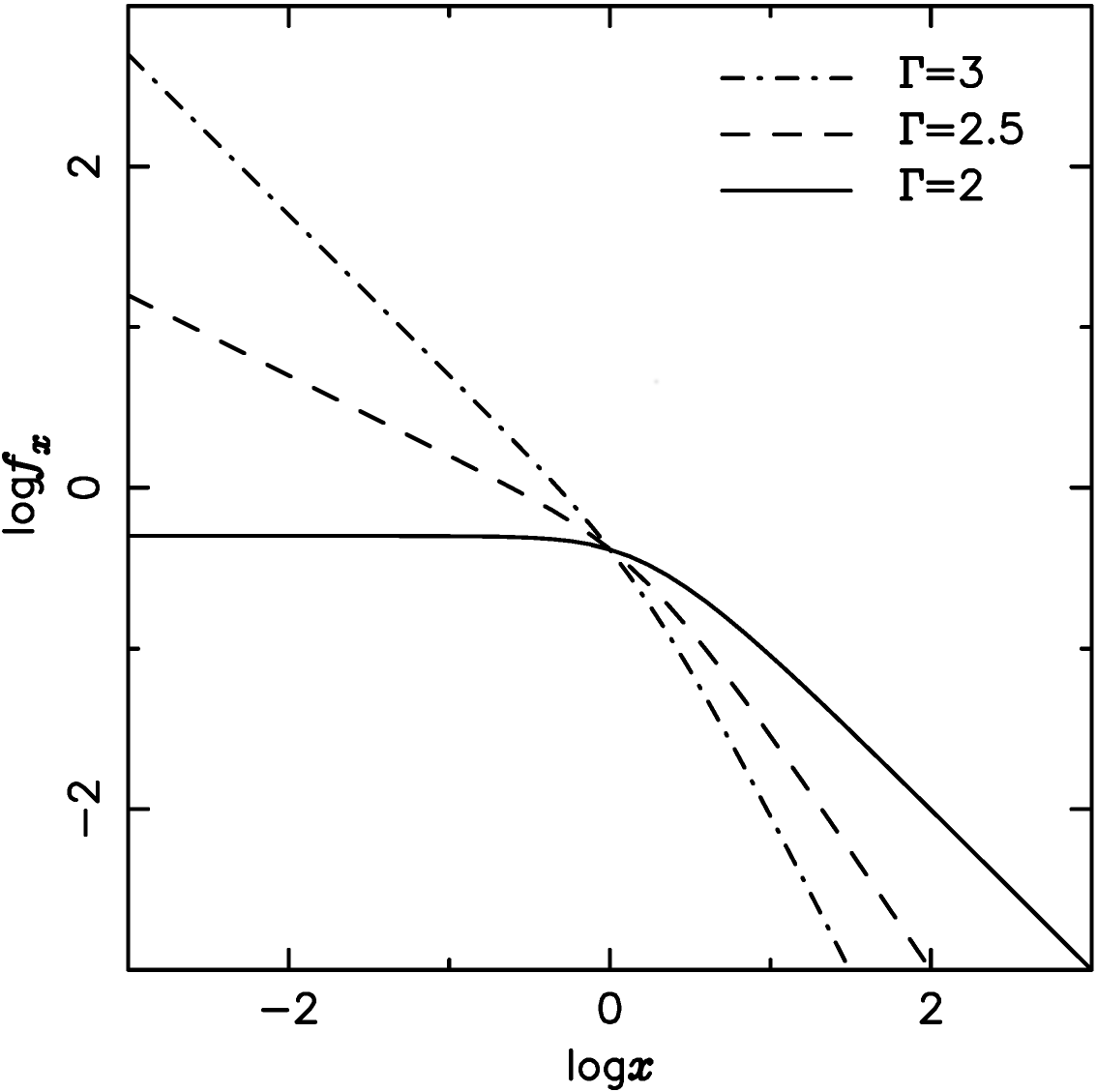}~~~
	\includegraphics[width=0.322\textwidth]{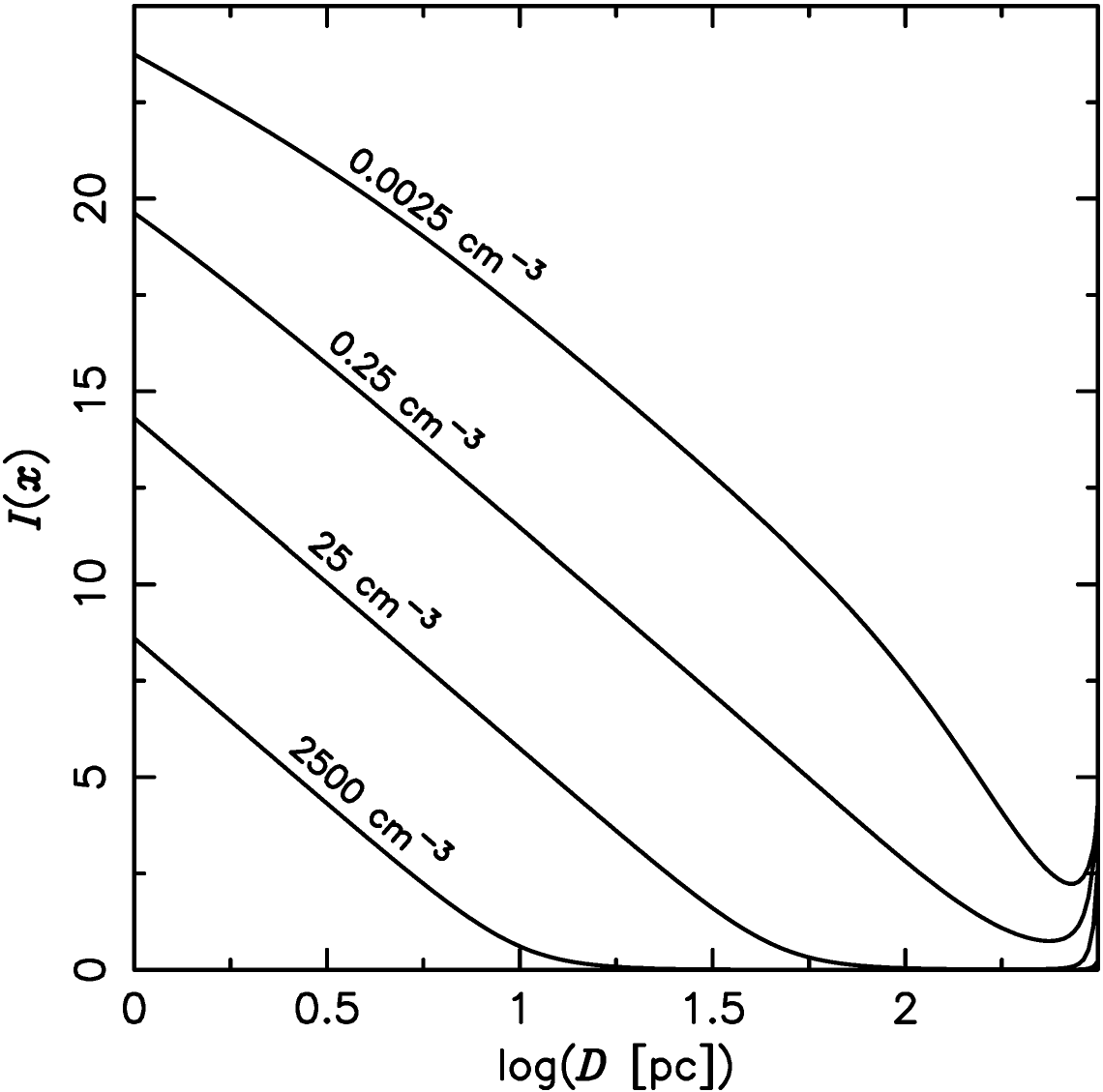}~~~
	\includegraphics[width=0.32\textwidth]{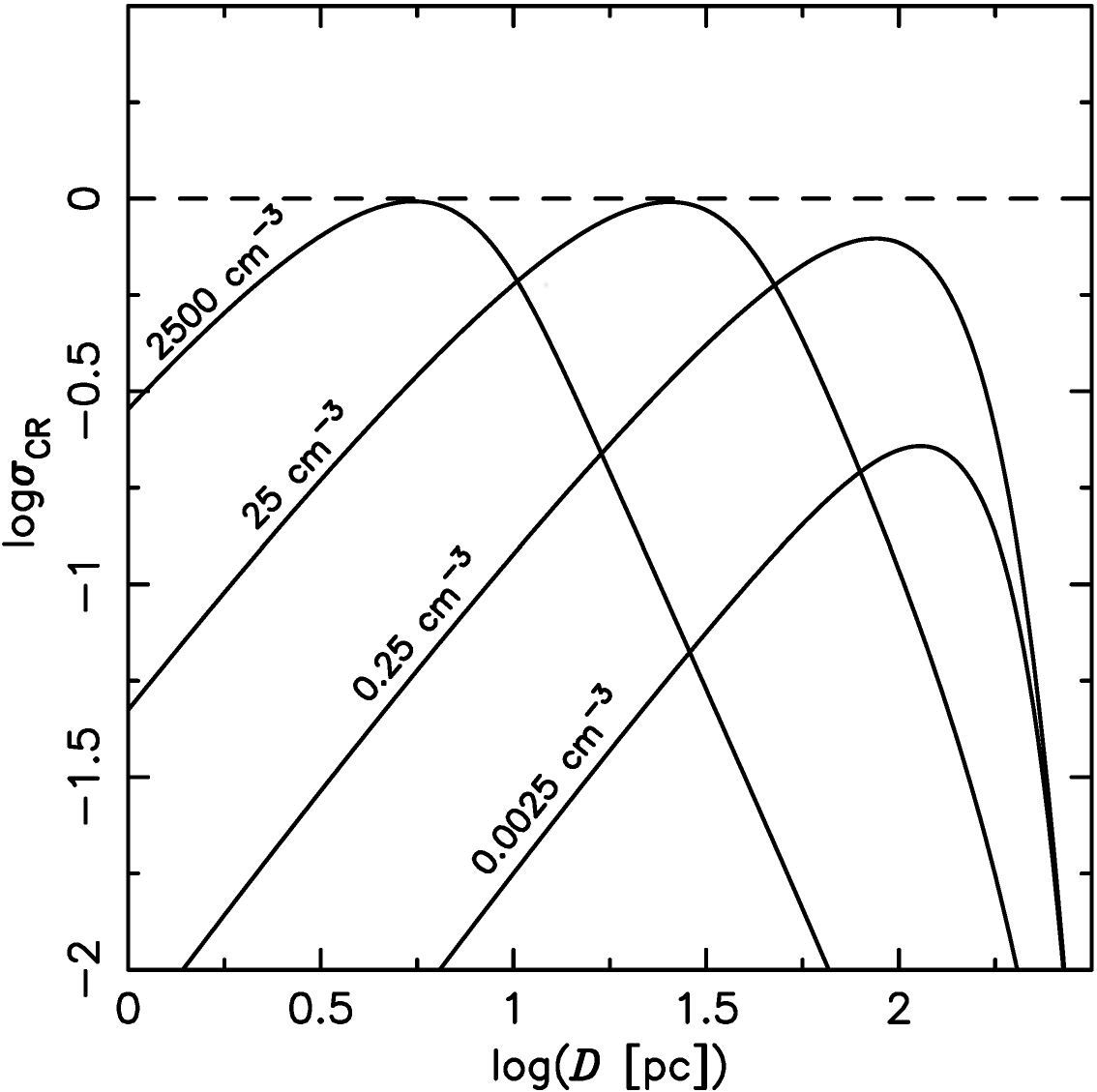}
	\caption{\textit{Left}. Function under the integral from Eq.~(\ref{eq:I(x)}), $f_x=x^{-\Gamma}\left(\sqrt{x^2+1}-1\right)$, for three values of $\Gamma$. The function represents the power-law momentum distribution of accelerated protons. The two slope regimes differentiate non-relativistic (slope $2-\Gamma$) and relativistic ($1-\Gamma$) particles. For a strong shock case ($\Gamma=2$) the shock accelerates more particles to relativistic energies than in weaker shocks. \textit{Middle}. Integral $I(x)$ as a function of SNR diameter, for four values of $n_0$. Its value mainly depends on the shock velocity (i.e. $x_{\text{max}}$). \textit{Right}. Ratio between CR and shock energy density, $\sigma_{\text{CR}}$, as a function of SNR diameter, for four values of $n_0$.}
	\label{fig:fx_ix_sig}
\end{figure*}

Assuming that protons dominate the CRs, their energy density is \citep{Arbutina2017}:
\begin{equation} \label{eq:e_cr}
	\epsilon_{\text{CR}} \approx \epsilon_{\text{p}} = \eta_{\text{p}} n_2(\Gamma-1)x_{\text{inj}}^{\Gamma-1}m_{\text{p}}c^2I(x),
\end{equation}
where $I(x)$ is the integral of the proton momentum in units of $m_{\text{p}}c$ with limits $x_{\text{inj}}$ and $x_{\text{max}}$,
\begin{equation} \label{eq:I(x)}
	I(x) = \int_{x_{\text{inj}}}^{x_{\text{max}}} x^{-\Gamma}\left(\sqrt{x^2+1}-1\right)\text{d}x,
\end{equation}
which we calculate numerically. The limits are:
\begin{eqnarray}
	\label{eq:x_inj}
	&&x_{\text{inj}} = \frac{p^{\text{p}}_{\text{inj}}}{m_{\text{p}}c}, \\
	\label{eq:x_max}
	&&\log x_{\text{max}} = 1+\frac{5}{2}\log \left( \frac{v_{\text{s}}}{100~{\text{km~s}}^{-1}} \right).
\end{eqnarray}
The approximation (\ref{eq:x_max}) for maximum proton momentum dependence on shock velocity is based on the work of \citet{PtuskinZirakashvili2003} and ensures that the fraction of CR pressure in shock ram pressure, $\sigma_{\text{CR}}\equiv\epsilon_{\text{CR}}/(\rho_0v_{\text{s}}^2)$, stays less than one. The left panel in Figure~\ref{fig:fx_ix_sig} shows the function under the integral, $f_x=x^{-\Gamma}\left(\sqrt{x^2+1}-1\right)$, that represents the momentum distribution of accelerated protons. The middle and right panel in Figure~\ref{fig:fx_ix_sig} show the evolution of $I(x)$ and $\sigma_{\text{CR}}$ with SNR diameter.

\begin{figure} \label{fig:emissivity}
	\centering
	\includegraphics[width=0.45\columnwidth]{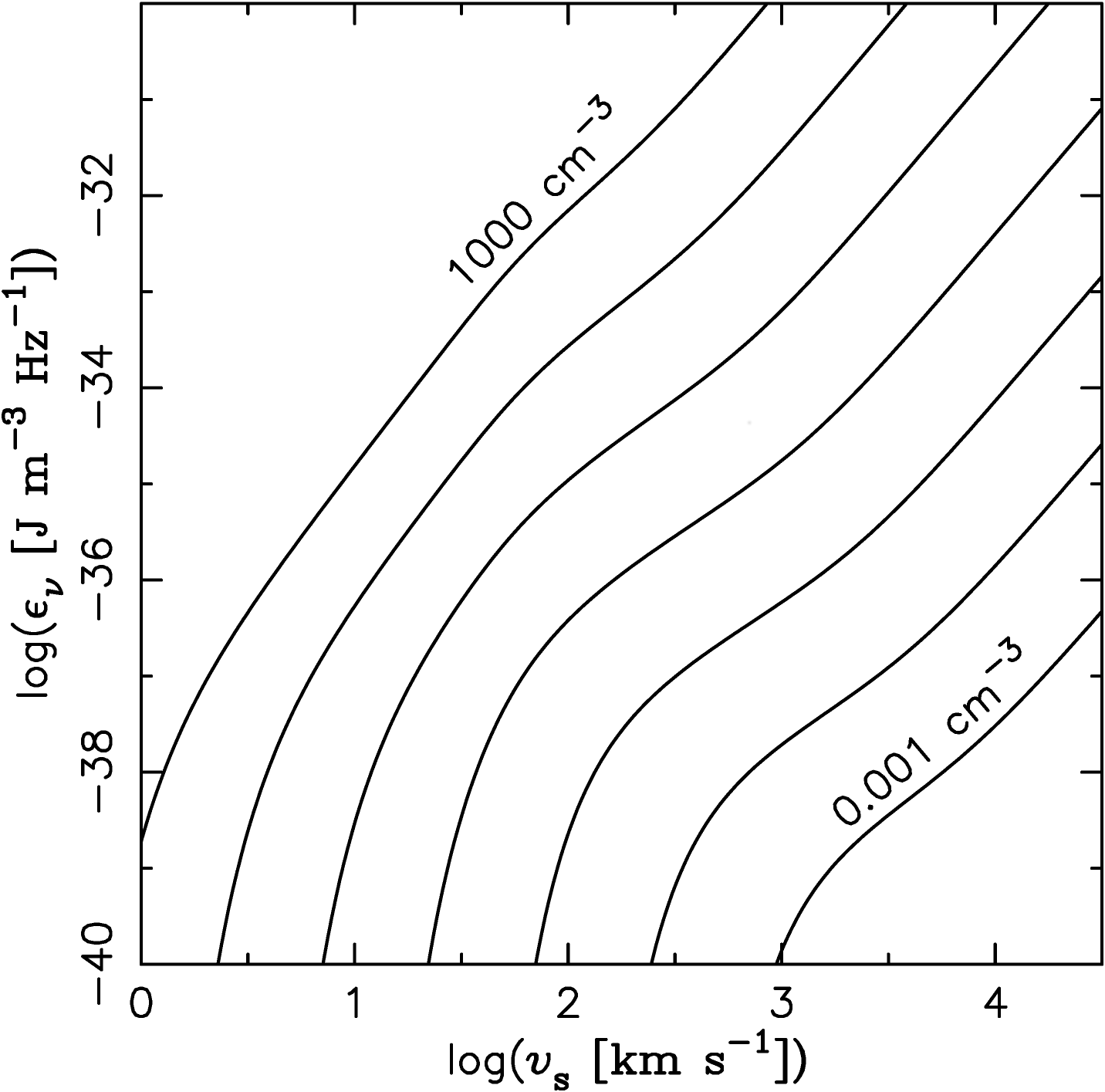}
	\caption{Emissivity dependence on the shock velocity for various ambient concentrations ($n_0$ for every neighboring line differs for one order of magnitude). The velocity range from 100 to 20\,000~km s$^{-1}$ covers the typical SNR shock velocities. For this calculation we use ambient pressure $p = 2500k_{\text{B}}$.}
\end{figure}

As can be seen from presented method, the calculation of emissivity only uses shock velocity $v_{\text{s}}$, particle concentration $n_2$, and Mach number $M$. Section~\ref{sec:shock_detection} presents the method of local detection of shock and its velocity. Figure~\ref{fig:emissivity} shows the emissivity dependence on $v_{\text{s}}$ for various values of ambient concentration $n_0$.

\section{Simulation setups}
\label{sec:simulation_setups}
All simulations are carried out using the HD code by \citet{Kostic2019}. The code uses MUSCL-Hahncock scheme with HLLC Riemann solver for upgrading states. It is purely hydrodynamic and conserves the values of density, momentum, and energy, without inclusion of the MF, nor the gas physics such as radiative cooling, thermal conduction or ionization.

\subsection{Initial conditions for supernova remnant} \label{subsec:init_SNR}

The 3D computational box has a 40~pc side with a resolution of 320 cells in all runs. The remnant is initially placed as an octant at the box corner and is initiated as in \citet{Kostic2019}, by placing the predefined Sedov profiles of density, momentum, and energy surrounded by a given model of the ISM. Considering the task of tracking the Sedov evolution of the SNRs (diameters of $\sim10$--80~pc) and finding the analytical model that supports the simulations, the ejecta mass is not included.\footnote{Although this approximation is poor at early stages in low densities, we use it from purely practical reasons of comparing the simulations to analytical HD model. The omitting of ejecta from the final analysis will be discussed in Section~\ref{sec:discussion}.} Although the observed SNR energies span more than two orders of magnitude \citep{Leahyetal2020}, we use the fiducial value of $E_0=10^{51}~\text{erg} = 10^{44}$~J in all simulations and models. The ambient density and initial SNR radius in simulations are set to $\rho_0=0.05~\text{cm}^{-3}$  and $R_0=2.5$~pc, respectively.

\subsection{Models of ISM}
\label{subsec:models_ISM}
The used ISM environments include uniform and clumpy medium of different density contrasts between clumps and interclump medium. Their physical properties are presented in the next three subsections.

\subsubsection{Uniform medium}
\label{sunsec:uni_model}
The density of the uniform medium is $\rho_0=0.05~\text{cm}^{-3}$. The pressure of the medium is set to a universal value of $p_0/k_{\text{B}}=2500~\text{K}~\text{cm}^{-3}$, which suits to the temperature of $50\,000$~K for $\rho_0$ given here. The medium velocity (momentum) is zero, as is for all other ISM models. This simulation, named U1, is used as a calibration model in all other simulations, for comparing simulated and theoretical results.

\subsubsection{Low-density bubble}
\label{subsec:LDB_model}
The low-density bubble (LDB) is a spherically symmetric construct and consists of two uniformly distributed density zones: one inside the radius $r=5$~pc ($\rho_{\text{l}}$), and the other outside ($\rho_{\text{h}}$), with the density 2, 5, and 10 times the inner value. The pressure is set to universal $p_0$. These simulation setups, named LDB2, LDB5, and LDB10, are used as hydrodynamic tests of the remnant, especially for the states of density, velocity, and energy at the shock. The results of these simulations will lead to analytical description of what is happening at shock after the discontinuity in density.

\subsubsection{Clumpy medium}
\label{sec:clumpy_model}
For the purpose of simple analytic modeling of the shock propagating through the clumpy medium, the clumps are modeled as non-overlapping spheres of uniform density. They are distributed randomly over a computational box and then a central sphere of 5~pc in radius around the star is cut out and replaced with a uniform density $\rho_0=0.05~\text{cm}^{-3}$. This is done to avoid the violation of rough spherical symmetry. However, the strong discontinuity between the uniform and clumpy environment is also needed to make the simulation results comparable to semi-analytical model. The radius value of 5~pc is arbitrary, as it will be the free parameter in modeling of the samples in Section~\ref{sec:discussion}.

The clumpy medium consists of two phases: the interclump medium (ICM) of the same density as the uniform medium at $<5$~pc radius ($\rho_{\text{icm}}=\rho_0$), and the ensemble of identical clumps of density $\rho_{\text{c}}>\rho_{\text{icm}}$. The three setups that differ in clump density will be presented: C10, C30, and C100 that have clump densities 10, 30, and 100 times higher than the ICM, respectively. These values are chosen so the wide range of possible density contrasts is covered. We run the simulations with all clump densities for $f_{\text{c}}=0.1$ and 0.25 to observe the dependence on volume-filling factor, whereas these setups get the suffix F10 and F25, respectively. The number of clumps is chosen to be high enough, so their random distribution could fill the volume of the computational box as even as possible, with the goal of having a nearly constant radial distribution of the average density. However, the clump's radius should be large enough so the individual clumps cover enough computational cells. With the clump's radius of $r_{\text{c}}\approx1.03$~pc (or $\approx8.24$ cells), the total number of the clumps in the computational box is 1400 in F10 mode, and 3500 in F25 mode. Although the number of clump covering cells seems small for simulating shock-clump interaction, we are not interested in resolving the hydrodynamic instabilities at the clump boundaries, but hold to a simpler approach where clumps are just being compressed by the shock, for which the given resolution is sufficient. The setup parameters are listed in Table~\ref{tab:setups}. For the purpose of determination of the variables from Sects.~\ref{sec:vjeff} and \ref{sec:exponent_m} the simulations with setups C3 and C300 were carried out, but these results are omitted for the sake of brevity.

\begin{table}
	\label{tab:setups}
	\centering
	\caption{Simulation setups.}
	\begin{tabular}{llcc}
		\hline\hline
		Model & Medium & \multicolumn{2}{c}{Density/[cm$^{-3}$]} \\
		\hline
		U1 & Uniform & \multicolumn{2}{c}{$\rho_0 = 0.05$} \\
		\hline
		& & $\rho_{\text{l}}~(R<5~\text{pc})$ & $\rho_{\text{h}}~(R\ge5~\text{pc})$ \\
		LDB2 &			& $0.25$ & $2\rho_{\text{l}}$ \\ 
		LDB5 & Uniform	& $0.25$ & $5\rho_{\text{l}}$ \\ 
		LDB10 & 	 	& $0.25$ & $10\rho_{\text{l}}$ \\ 
		\hline
		& & $\rho_{\text{icm}}~(\text{and}~R<5~\text{pc})$ & $\rho_{\text{c}}~(R\ge5~\text{pc})$ \\
		C3 &						 & $0.05$ & $3\rho_{\text{icm}}$ \\
		C10 & 						 & $0.05$ & $10\rho_{\text{icm}}$ \\
		C30	&	Clumpy				 & $0.05$ & $30\rho_{\text{icm}}$ \\
		C100	&					 & $0.05$ & $100\rho_{\text{icm}}$ \\
		C300	&					 & $0.05$ & $300\rho_{\text{icm}}$ \\
		\hline
	\end{tabular}
	\tablecomments{All clumpy medium setups have a suffix, depending on volume-filling factor of the clumps: F10 for $f_{\text{c}}=0.1$, or F25 for $f_{\text{c}}=0.25$.}
\end{table}

\subsection{Shock wave detection} \label{sec:shock_detection}
The total SNR luminosity is calculated as a sum of local cell luminosities covered by the shock wave. The algorithm detects the representative shock cells and use their velocity to find the local shock velocity (Section~\ref{sec:shock_velocity}). The emission comes from local volume elements, which are derived in Section~\ref{sec:shock_volumes}.

\subsubsection{Local shock velocity} \label{sec:shock_velocity}
The shock wave in simulations is found locally at cells where the momentum is maximal in direction of velocity vector. Numerically, this is done by finding (marking) the cells across which the quantity $\text{d}p/\text{d}v$ changes sign. If the marked cells are side-neighboring, and their velocity directions partially point to each other, then the cell with lower velocity is unmarked. The point of this clearing is to rarefy the shock points (marks) as much as possible, which is needed for the shock volume estimation method. Also, this keeps only the shock points with maximal $v_2$, to ensure the higher precision in evaluating the shock velocity. The cells from the interior of the SNR volume are left out from the algorithm, so the reflected shocks could not be detected. Thus, only the ICM and CL shocks (belonging to forward shock) are detected.

From Rankine-Hugoniot condition:
\begin{equation}
	v_{\text{s}} = \frac{X}{X-1}v_2,
\end{equation} 
we can derive the shock compression as a function of $v_2$ and ambient sound speed $a_0$, using Eq.~(\ref{eq:X}) and $M = v_{\text{s}}/a_0$:
\begin{equation}
		X = \frac{f^2+2/3+\sqrt{f^2+4/9}}{f^2+1/3},~~~ f = a_0/v_2,
\end{equation}
where $\gamma=5/3$ is implied. Due to numerical cutting of $v_2$, the precision of this method depends on the resolution of simulation.

\subsubsection{Volume of detected shock points} \label{sec:shock_volumes}
\begin{figure}
	\centering
	\includegraphics[width=0.95\columnwidth]{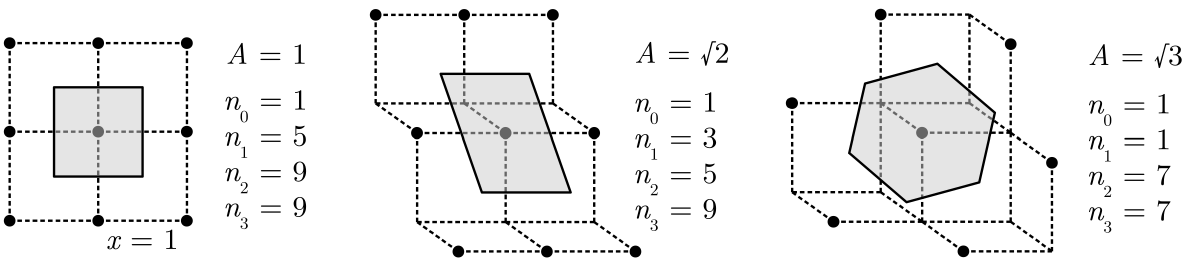}
	\caption{Illustration of shock points in the three referent cases of shock propagation relative the grid. The dashed lines connect the centers of the neighboring grid cells and the black dots represent the detected shock cells. The surface of the shock is divided between the shock points, where the area belonging to one point ($A$, in units of $x^2$) is shaded gray. The number of shock points inside the circles of radii 0, 1, $\sqrt{2}$ and $\sqrt{3}$ are presented as $n_0$, $n_1$, $n_2$ and $n_3$.}
	\label{fig:shock_points}
\end{figure}

The emitting volume around the detected shock points is calculated as the product of the estimated area and the thickness of the emitting shell at that point (for the purpose of obtaining the observed span of surface brightnesses, we use $\Delta R=10^{-2}R$; the explanation for this value is given in Section~\ref{sec:mw_models}). The sum of local areas over the entire shock should give its total  surface. The local area of certain shock point (cell) is estimated using the number of neighboring points at distance of $x\sqrt{3}$, $x$ being the cell's edge length. The concept is illustrated on Figure~\ref{fig:shock_points}, showing three representative cases of the shock traveling in characteristic directions relative to the grid.

The first case is a flat shock traveling in parallel direction to a grid axis (e.g. $z$-axis). Here, every cell in the $x$-$y$ plane is marked as a shock point. The belonging area to every shock point is $x^2$. In the second case the shock travels in the direction normal to one of the grid axes, say $z$-axis, but by the angle of 45 degrees from $x$- and $y$-axis, so the belonging area is $\sqrt{2}x^2$. For the case where the direction of the propagation forms equal angles with all three axes, the area is $\sqrt{3}x^2$. From now on, $x=1$ for easier derivation.

The mathematical concept is that the spatial frequency of points that are up to $\sqrt{3}$ away (including the examined point) can give us the sought area. The problem is that the number of points can be the same, e.g. in cases 1 and 2 (nine points), with different areas. So, we need to calculate the weighted frequency among the distance from the examined point. These distances take only discrete values, namely $1$, $\sqrt{2}$, and $\sqrt{3}$. Assuming that points lie in the same plane (which is not generally true but the deviations, when exist, are small), the weighted surface frequency of the points is:
\begin{equation}
	\label{freq_int}
	f_{\text{w}} = \int_{A_{\text{min}}}^{\pi}\frac{n_0}{A}\text{d}A + \int_{\pi}^{2\pi}\frac{n_1}{A}\text{d}A + \int_{2\pi}^{3\pi}\frac{n_2}{A}\text{d}A + \int_{3\pi}^{A_{\text{max}}}\frac{n_3}{A}\text{d}A,
\end{equation} 
where $n_0=1$ is the examined point, while $n_1$, $n_2$ and $n_3$ are the total number of points inside the radii $1$, $\sqrt{2}$ and $\sqrt{3}$. The boundary values for the area $A$ are the encircled areas for the given radii, $A_{\text{min}}$ and $A_{\text{max}}$ being unknown. This integral reduces to:
\begin{equation}
	f_{\text{w}} = f_0 + l_1n_1 + l_2n_2 + l_3n_3,
\end{equation} 
where $f_0$ and $l_3$ are obtained numerically by minimizing the product of absolute deviations:
\begin{equation}
		\min\left(
		\left|\frac{f_1/f_2 - \sqrt{2}}{\sqrt{2}}\right|
		\left|\frac{f_1/f_3 - \sqrt{3}}{\sqrt{3}}\right|
		\right),
\end{equation} 
calculated for the system:
\begin{eqnarray}
		&&f_1 = f_0 + 5l_1 + 9l_2 + 9l_3, \\
		&&f_2 = f_0 + 3l_1 + 5l_2 + 9l_3, \\
		&&f_3 = f_0 + l_1 + 7l_2 + 7l_3,
\end{eqnarray} 
where $f_1$, $f_2$ and $f_3$ are the weighted frequencies for the cases 1, 2 and 3. The result is:
\begin{equation}
	f_0 = -0.252500, \text{ }
	l_3 = 0.378700,
\end{equation} 
for which the ratios of frequencies (that are inverse to the ratios of areas) are:
\begin{eqnarray}
		&&f_1/f_2 = 1.414200\text{ }(1.414214),\\
		&&f_1/f_3 = 1.732051\text{ }(1.732051),\\
		&&f_2/f_3 = 1.224757\text{ }(1.224745). 
\end{eqnarray} 
In the parentheses are the values $\sqrt{2}$, $\sqrt{3}$ and $\sqrt{3/2}$, the ratios for the three cases. The frequencies for all other directions of shock propagation are in between these three cases. After calibrating the frequencies to the value of $f_1$, we have the belonging area at $i$-th shock point (in units of $x^2$):
\begin{equation}
	A_i = \frac{f_1}{f_0 + l_1n_{1i} + l_2n_{2i} + l_3n_{3i}},
\end{equation} 
which multiplied by the thickness of the shell gives the volume for the given shock point:
\begin{equation}
	V_i = A_ix^2\Delta R_i.
\end{equation} 
The values of $l_1$, $l_2$ and $f_1$ are: $l_1 = \ln2$, $l_2 = \ln(3/2)$, and $f_1 = 10.270722$. 

\section{Semi-analytical model of hydrodynamic evolution and emission} \label{sec:semi-analytical_model}

The semi-analytical model will be derived and tested against the simulation results for the ISM setups from Section~\ref{sec:simulation_setups}:
\begin{enumerate}
	\item Uniform medium. This is the trivial case, where the Sedov law of expansion in a constant ambient density is used.
	\item Low-density bubble (LDB). This is the test of the HD analytics after the discontinuity in density.
	\item Clumpy medium. These are the main simulations of this paper that are given the most attention.
\end{enumerate}
The prefix ``semi-'' in semi-analytical stands because some variables of the model must be calculated numerically, such are the outputs from Riemann solver.

\subsection{Transition from the lower to higher density -- LDB model}
\label{sec:LDB}
During the ST phase, in the case of the strong shock ($X=4$) and $\gamma=5/3$, the energy density at the shock is:
\begin{eqnarray}
	\label{E_2a}
	E_2 &=& \frac{1}{2}\rho_2v^2_2 + \frac{p_2}{\gamma-1} \\
	\label{E_2b}
	&=& \frac{9}{8}\rho_0v^2_{\text{s}} + \frac{9}{8}\rho_0v^2_{\text{s}} \\
	\label{E_2c}
	&=& \frac{9}{4}\rho_0v^2_{\text{s}},
\end{eqnarray}
or after some algebraic manipulations using the Sedov law of expansion:
\begin{equation}
	\label{eq:E_uni}
	E_{2}^{\text{uni}} = \frac{9}{25}1.15^5R^{-3}E_0.
\end{equation}
So, independently of the ambient density, as long as it is uniform (hence the index ``uni''), $E_{2}^{\text{uni}}$ depends only on $E_0$ and is proportional to $R^{-3}$. However, in transition from lower to higher density, the shock density, velocity, and pressure undergo a jump according to Riemann problem solution \citep[see][]{Toro2009}. Since $p_2=E_2/3$, the jump in the energy and pressure is the same, $E_{\text{j}}=p_{\text{j}}=k_{\text{j}}$,
\begin{equation}
	k_{\text{j}}E_2 = \frac{1}{2}\rho_{\text{j}}\rho_2(v_{\text{j}}v_2)^2 + \frac{k_{\text{j}}p_2}{\gamma-1}.
\end{equation}
where $\rho_{\text{j}}$ is the ambient (and shock) density jump at discontinuity and $v_{\text{j}}$ is the velocity jump. Combining this with Eq.~(\ref{E_2a}) we obtain the link between the three jumps:
\begin{equation}
	\label{eq:kj}
	k_{\text{j}}= \rho_{\text{j}} v_{\text{j}}^2.
\end{equation}
The velocity jump can also be obtained by using an analytic approximation from \citet{Borkowskietal1997}.

\begin{figure}
	\centering
	\includegraphics[width=0.45\columnwidth]{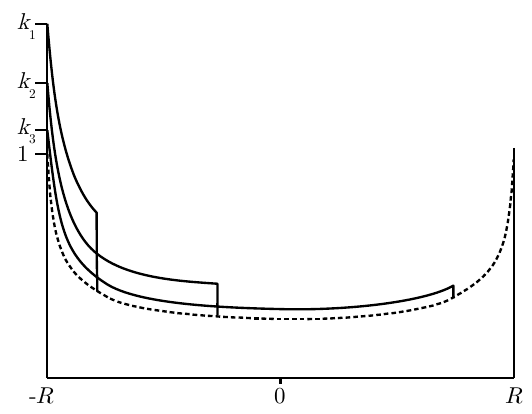}
	\caption{Illustration shows the wave that redistributes the compressed energy from the SNR forward shock back through its volume, after hitting the shell of higher density region (in LDB setup). Because of spherical symmetry, only the 1D radial energy distribution from one side is shown. While the SNR is in (lower) uniform ambient density, the energy is distributed as Sedov profile, depicted with the dashed line. Arrival of the discontinuity of a higher density brings a rise in energy density at the shock, $E_2=kE_2^{\text{uni}}$, which is maximal at the moment of hitting the discontinuity, but gradually decreases as the energy redistributes through the remnant's volume. The identical wave from the other side (not depicted in the picture) interferes with the one presented. At some point the radial energy distribution restores to the self-similar Sedov profile, at least near the forward shock, and so the factor $k$ returns to 1.}
	\label{fig:energy_distribution}
\end{figure}

Eq.~\ref{eq:kj} enables us to find the shock velocity from the evolution of shock energy, which is determinable. By entering the higher density ($\rho_{\text{h}}$), the forward shock slows down ($v_{\text{j}}<1$) and its energy and density rise ($k_{\text{j}},\rho_{\text{j}}>1$). Behind the shock, the compressed energy expands inwards, tending to restore the self-similar Sedov profile. This is illustrated on Figure~\ref{fig:energy_distribution} by the three successive moments after the transition. Following the remnant's expansion into higher density, the initial energy jump $kE_2^{\text{uni}}$ gradually restore to the $E_2^{\text{uni}}$ value. 

However, according to Riemann problem (RP), when shock wave hits the higher density, two shock waves arise from this collision: the one that travels forward, which is described by the so-called right star state, $E_{\text{R}}^*=k_{\text{j}}E_{\text{L}}$, and the reflected one that travels backward (the left star state, $E^*_{\text{L}}$). The two RP shocks move to the left and right from contact discontinuity as constant states, because the left state (the initial flow) is also constant. However, the flow in SNR is not constant along the radius, but velocity and energy have a steep fall behind the shock, because there is a finite amount of energy that fills the expanding SNR volume. As a consequence, the reflected shock rapidly accelerates inward taking the part of the shock energy density with itself, while the forward shock energy becomes reduced by that part. If we denote the reflected part of the shock energy with $E_{\text{ref}}$ and the part remained in forward shock with $E_{\text{for}}=1-E_{\text{ref}}$, then the true energy jump at the discontinuity is $E_{\text{for}}k_{\text{j}}$.

The shock energy, first increased by compression, gradually decreases, but not to the $E_2^{\text{uni}}$ value, which would be expected, but below it because of the removed (reflected) energy. Nevertheless, when the reflected shock reaches the other side of the remnant, it brings the removed energy back, so the energies reunite at the forward shock to $E_2^{\text{uni}}$ value. The evolution continues as in the uniform case of $\rho_{\text{h}}$.

In a simple approximation, we can assume that the reflected shock stays at the position where the density discontinuity was, and that the filling of the remnant's volume with compressed shock energy follows merely from the expansion of the remnant. Assumingly, before the shocks reunite, the forward shock energy should decrease from $E_{\text{for}}k_{\text{j}}E_2^{\text{uni}}$ toward $E_{\text{for}}E_2^{\text{uni}}$, with excess energy density being proportional to the volume ratio $(R_{\text{d}}/R)^3$. However, the reflected shock does not rest at $R_{\text{d}}$ but rapidly accelerates inward, creating even steeper energy and velocity fall behind the shock (some kind of ``velocity vacuum''). This leads to faster decrease in shock energy compared to the volume ratio $(R_{\text{d}}/R)^3$. Hence, we can express the factor $k=E_2/E_2^{\text{uni}}$ over the radius as:

\begin{equation}
	\label{eq:k(R)for}
	k(R)=
	\left\{
	\begin{array}{ll}
	&E_{\text{for}}+E_{\text{for}}(k_{\text{j}}-1)(R_{\text{d}}/R)^m,~R_{\text{d}}\le R <R_{\text{u}},\\
	&1,~R\ge R_{\text{u}},\\
	\end{array}\right..
\end{equation}
where $R_{\text{u}}$ is the radius at which reflected and forward shock reunite. The exponent $m>3$ depends on $\rho_{\text{j}}$, because higher density jumps result in faster reflected shock. The solutions for $m$ and $R_{\text{u}}$ are derived from simulations. The reflected part of energy is derived using the exact Riemann solver as $E_{\text{ref}}=(E^*_{\text{L}}-E_{\text{L}})/E^*_{\text{L}}$.

Now, the energy density after the discontinuity is:
\begin{equation}
	E_2(R) = k(R)E_{2}^{\text{uni}}= \frac{9}{4}\rho_{\text{h}}v_{\text{s}}^2(R),
\end{equation}
and the shock velocity:
\begin{equation}
	v_{\text{s}}(R) = \sqrt{\frac{4}{9}\frac{E_2(R)}{\rho_{\text{h}}}}.
\end{equation}

The emissivity of the shell at any $R$ is calculated using $\rho(R)$ and $v_{\text{s}}(R)$. The $\Sigma$--$D$ relation of this model is shown in Figure~\ref{fig:LDB} in Section~\ref{sec:results}.

\subsection{Shock propagation through the clumpy medium}

\subsubsection{Presumptions}
The semi-analytical model of the hydrodynamic evolution of SNRs in the presence of clumpy medium is a spherically-symmetric 3D model that deals with the environment properties and shock geometry. The model finds the density and shock velocity fractions over the shock surface. Then, by applying the synchrotron emission model over the entire shock wave, we can calculate the total luminosity of the remnant. The aim is to derive the $\Sigma$--$D$ relation of an SNR in a given environment, without running the HD simulations.

The clumpy medium is approximated with two homogeneous phases: clumps (CL) and interclump medium (ICM), as in simulations. In order to analytically approach the development of the model, the following assumptions/conditions are needed:
\begin{itemize}
	\item the clumps are distributed evenly and isotropically with respect to the SNR center,
	\item the clumpy medium starts after some radius $R_{\text{d}}$, sufficiently larger than the clump radius.
\end{itemize}
These conditions are needed to ensure pseudo-sphericity and symmetry of the SNR, because all model variables are linked to its radius. The SNR radius is related to its volume $V$ as $R=(3V/4)^{1/3}$. The same conditions are applied in simulations.

\subsubsection{The shock velocity in pseudo-uniform medium}
\label{sec:pseudo-uniform_velocity}
The ambient density fluctuations affect the local shock velocity, as well as its average expansion velocity. Let us imagine the shock wave propagating through the medium of large number of very small clumps. The material behind the shock is being homogenized by thermal conduction, clumps disintegration and evaporation, so the density, velocity and pressure distributions tend to remain in self-similar Sedov profiles. We refer to this type of medium as pseudo-uniform and assume the equivalence of the mean shock velocity in a uniform and pseudo-uniform medium of equal average density.

The average shock velocity $\bar{v}_{\text{s}}$ of pseudo-spherical SNR expanding in pseudo-uniform medium can be derived from the classical Sedov law of expansion:
\begin{equation} \label{eq:v_mean}
	\bar{v}_{\text{s}}(R) = \sqrt{\frac{4}{9}\frac{E_2^{\text{uni}}}{\bar{\rho}(\le R)}} = \frac{2}{5}1.15^{5/2}R^{-3/2}\sqrt{\frac{E_0}{\bar{\rho}(\le R)}},
\end{equation}
where $\bar{\rho}(\le R)$ is the average ISM density inside the volume of the remnant. By its definition, this approximation holds only at $R\gg R_{\text{d}}$, and it is used for determination of mean shock velocity in the limit $R\rightarrow\infty$, which is justified to a large degree by the simulations results.

\subsubsection{Model basics}
\label{sec:model_basics}
The basic idea of the model is explained using the trivial setup of clumps with density almost the same as ICM, but only slightly higher. In this case, the SNR shock is propagating both through the clumps and the ICM, approximately as a sphere. The fraction of the shock surface ``occupied'' by the clumps (CL shock) is equal as the volume-filling factor of the clumps, $f_{\text{c}}$. Therefore, the fraction occupied by the ICM (ICM shock) is $f_{\text{icm}}=1-f_{\text{c}}$. 

The mean CL shock velocity $v_{\text{c}}$ is lower than the mean ICM shock velocity $v_{\text{icm}}$, and their effective ratio $v^*_{\text{j}}$ is lower than the velocity jump $v_{\text{j}}$ from the Riemann solver. The average shock velocity $\bar{v}_{\text{s}}$ is the function of these two velocities and their corresponding fractions over the shock surface:
\begin{equation}
	\label{v1v2}
	\bar{v}_{\text{s}} = f_{\text{icm}}v_{\text{icm}} + f_{\text{c}}v_{\text{c}} = v_{\text{icm}}(f_{\text{icm}} + f_{\text{c}}v^*_{\text{j}}).
\end{equation}
Knowing the CL and ICM velocities, the emissivities of the CL and ICM shock, $\varepsilon_{\nu\text{c}}$ and $\varepsilon_{\nu\text{icm}}$, can be found. The total luminosity of such SNR is:
\begin{equation}
	\label{eq:luminosity}
	L = 4\pi R^2 \Delta R \left( f_{\text{c}}\varepsilon_{\nu\text{c}} +  f_{\text{icm}}\varepsilon_{\nu\text{icm}} \right),
\end{equation}
where $\Delta R$ is the thickness of the emitting region, for which we use $\Delta R=10^{-2}R$.

Equation (\ref{eq:luminosity}) demonstrates the main idea of how the luminosity in our model is calculated. However, in the case of high density contrast ($\rho_{\text{j}}\gg 1$), certain processes significantly alter the CL and ICM velocities, as well as their shock fractions, so they have to be included into a model. The most important processes are: 
\begin{itemize}
	\item Shock wave stays longer at dense clumps, forming shock ``islands'' around the clumps. As this leads to growth of total shock surface and its CL fraction, it also affects the CL shock velocity, which varies over the island surface.
	\item Reflection and redistribution of energy over the shock surface. Part of the energy that is reflected from the arriving clumps is being deposited both at the ICM shock and the surviving CL shock islands. This complex process also impacts the introduced effective velocity ratio $v^*_{\text{j}}$, which have to be determined semi-empirically. 
\end{itemize}
All of these effects eventually affect the final SNR luminosity, so the model must include: determination of CL and ICM shock surfaces, effective velocity ratio, and expansion exponent $m$. Some of them are derived analytically and some are determined empirically from simulations.

\subsubsection{Surface of the forward shock}
\label{sec:surface}
When the shock hits the spherical clump, it keeps moving through the clump with lower velocity, compressing it and simultaneously bending around it. In case of density contrast $\rho_{\text{j}}>10$, the shock eventually envelopes the clump from all sides, creating the CL shock island. These events increase its surface above the simple relation to the radius $S_0=4\pi R^2$ (hereafter, $S_0$ stands for a spherical surface). Also, the CL shock fraction becomes higher than $f_{\text{c}}$. Passing the clump, the ICM shock merges and continue to spread forward, tending to keep the spherical shape. It is important to notice that the SNR shock has a steep fall in density and velocity behind the shock front, so the destruction process of relatively large clumps (which are of interest here) is slower than in the case of planar shocks of constant distribution of density and velocity. Instead, the clumps that have sunk into a remnant's volume are further being shocked by high pressure and less energetic internal flows, until eventually overrun. This is also known as the clump/cloud evaporation and was discussed by \citet{McKeeOstriker1977}.  

\begin{figure*}
	\centering
	\includegraphics[width=0.47\columnwidth]{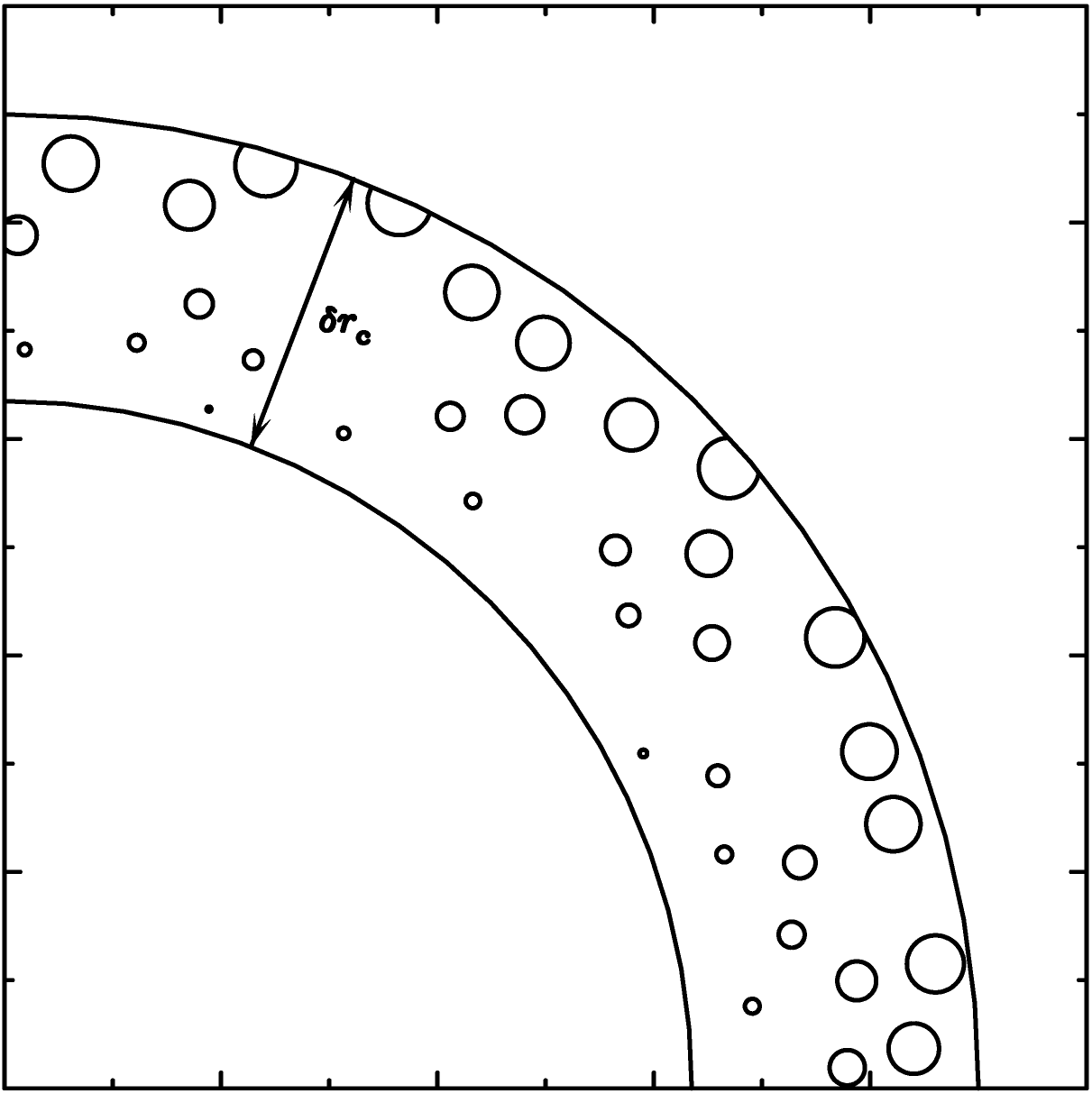}~~~~~
	\includegraphics[width=0.47\columnwidth]{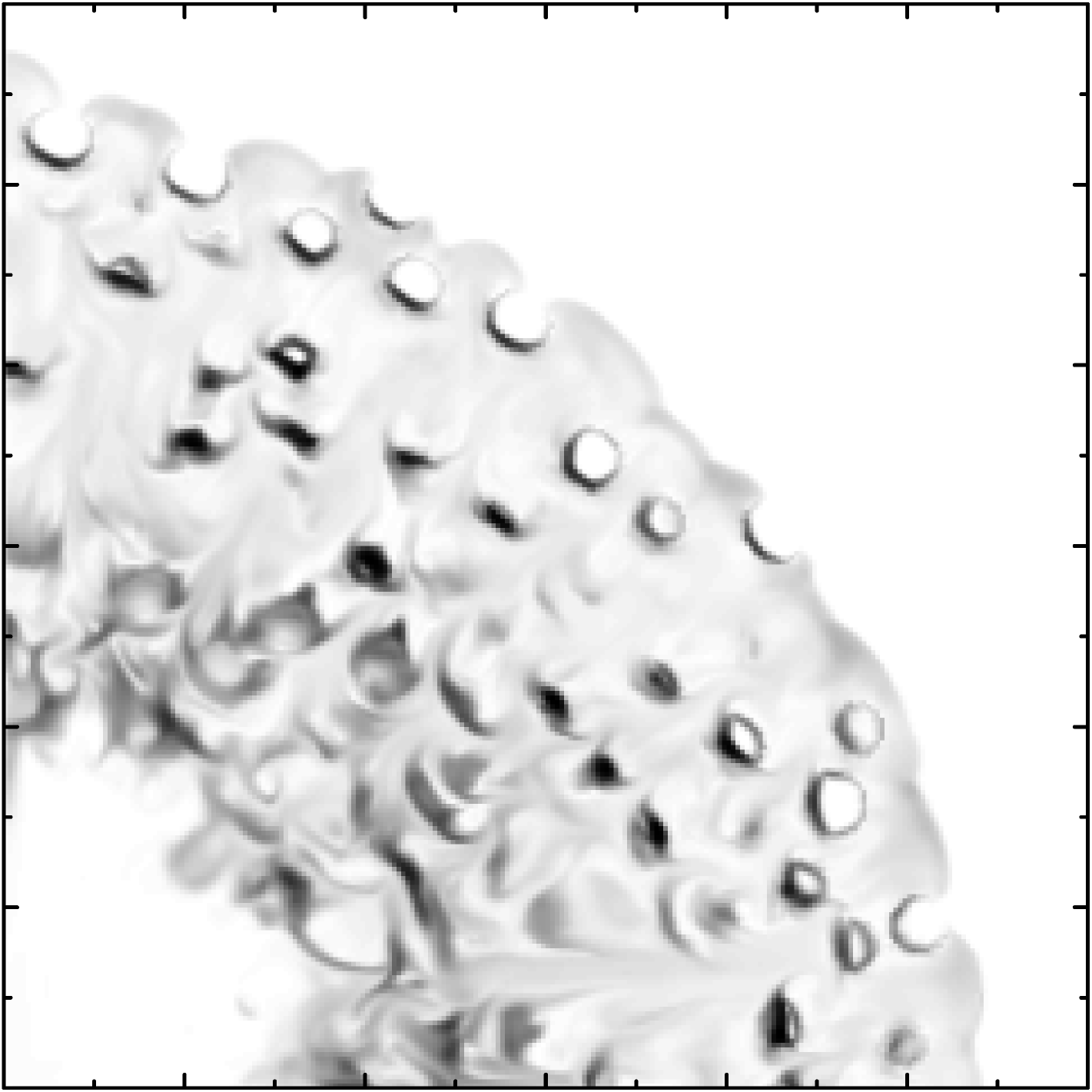}
	\caption{\textit{Left.} Illustrated model of surviving CL islands. Behind the ICM shock there is a shell containing clumps that are still not overrun by the shock. As the clumps sink deeper through the shell, their intact part gets smaller until it evaporates. The thickness of the shell is given by Eq.~(\ref{eq:delta}). \textit{Right.} C100F25 simulation snapshot showing the total momentum. The CL shock islands closely resemble the model at the left panel.}
	\label{fig:SCmodel}
\end{figure*}

Here, we will present the theoretical model for the calculation of the total shock surface $S$ and its modified CL fraction $F_{\text{c}}$, based on the simulation results. The important variable for this derivation is the maximal depth of the surviving CL islands within the SNR. In units of clump's radius it can be approximated as:
\begin{equation}
	\label{eq:delta}
	\delta = \frac{2(1-v_{\text{j}})}{v_{\text{j}}}.
\end{equation}
The length $\delta r_{\text{c}}$, which monotonically grows with the density contrast, represents the thickness of the shell containing the surviving CL islands (see Figure~\ref{fig:SCmodel}).

The increase in shock surface can be given by a factor $\kappa \equiv S/S_0$, which is the sum of the increased values of fractions $f'_{\text{c}}>f_{\text{c}}$ and $f'_{\text{icm}}>f_{\text{icm}}$ (which are related to $S_0$),
\begin{equation}
	\label{eq:kappa}
	\kappa = f'_{\text{c}} + f'_{\text{icm}}.
\end{equation}
At radius $R_{\text{d}}$, when the spherical shock hits the clumpy medium, $f'_{\text{icm}}=f_{\text{icm}}$ and $f'_{\text{c}}=f_{\text{c}}$. If $\delta\ll1$ the fractions keep these values through the evolution. However, in the case of higher density contrasts ($\delta>1$) the simulations show that both $f'_{\text{icm}}$ and $f'_{\text{c}}$ grow as the shock moves through the clumpy medium. The first one grows until it reaches $f'_{\text{icm}}\approx1$, because the ICM shock tends to be a sphere. This growth can be described with the empirical equation:
\begin{equation}
	\label{eq:ficm}
	f'_{\text{icm}} = f_{\text{icm}} + C_1Af_{\text{c}},
\end{equation}
where $C_1 = \min \left( \delta, 1 \right)$ is the empirical constant that has influence only at low density contrasts ($\delta<1$). The number $A$ is a cumulative fraction of the shock surface that has been in interaction with clumps before. At $R_{\text{d}}$ it is 0, but as the shock advances through the clumps, this number rises until it saturates to 1. With random clump distribution, the growth of $A$ is proportional to the rate of shock--clump interaction events ($n_{\text{c}}S_0\text{d}r$), the clump's cross-section ($r_{\text{c}}^2\pi/S_0$) and the unaffected fraction $1-A$,
\begin{equation}
	\label{eq:dA}
	\text{d}A = n_{\text{c}} r_{\text{c}}^2\pi (1-A) \text{d}r,
\end{equation}
where $n_{\text{c}}$ is the number of clumps per volume. The integration of this equation from $R_{\text{d}}$ to $R>R_{\text{d}}$,
\begin{equation}
	\label{eq:intA}
	\int_{0}^{A} \frac{\text{d}A}{1-A} = C_0\int_{R_{\text{d}}}^{R} \text{d}r,
\end{equation}
with $C_0=n_{\text{c}}r_{\text{c}}^2\pi = 3f_{\text{c}}/(4r_{\text{c}})$, yields:
\begin{equation}
	\label{eq:A}
	A = 1-\text{e}^{-C_0(R-R_{\text{d}})}.
\end{equation}

For the derivation of $f'_{\text{c}}$, we must describe the model of the CL islands. Figure~\ref{fig:SCmodel} illustrates a shell of surviving CL islands with high density contrast ($\rho_{\text{j}}=100,~\delta \approx 7.5$). The CL islands are approximated as spheres of radius that linearly decreases from $r_{\text{c}}$, at the first moment of interaction, to zero, at depth of $\delta r_{\text{c}}$ within a remnant. They are positioned at the places od progenitor clumps (i.e. motionless). Therefore, the total surface of CL islands can be calculated by integrating the product of corresponding island surface and the number of clumps through all the layers of the $\delta r_{\text{c}}$ shell:
\begin{equation}
	\label{eq:int_fc}
	S_0 f'_{\text{c}} = R\int_{r_0}^1 \frac{(r-r_0)^2}{(1-r_0)^2}4\pi r_{\text{c}}^2 n_{\text{c}} r^2 S_0 \text{d}r = \frac{3 f_{\text{c}} \delta S_0}{(1-r_0)^3} \int_{r_0}^1 (r-r_0)^2 r^2 \text{d}r,
\end{equation}
where $r_0 = 1-\delta r_{\text{c}}/R$ is the inner shell boundary. The solution to the last equation is:
\begin{equation}
	f'_{\text{c}} = 3f_{\text{c}} \delta \frac{r_0^2 + 3r_0 +6}{30}.
\end{equation}
This is a valid approximation for $\delta\gg1$, but to be valid for $\delta\ll1$, where $f'_{\text{c}}=f_{\text{c}}$, the factor $1+1/\delta$ must be added:
\begin{equation}
	\label{eq:fcII}
	f'_{\text{c}} = f_{\text{c}} (\delta+1) \frac{r_0^2 + 3r_0 +6}{10}.
\end{equation}

The evolution of $f'_{\text{c}}$ has two regimes: 1) $\delta r_{\text{c}} > R-R_{\text{d}}$, and 2) $\delta r_{\text{c}} \le R-R_{\text{d}}$. Equation (\ref{eq:fcII}) stands for the second regime. The first regime is characterized by faster growth of $f'_{\text{c}}$ because all CL islands are still surviving while new clumps are being shocked. The calculation of the total CL shock surface in this case is the same as in Eq.~(\ref{eq:int_fc}), but with integral limits from $r_{\text{d}} \equiv R_{\text{d}}/R$ to 1. The final solution for the regime 1) is:
\begin{equation}
	\label{eq:fcI}
	f'_{\text{c}} = f_{\text{c}} (\delta+C_2) \frac{10r_0^2\left(1-r_{\text{d}}^3\right)-15r_0\left(1-r_{\text{d}}^4\right)+6\left(1-r_{\text{d}}^5\right)}{10\left(1-r_0\right)^3},
\end{equation}
where $C_2=(1/3) (1-r_0)/(1-r_{\text{d}})+(2/3) (1-r_{\text{d}})/(1-r_0)$. For $r_{\text{d}} = r_0$, this equation becomes Eq.~(\ref{eq:fcII}).

Finally, the CL shock fraction is:
\begin{equation}
	\label{eq:Fc}
	F_{\text{c}} = \frac{f'_{\text{c}}}{\kappa}.
\end{equation}
Far away from discontinuity ($R\rightarrow \infty, r_0\rightarrow 0$), $F_{\text{c}}$ reaches its asymptotic value:
\begin{equation}
	\label{eq:Fc_infty}
	F_{\text{c}}^{\infty} = \frac{f_{\text{c}}(\delta+1)}{f_{\text{c}}(\delta+1)+f_{\text{icm}}+C_1f_{\text{c}}}.
\end{equation}

The SNR radius is defined as the ICM shock radius. However, as the CL islands are distributed at different radii, their effective radius must be calculated to scale the thickness of the emitting CL shock region. This is done numerically by finding $r_{\text{d}}$ in Eq.~(\ref{eq:fcI}), such that gives the result $f'_{\text{c}}/2$. Designating it with $r_{\text{eff}}$, the thickness of emitting CL shock is $r_{\text{eff}}\Delta R$. The comparison of the model for $\kappa(R)$ and $F_{\text{c}}(R)$ with simulation data is shown in Section~\ref{sec:results}.

\subsubsection{Effective velocity jump $v_{\text{j}}^*$}
\label{sec:vjeff}
\begin{figure*}
	\centering
	\includegraphics[height=0.33\textwidth]{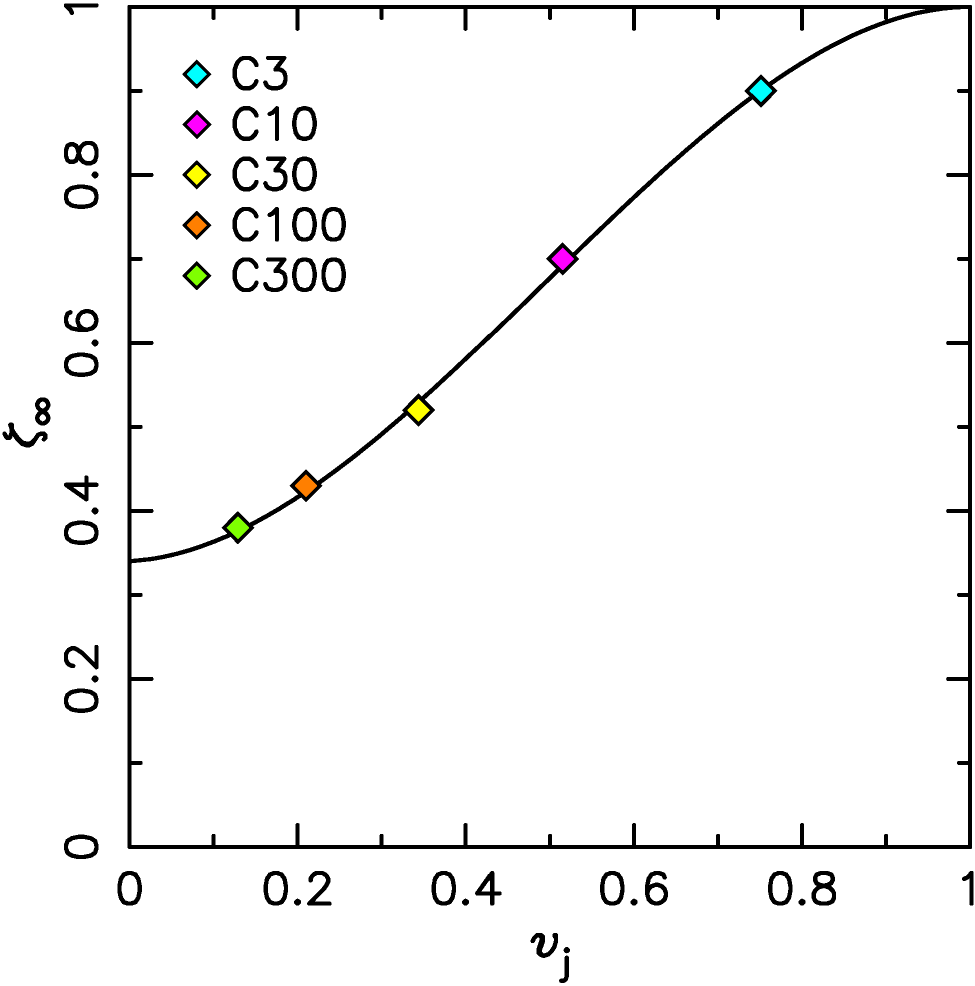}~
	\includegraphics[height=0.33\textwidth]{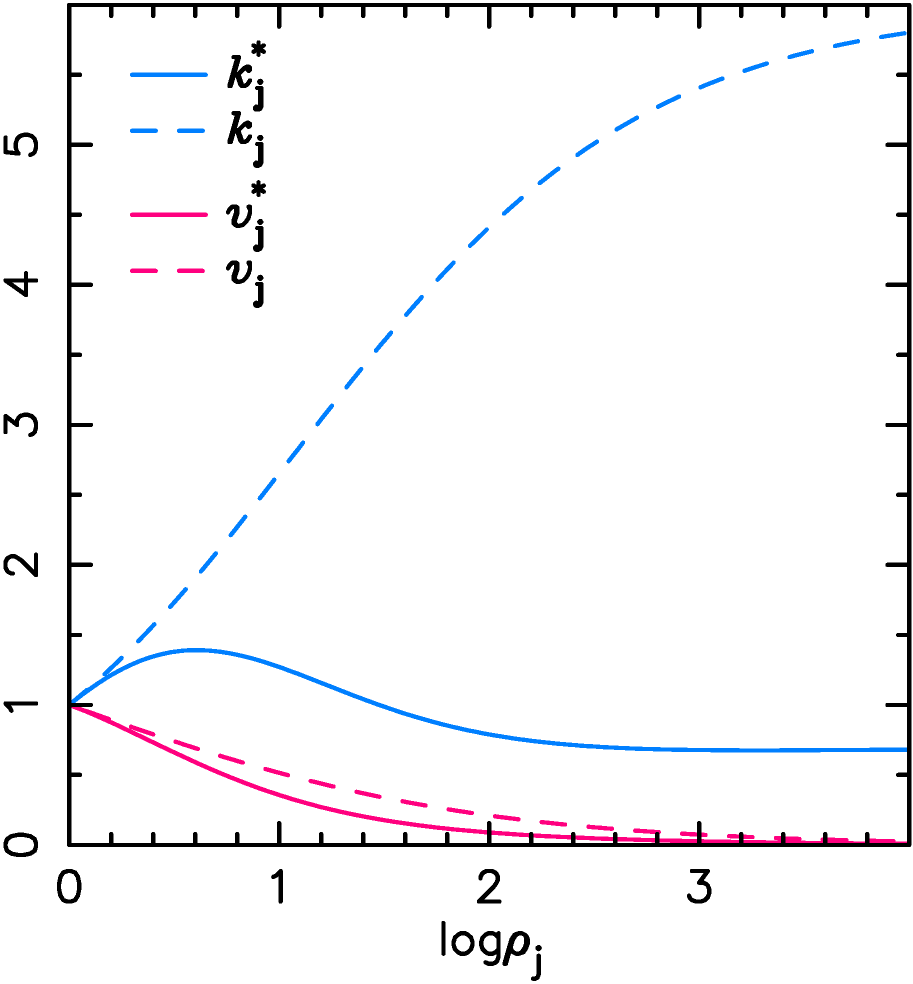}
	\includegraphics[height=0.33\textwidth]{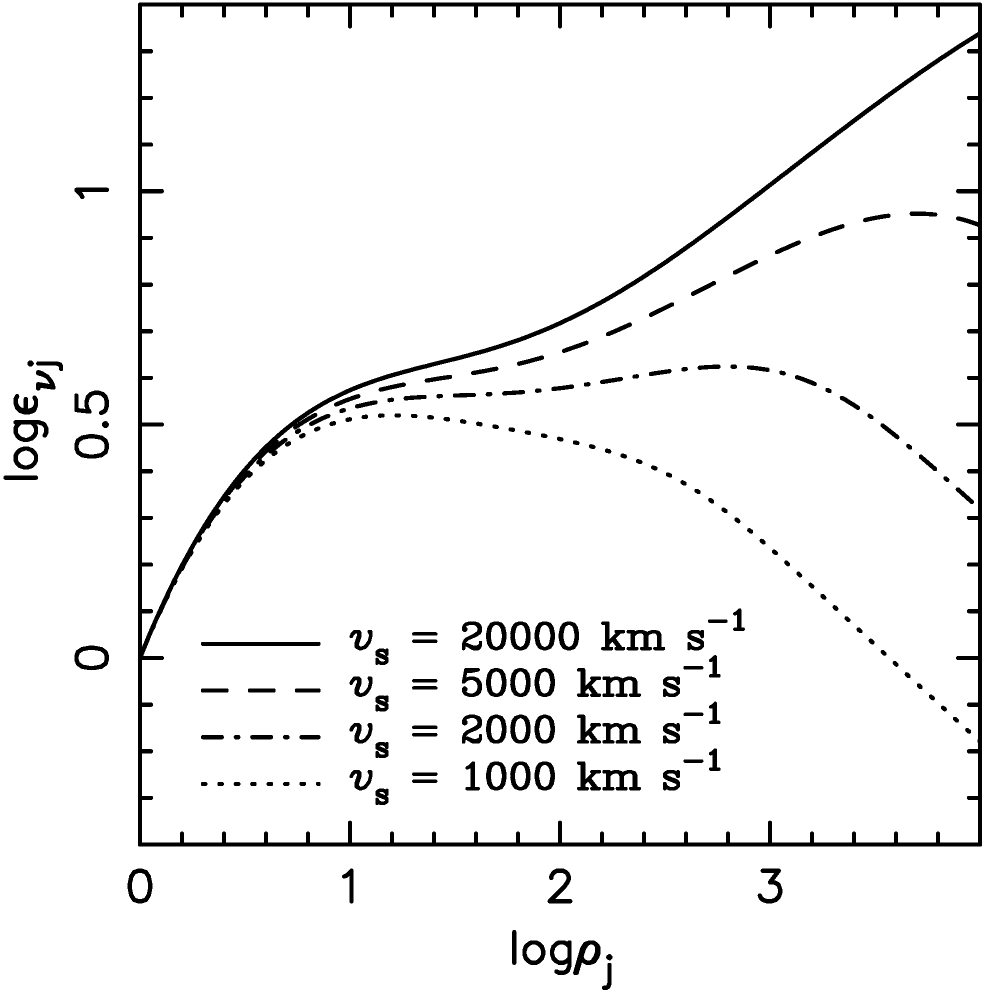}
	\caption{\textit{Left}. Fit for $\zeta_\infty$ from Eq.~(\ref{eq:zeta_inf}) as a function of $v_{\text{j}}$. \textit{Middle}. Velocity and energy jump as a function of density jump. The effective values are presented with solid lines, while dashed lines present theoretical (RP) values. \textit{Right}. Emissivity jump $\varepsilon_{\nu \text{j}}$ at clumps as a function of density jump and ICM shock velocity, in high-Mach regime (for both ICM and CL shocks).}
	\label{fig:jumps_eff}
\end{figure*}

The effective velocity jump $v_{\text{j}}^*$, representing the mean ratio between CL and ICM shock velocities, is lower than the theoretical (RP) one due to few effects taking place during the shock--clump interaction. At the point of hitting the clump (point 1 - inner side), $v_{\text{c}} = v_{\text{icm}}v_{\text{j}}\sqrt{E_{\text{for}}}$, as the reflected shock takes away the fraction of the energy density. In case of low density contrasts ($\delta\ll1$), $v_{\text{j}}^* = v_{\text{j}}\sqrt{E_{\text{for}}}$ is valid approximation. Otherwise, the CL shock surface bends around the clump's spherical shape, and its velocity decreases according to the inclination angle. So, $v_{\text{j}}^*$ effectively decreases with density contrast due to more inclined CL shocks. Soon after enveloping the clump (in case $\rho_{\text{j}}>10$) the CL shock velocity is distributed from $v_{\text{icm}}v_{\text{j}}\sqrt{E_{\text{for}}}$ at point 1, to almost zero at the other side of the clump (point 2 - outer side), being approximately $0.5v_{\text{icm}}v_{\text{j}}\sqrt{E_{\text{for}}}$ on average. As the clump sinks deeper into the remnant, it contracts in radial direction, at both sides. The inner side is being carved by the flow, increasing the surface perpendicular to the flow. The outer side is being compressed by the pressure behind the merging ICM shock. Both effects increase the shock velocity on each side, boosting the effective velocity jump. Another booster comes when the energy reflected from the initially crushed clumps starts returning to currently surviving clumps. Hence, the effective velocity jump can be defined as:
\begin{equation}
	\label{eq:vj*}
	v_{\text{j}}^* = \zeta(R) v_{\text{j}},
\end{equation}
where $\zeta(R)\le 1$ sums up all these effects and evolves with SNR radius. We approximate two regimes in modeling this variable: a) its exponential decrease from $\zeta(R_{\text{d}})=\sqrt{E_{\text{for}}}$ until merging with second regime, and b) its gradual increase due to absorbing of the reflected energy by the CL islands. The regime a) dominates until the CL shell grows to the first few clump radii, when the regime b) takes over. In regime b) we simply assume $\zeta(R)\propto \sqrt{F_{\text{c}}(R)}$ because the higher CL fraction results in higher chance that CL shock will absorb more reflected energy (instead of ICM shock). Just as $F_{\text{c}}(R)\rightarrow F_{\text{c}}^\infty$ far away from discontinuity, similarly $\zeta(R)\rightarrow\zeta_\infty$ at $R\rightarrow \infty$. So, if we define the second regime as: 
\begin{equation}
	\zeta_2 = \zeta_\infty \sqrt{F_{\text{c}}/F_{\text{c}}^{\infty}},
\end{equation}
then the final equation, including the first regime, can be written as:
\begin{equation}
	\zeta(R) = \zeta_2 + \left(\sqrt{E_{\text{for}}}-\zeta_2 \right)\text{e}^{-(R-R_{\text{d}})/r_{\text{c}}}.
\end{equation}
The value of $\zeta_\infty$ is obtained from simulations and fitted with third-order polynomial:
\begin{equation}
	\label{eq:zeta_inf}
	\zeta_\infty = 0.34 + 0.06v_{\text{j}} + 1.86v_{\text{j}}^2 - 1.26v_{\text{j}}^3.
\end{equation}
This fitting function, as well as the jumps $v_{\text{j}}^*$ and $k_{\text{j}}^*$, are shown in Figure~\ref{fig:jumps_eff}.

Considering the reflected energy, the crucial difference from the LDB case is that here reflections happen all the time, not at once. From the moment the first reflected energy returns to the shock, a quasi-equilibrium between energy departures and returns establishes (which happens approximately in one doubling of the radius). However, before that happens the shock evolves with reduced energy, similar as in LDB case. This regime is not modeled due to its complexity.

The weakening of the average shock speed at dense clumps puts a limit on effective emissivity jump, $\varepsilon_{\nu \text{j}}^* = \varepsilon_{\nu \text{c}}/\varepsilon_{\nu \text{icm}}$, as is shown in Figure~\ref{fig:jumps_eff} for strong shocks ($M\gg1$). The emissivity jump at clumps grows almost exponentially with density jump up to $\rho_{\text{j}} \approx 5$, then stagnates at $\varepsilon_{\nu \text{j}}^* \approx 4$ up to $\rho_{\text{j}} \approx 50$ (even for highest shock velocities). The fact that $I(x)$ depends significantly on $v_{\text{s}}$ causes the fall of emissivity jump with SNR radius, but in case the Mach number of the ICM shock starts to fall, the jump rises again due to lower sound speed (and higher Mach number) in clumps.

\subsubsection{Shock velocity in clumpy medium}
\label{sec:velocity_clumps}
The derivation of shock velocity in clumpy medium is more complex than in LDB case, but it will rely on the concept from Section~\ref{sec:LDB}. From now on, we make a distinction between volume-filling factor of the clumps, $f_{\text{c}}$, and the clump coverage at the shock surface, $F_{\text{c}}$. Also, all calculations include the effective velocity jump $v^*_{\text{j}}$, as well as  $k^*_{\text{j}}=\rho_{\text{j}}v^{*2}_{\text{j}}$.

The velocity will be calculated from the shock energy density, although in this case each variable has CL, ICM, and average value. The average energy density over the shock surface is weighted sum over the clumps and ICM,
\begin{equation}
	\label{E2_clumps}
	\bar{E}_2 = \frac{9}{4} \left( F_{\text{icm}}\rho_{\text{icm}} v_{\text{icm}}^2 + F_{\text{c}}\rho_{\text{c}} v_{\text{c}}^2 \right) = 
	\frac{9}{4} \rho_{\text{icm}} \bar{v}_{\text{s}}^2 \frac{F_{\text{icm}} + F_{\text{c}} k_{\text{j}}^*}{\left( F_{\text{icm}} + F_{\text{c}} v^*_{\text{j}} \right)^2 },
\end{equation}
where we used identity $\bar{v}_{\text{s}} = v_{\text{icm}} \left( F_{\text{icm}} + F_{\text{c}} v^*_{\text{j}} \right)$. Similarly as in LDB case, the energy evolution is calculated using the energy ratio $k(R)=\bar{E}_2(R)/E_2^{\text{uni}}$:
\begin{equation}
	\label{eq:k_CL}
	k(R) = k_\infty + (\bar{k}_{\text{j}}-k_\infty) \left( \frac{R_{\text{d}}}{R} \right)^m,
\end{equation}
where $k_{\infty}=k(R\rightarrow\infty)$. Determination of exponent $m$ is described in Section~\ref{sec:exponent_m}.
The average energy jump at $R_{\text{d}}$ is:
\begin{equation}
	\bar{k}_{\text{j}} = f_{\text{icm}} + f_{\text{c}}k_{\text{j}}E_{\text{for}},
\end{equation}
We assume that far away from discontinuity the remnant evolves in a pseudo-uniform medium, i.e. it ``forgets'' the discontinuity ($F_{\text{c}}$ and $v^*_{\text{j}}$ stay constant). Then, the mean shock velocity from equation:
\begin{equation}
	E_2^{\text{uni}} = \frac{9}{4} \bar{\rho} \bar{v}_{\text{s}}^2 = \frac{9}{4} \rho_{\text{icm}} \bar{v}_{\text{s}}^2 \left( f_{\text{icm}} + f_{\text{c}}\rho_{\text{j}} \right).
\end{equation}
can be used as a boundary condition to find the factor $k_\infty$:
\begin{equation}
	k_\infty = \frac{\bar{E}_2(R\gg R_{\text{d}})}{E_2^{\text{uni}}} = 
	\frac{F_{\text{icm}}^{\infty} + F_{\text{c}}^{\infty} k_{\text{j}}^{*\infty}}{\left( f_{\text{icm}} + f_{\text{c}}\rho_{\text{j}} \right) \left( F_{\text{icm}}^{\infty} + F_{\text{c}}^{\infty} v_{\text{j}}^{*\infty} \right)^2}.
\end{equation}
Here we use the far-away values for $F_{\text{icm}}$, $F_{\text{c}}$, $k_{\text{j}}^*$, and $v_{\text{j}}^*$ ($v_{\text{j}}^{*\infty} = \zeta_\infty v_{\text{j}}$). Unlike the LDB case, where all the reflected energy eventually reunite and $k(R>R_{\text{u}})$ becomes unity, here the continuous shocking of the new clumps keeps $k_\infty<1$, correlating with $v_{\text{j}}$.

Finally, the average shock velocity at $R>R_{\text{d}}$ is:
\begin{equation}
	\bar{v}_{\text{s}}(R) = \sqrt{\frac{4}{9}\frac{k(R)E_2^{\text{uni}}}{\rho_{\text{icm}}} \frac{\left( F_{\text{icm}} + F_{\text{c}} v_{\text{j}}^* \right)^2 }{F_{\text{icm}} + F_{\text{c}} k_{\text{j}}^* }},
\end{equation}
where all variables except $\rho_{\text{icm}}$ depend on $R$.

\subsubsection{Expansion exponent $m$}
\label{sec:exponent_m}
The remnant expansion rate is determined by the exponent $m$ in Eq.~(\ref{eq:k_CL}). In contrast to the LDB setup, where the energy reflection propagates in purely radial direction, the clumps reflect bow shocks that to a much lesser extent travel towards SNR center. Whereas much of the reflected energy ends up at shock again (both on its CL and ICM fraction), the part traveling inwards distributes slower. Because of that, the shock deceleration is more gradual, so the exponent $m$ takes much lower values than in LDB model. All simulations give the values $m\le3$, meaning that the compressed energy fails to redistribute over the whole SNR volume, but effectively stays in the region behind the shock.

The exponent $m$ is mainly affected by the density contrast and volume filling factor of the clumps. For a fixed $\rho_{\text{j}}$, if $f_{\text{c}}\rightarrow0$, the ICM shock dominates in collecting the energy from bow shocks, so $m\rightarrow0$. Otherwise, for high $f_{\text{c}}$, the CL shock absorbs more of the reflected energy, so $m\rightarrow3$. For a fixed $f_{\text{c}}$, if $\rho_{\text{j}}\rightarrow1$, in absence of CL islands, the reflected energy distributes easier over the SNR volume, so $m\rightarrow3$. However, at high $\rho_{\text{j}}$, the CL islands succeed to collect larger part of reflected energy, preventing its redistribution and lowering the value of $m$. In this case, $m$ approaches some minimum value, not zero (except for $f_{\text{c}}\rightarrow0$).

\begin{figure}
	\centering
	\includegraphics[width=0.45\columnwidth]{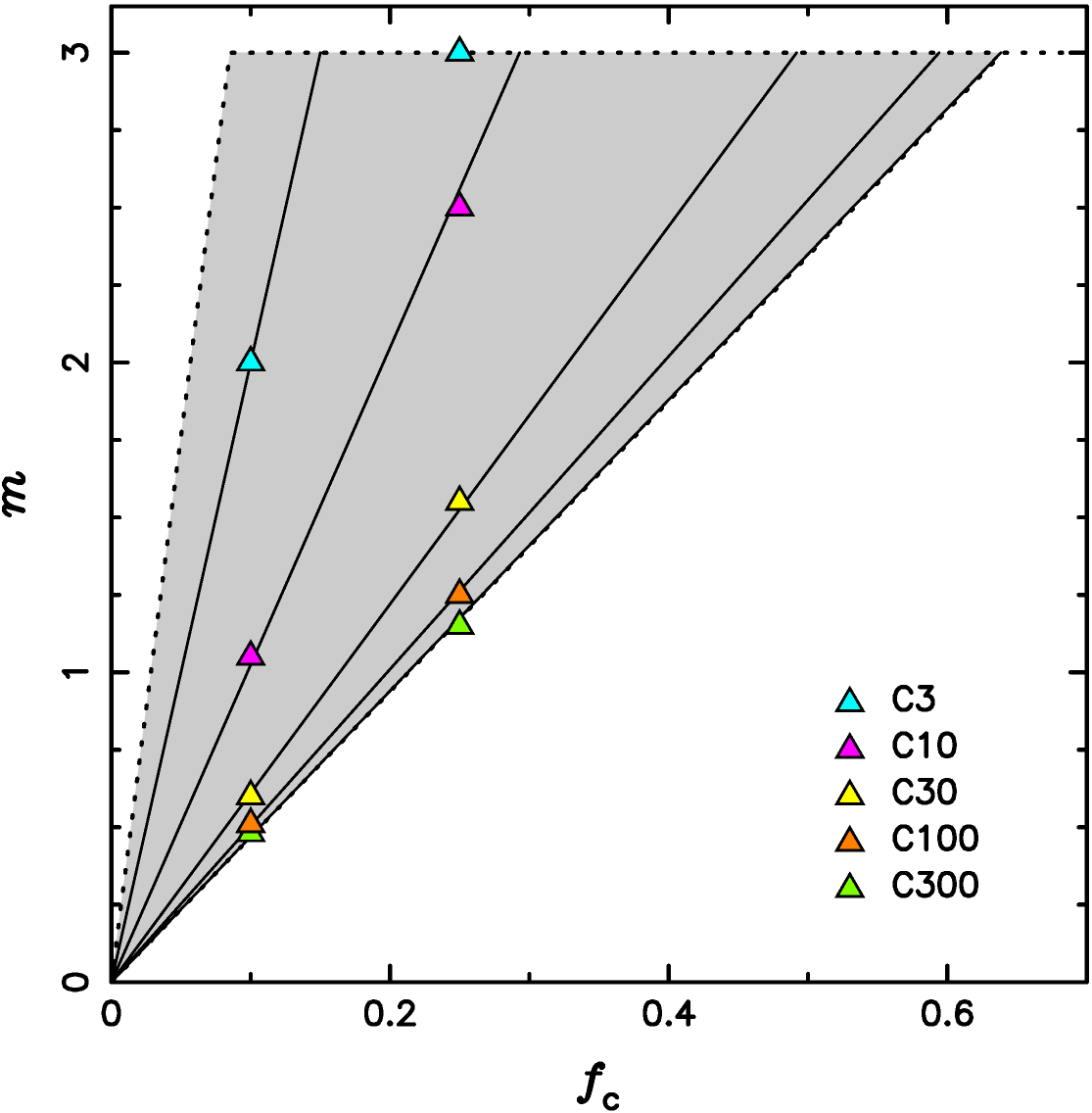},
	\caption{Map of predicted values of the expansion exponent $m$, calculated by the empirical Eq.~(\ref{eq:exp_m}). The symbols designate the best-fitted values for the individual simulations. The solid lines are linear fits to the simulation data and the shaded area with dotted-line border represents the domain of possible values of $m$. The maximum value of $m$ is 3.}
	\label{fig:exp_m}
\end{figure}

Although the dependency of $m$ on these two properties is complex, a fairly simple empirical relation is found:
\begin{equation} \label{eq:exp_m}
	m = \min \left( f_{\text{c}} \sqrt{22.4+1219v_{\text{j}}^{4.08} }, 3 \right),
\end{equation}
and schemed in Figure~\ref{fig:exp_m} as a map of predicted values of $m$.

\section{Results} \label{sec:results}

\subsection{Uniform medium - U1. Calibration of data} \label{subsec:results_uni}
\begin{figure*}
	\centering
	\includegraphics[width=0.32\textwidth]{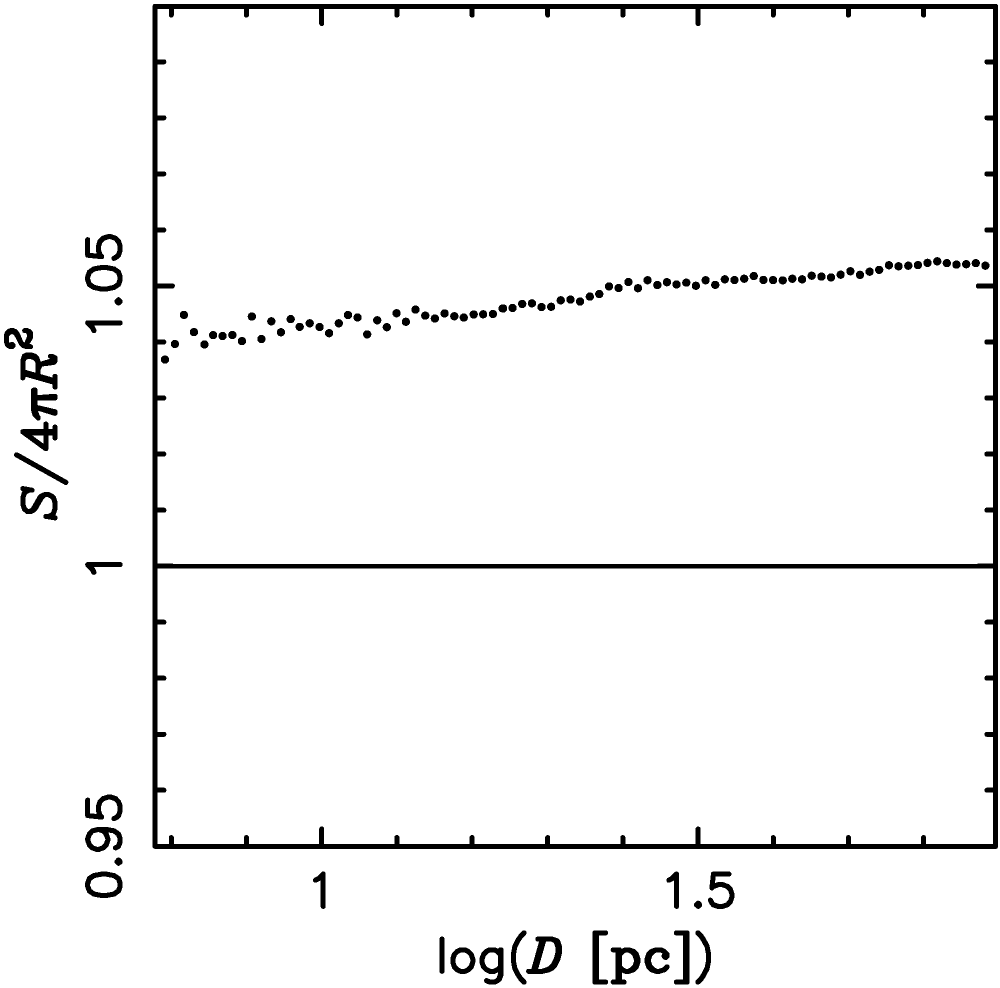}~
	\includegraphics[width=0.32\textwidth]{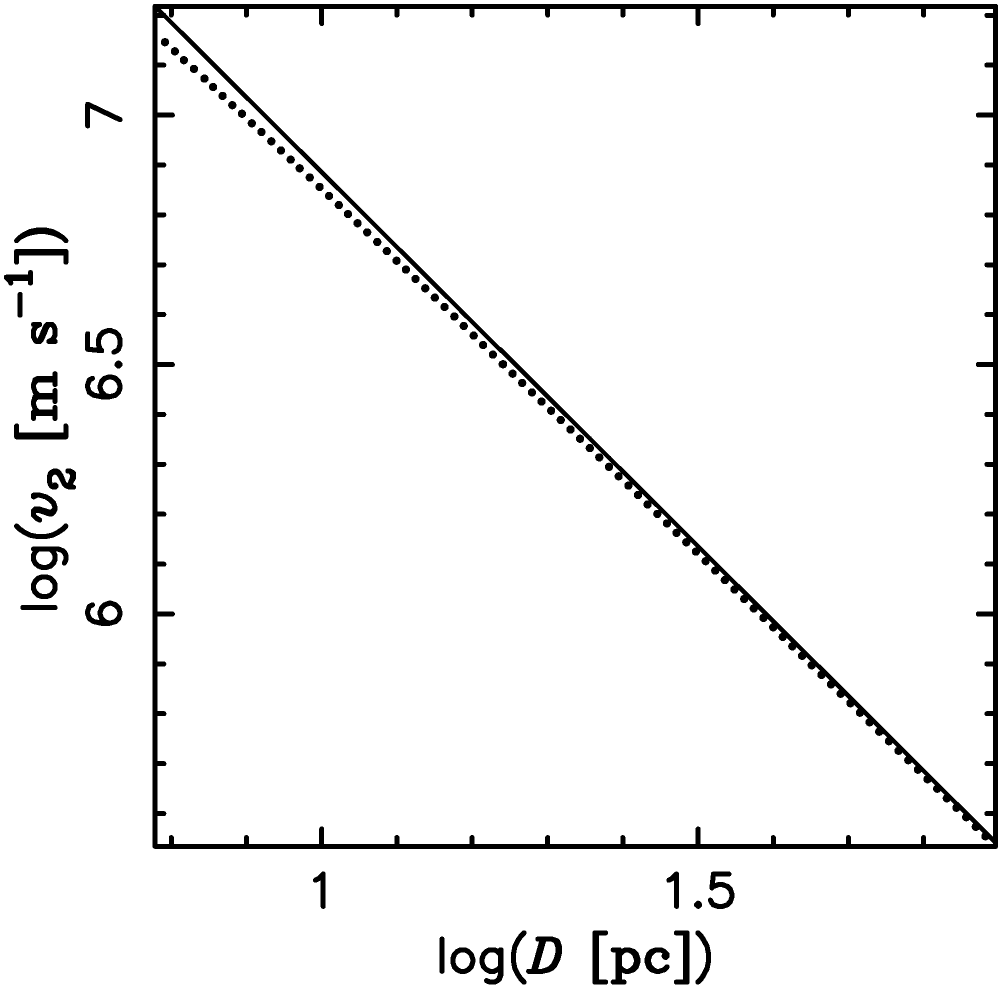}~
	\includegraphics[width=0.32\textwidth]{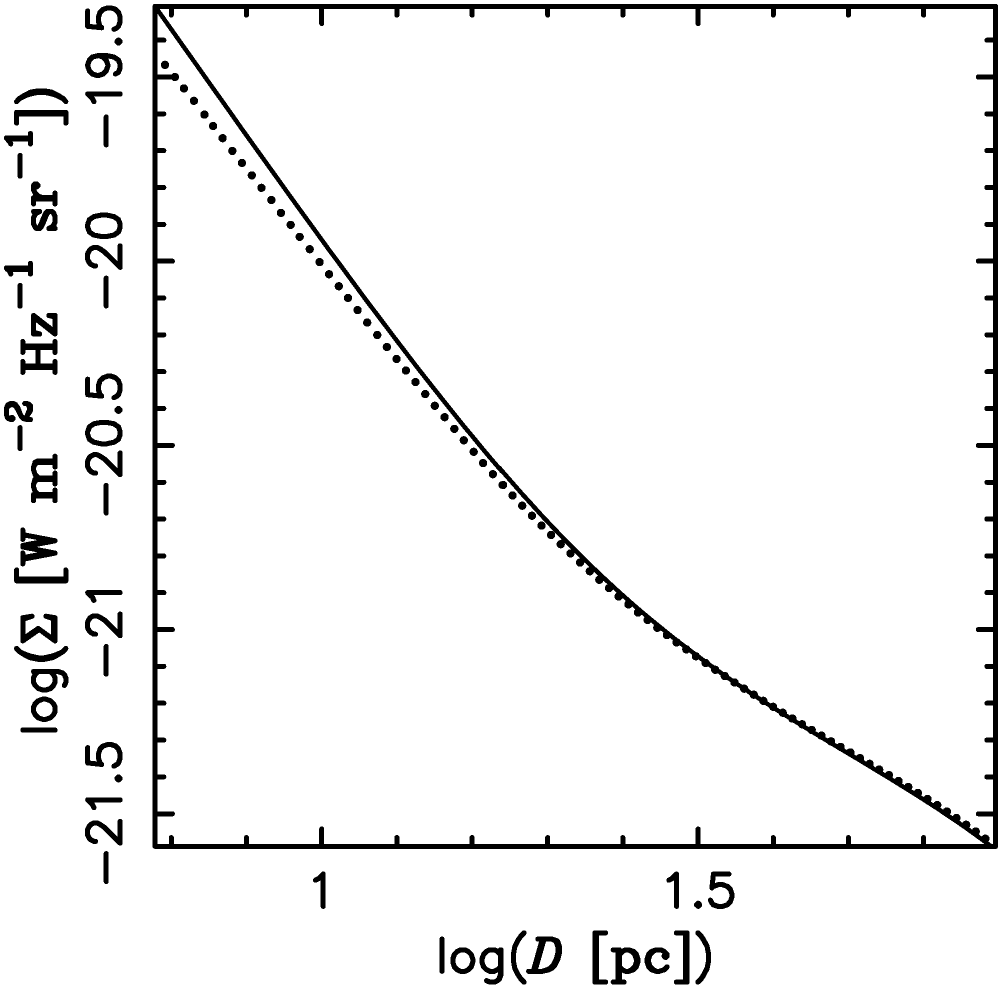}
	\caption{Comparison of theoretical (line) and uncalibrated simulation results (points) for the model U1. The graphs show the surface area of the shock $S$ (\textit{left}) and the mean velocity behind the shock $v_2$ (\textit{middle}) as a function of SNR diameter, and the $\Sigma$--$D$ relation (\textit{right}). Apart from the left side graph, where the systematic error of $\approx5\%$ is present, it is clearly seen that, as the SNR covers more cells, the values converge to theoretical ones.}
	\label{fig:calibration}
\end{figure*}

Due to the finite resolution and numerical method, the density, momentum and energy profiles of the shock are smeared. Consequently, the measured velocity behind the shock ($v_2$) is smaller (cut off) compared to the theoretical value, although the shock velocity is correct. The same stands for other variables that depend on $v_2$, such as emissivity or surface brightness. So, to compare the simulation results to the semi-analytical model, the former have to be calibrated. The calibration factors are obtained from the U1 setup, for which the theoretical solution is known, and they are used for adjusting the results of all other simulations. The differences in uncalibrated and theoretical values for shock surface, velocity, and surface brightness for the U1 simulation are shown in Figure~\ref{fig:calibration}.

\subsection{Low-density bubble - LDB2, LDB5, LDB10}
\begin{figure*}
	\centering
	\includegraphics[width=0.33\textwidth]{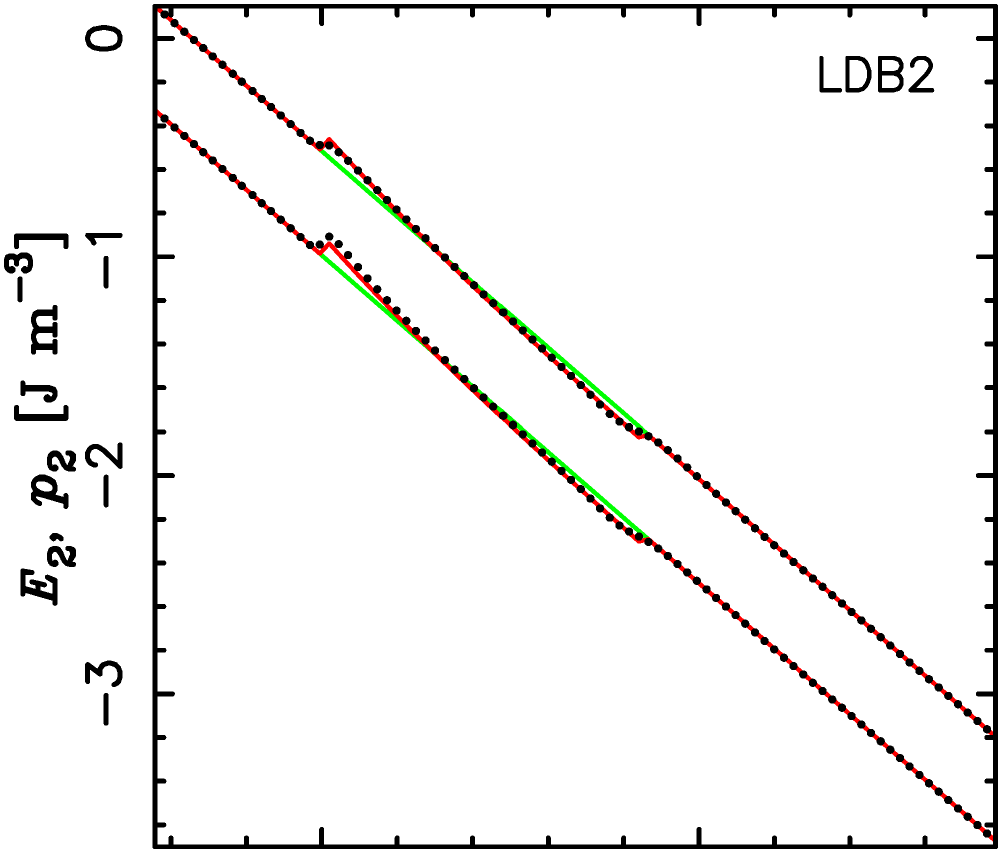}
	\includegraphics[width=0.2805\textwidth]{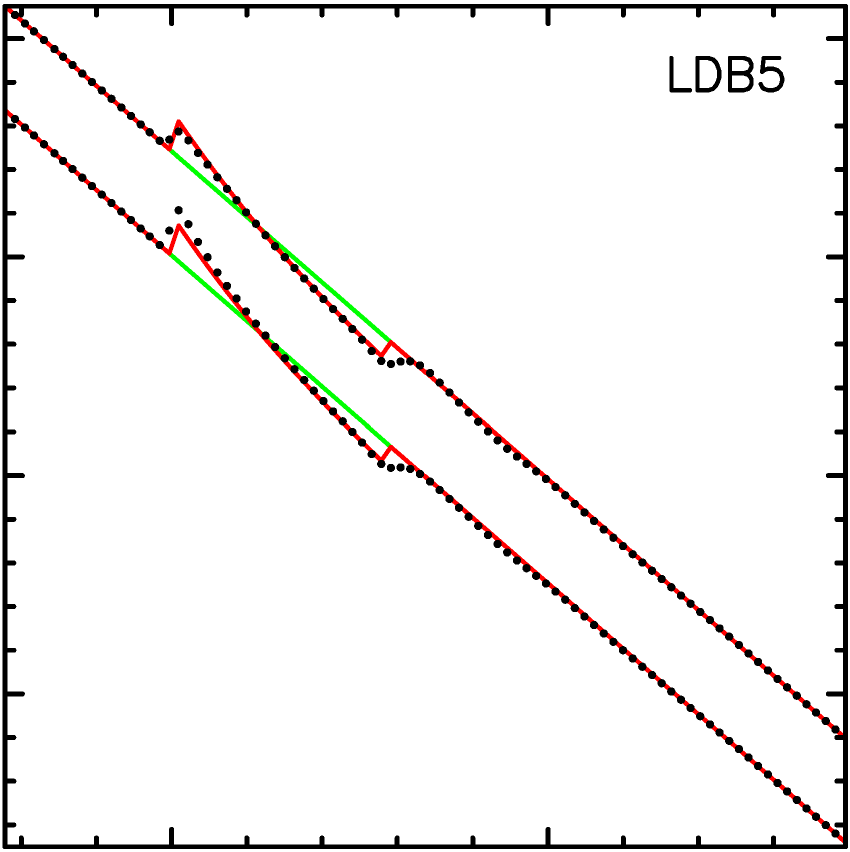}
	\includegraphics[width=0.2805\textwidth]{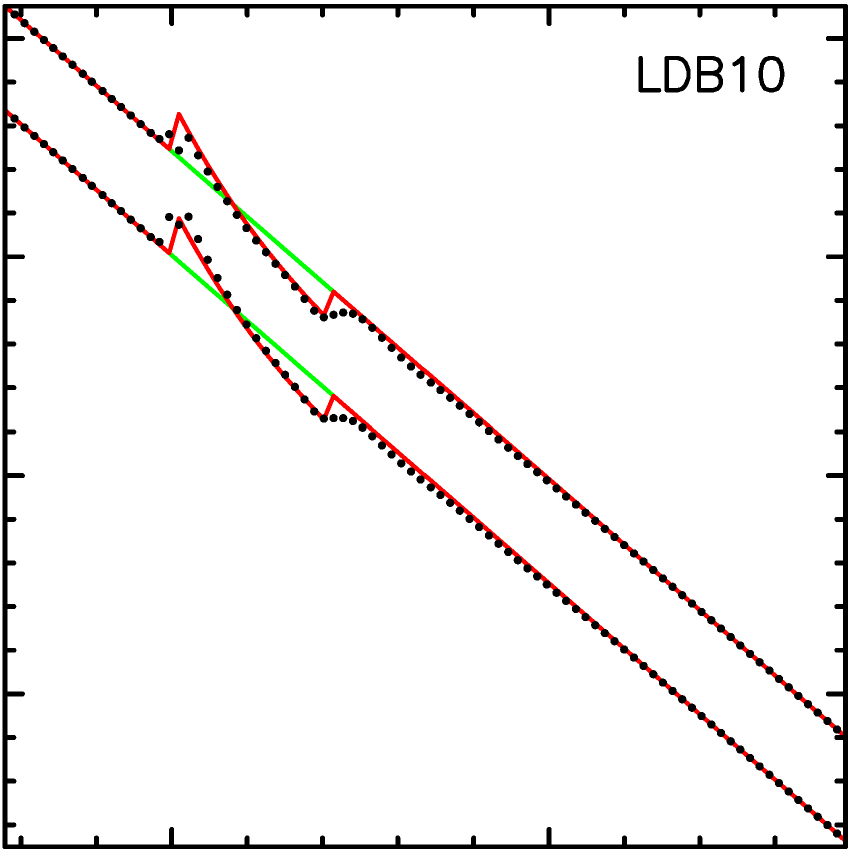}
	\linebreak
	\includegraphics[width=0.33\textwidth]{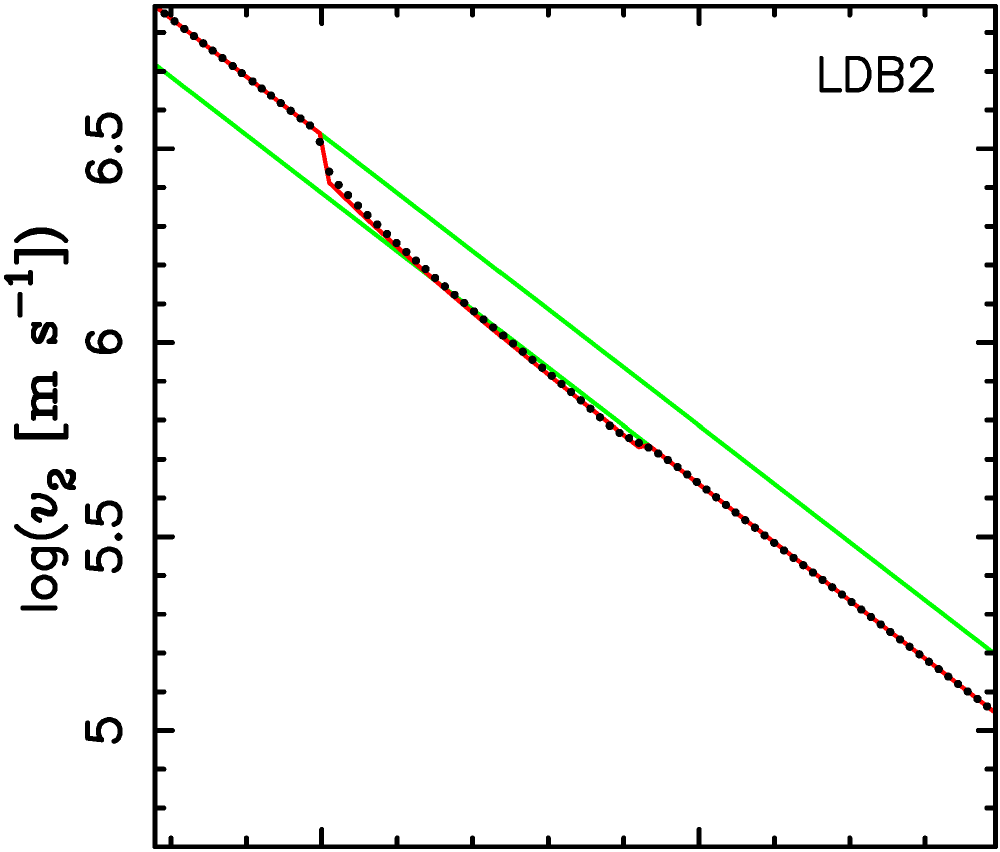}
	\includegraphics[width=0.2805\textwidth]{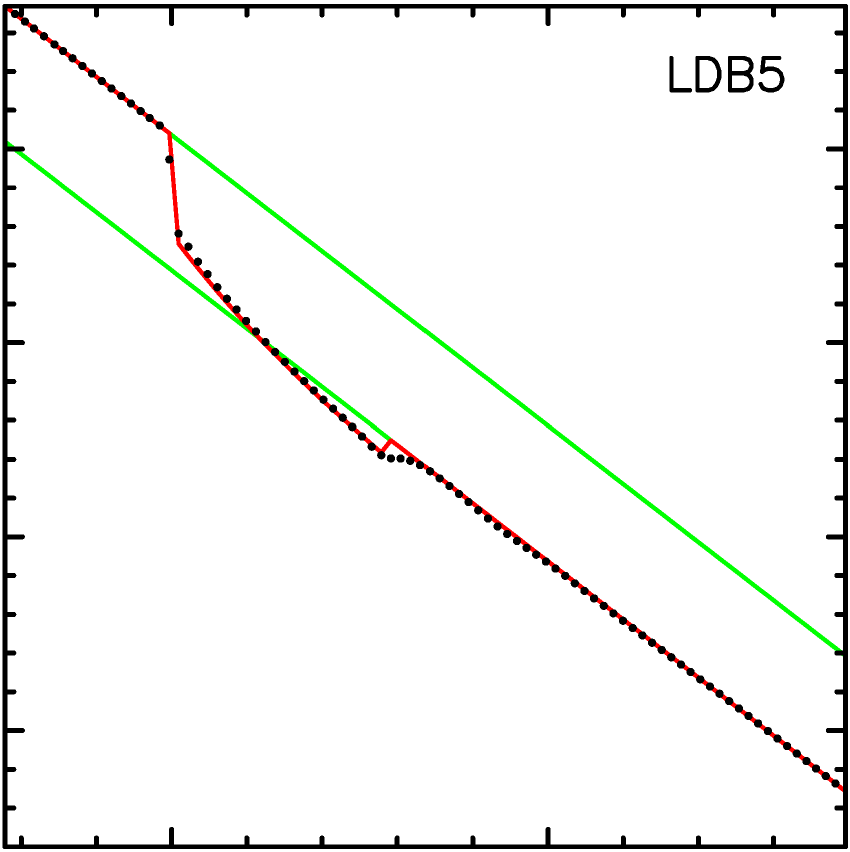}
	\includegraphics[width=0.2805\textwidth]{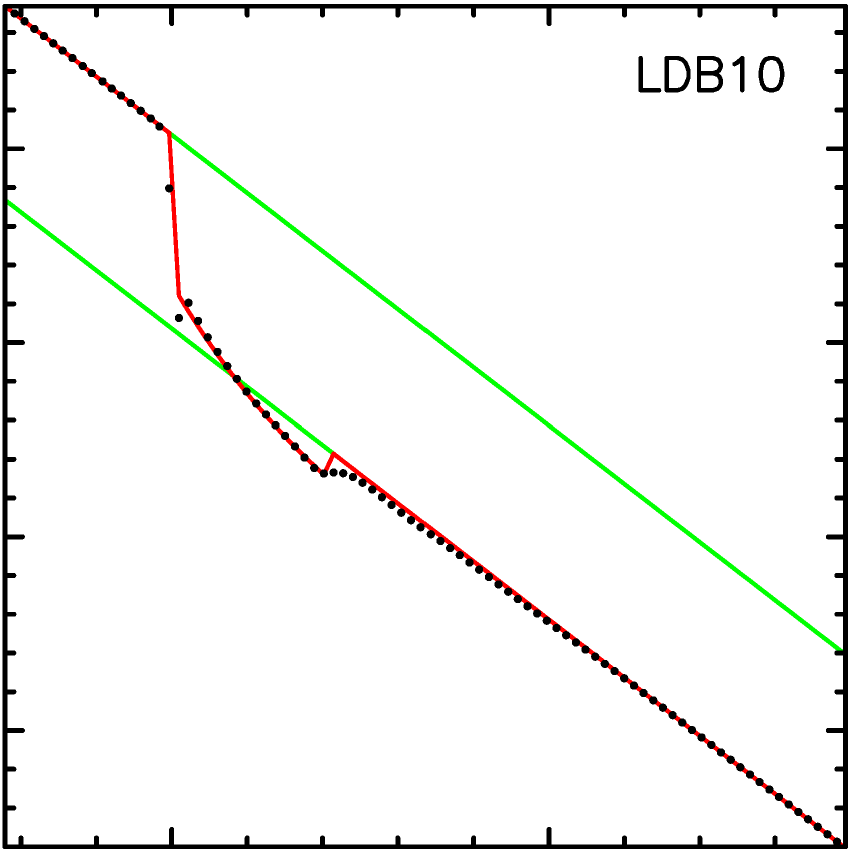}
	\linebreak
	\includegraphics[width=0.33\textwidth]{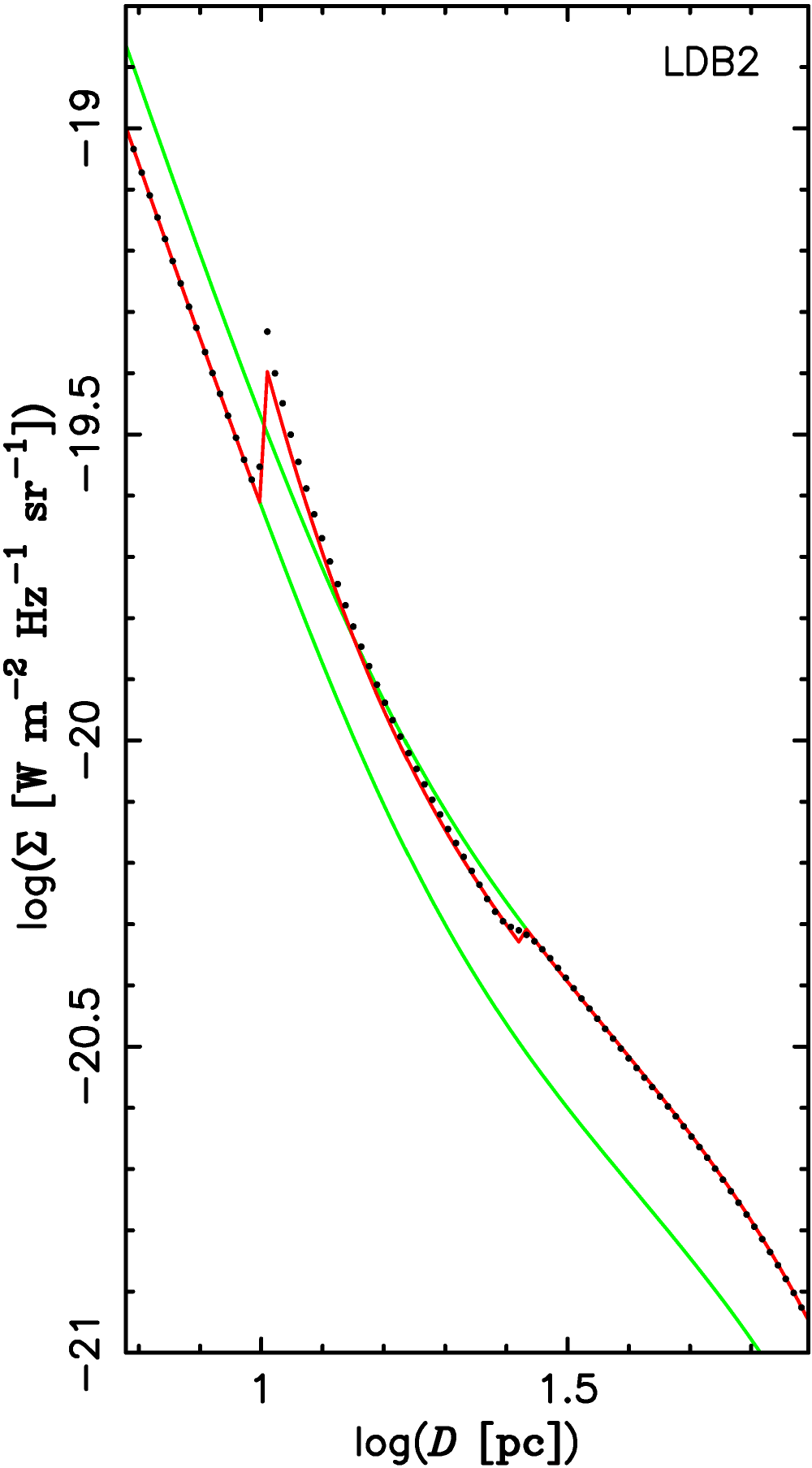}
	\includegraphics[width=0.2805\textwidth]{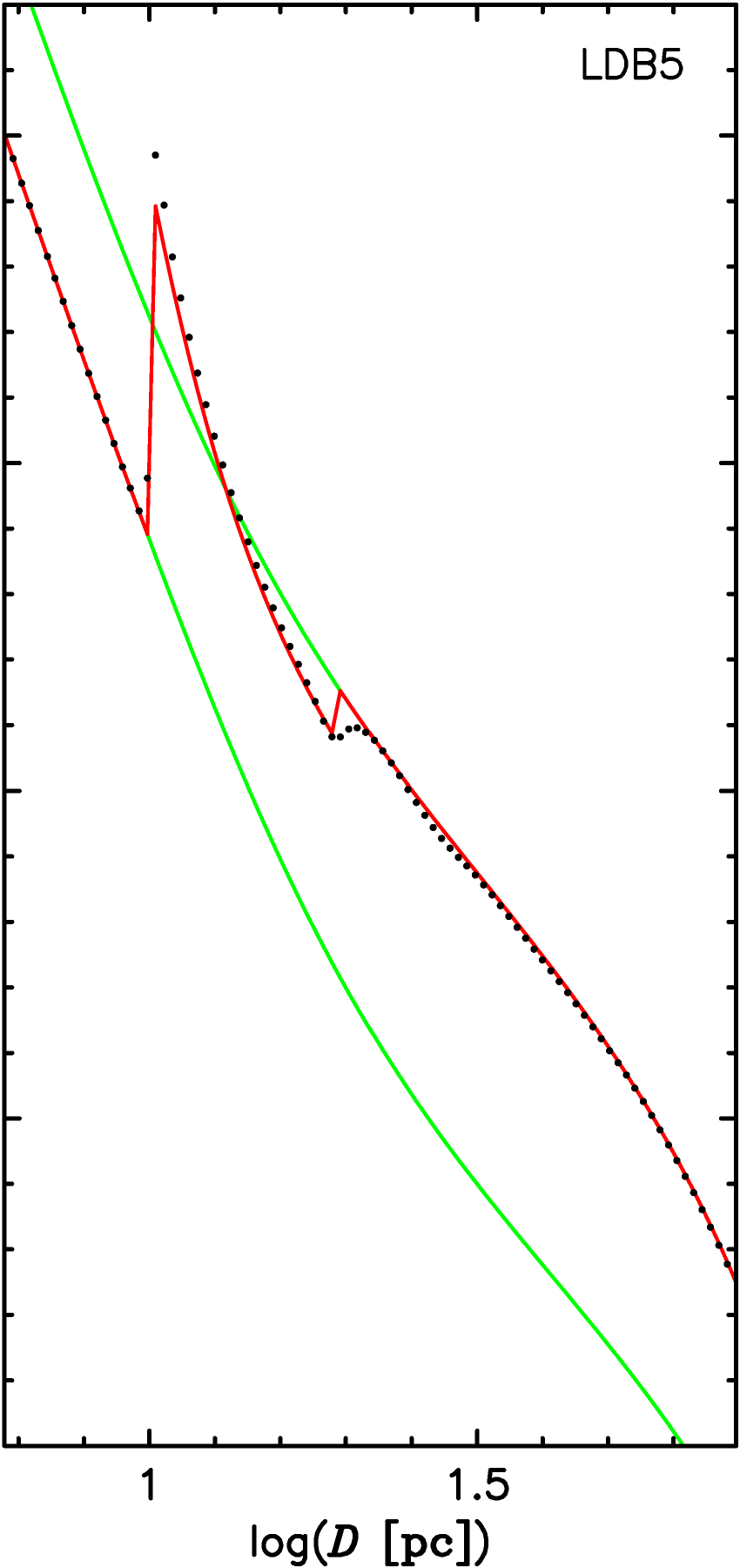}
	\includegraphics[width=0.2805\textwidth]{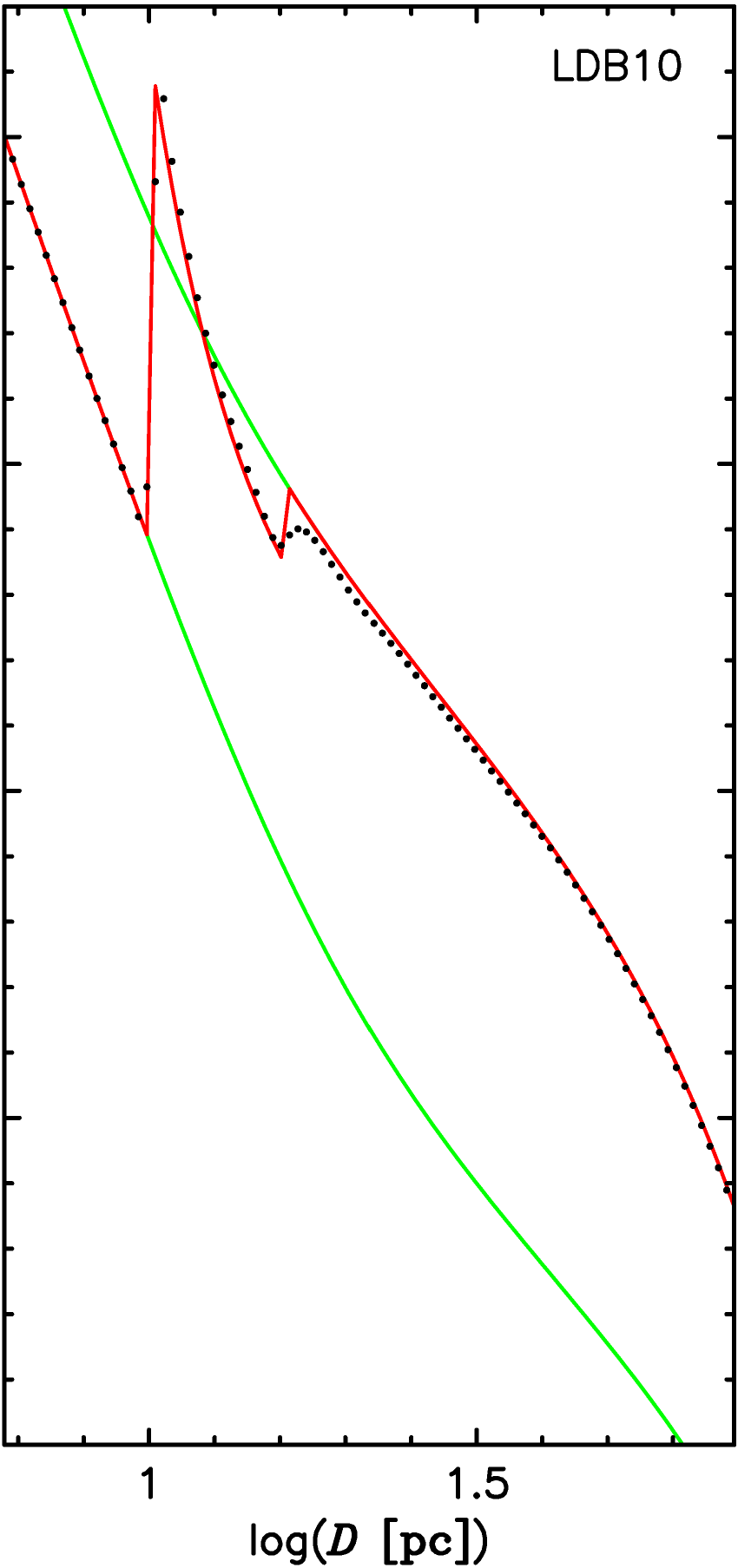}
	\linebreak
	\caption{Simulation results and semi-analytical model for LDB2, LDB5, and LDB10 setup. The black points and red line represent the simulation results and semi-analytical model, respectively. Green lines show the uniform evolutions in $\rho_{\text{l}}$ and $\rho_{\text{h}}$ ambient densities. From top to bottom row: evolutions of the energy (higher curves) and pressure (lower curves), velocity behind the shock and $\Sigma$--$D$ relation.}
	\label{fig:LDB}
\end{figure*}

The LDB results are presented in Figure~\ref{fig:LDB}. The energy/pressure jump is clearly seen at the discontinuity and at the radius of reunification of the reflected and forward shock. The latter radius is the one where the reflected shock wave, traveling all the way through the center of the remnant, merge with the forward shock at the opposite side, bringing the missing energy to the forward shock. As expected, the reduced forward shock energy falls below $E_2^{\text{uni}}$, until the reflected shock arrives from the other side, after which the evolution continues as in uniform medium of higher density. This behavior directly affects the shock velocity and surface brightness.

These tests were used to establish and approve the model of energy removal at the forward shock. The radius $R_{\text{u}}$ at which the shocks reunite and exponent $m$ that governs the rate of energy redistribution to Sedov profile are determined empirically:
\begin{equation}
	R_{\text{u}} = \frac{R_{\text{d}}}{1-0.75v_{\text{j}}},~~~
	m = \frac{\text{e}^{5.651v_{\text{j}}}+47.04}{\text{e}^{4.706v_{\text{j}}}}.
\end{equation}

\begin{figure*}
	\centering
	\includegraphics[width=0.358\textwidth]{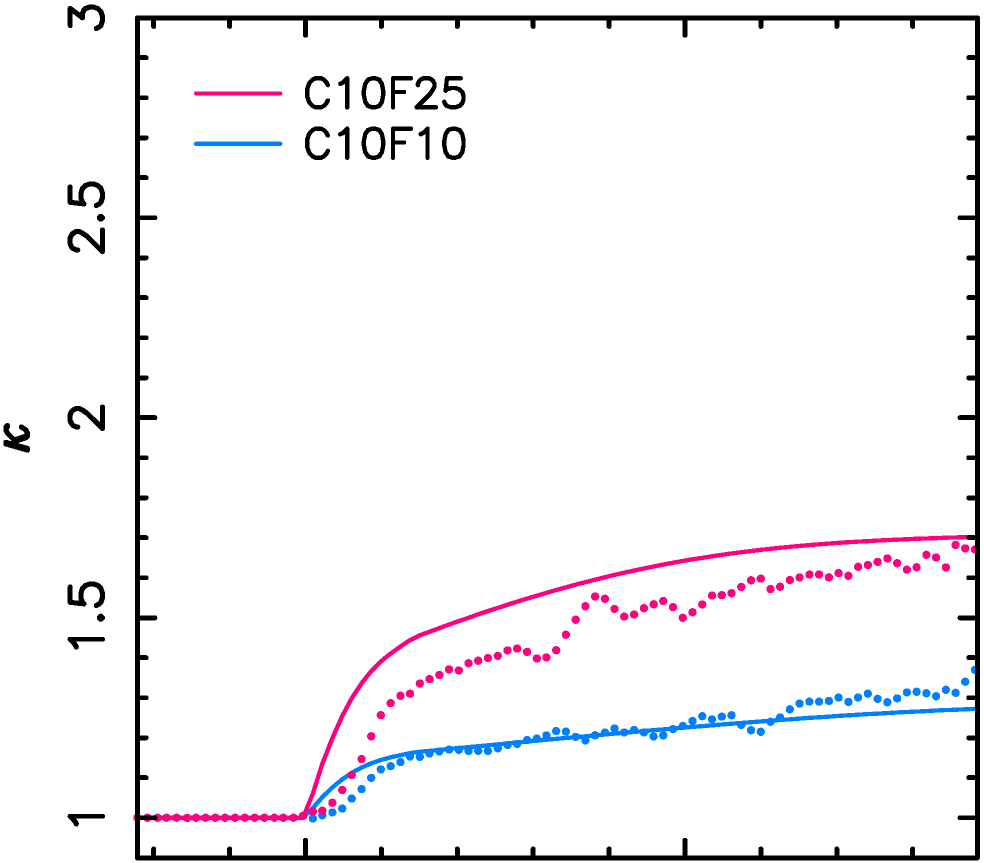}
	\includegraphics[width=0.31\textwidth]{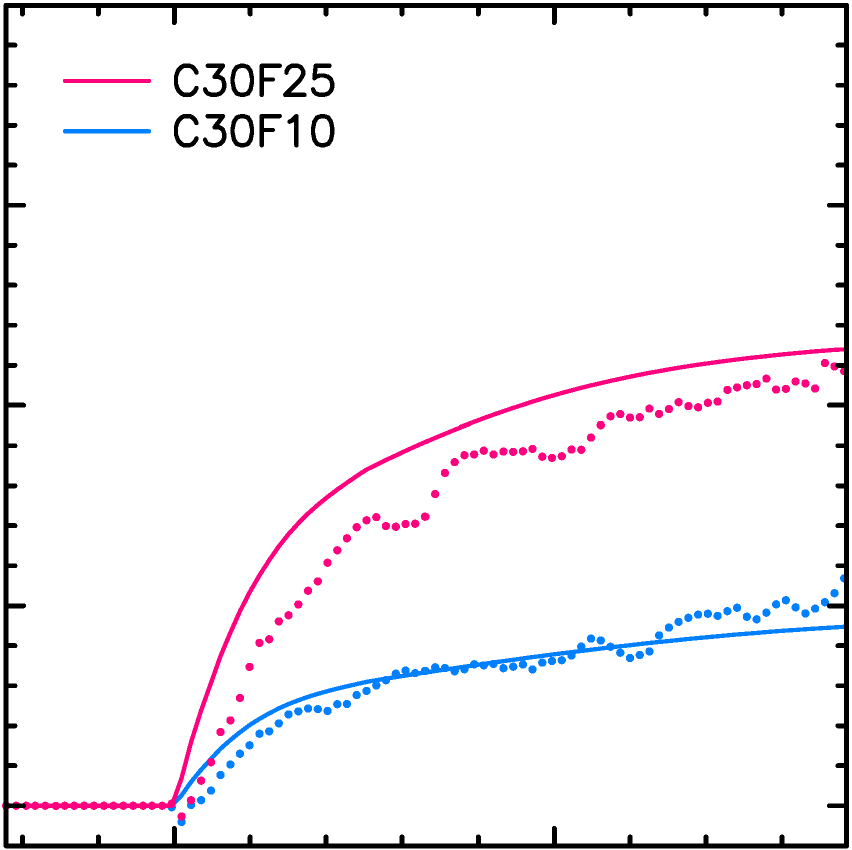}
	\includegraphics[width=0.31\textwidth]{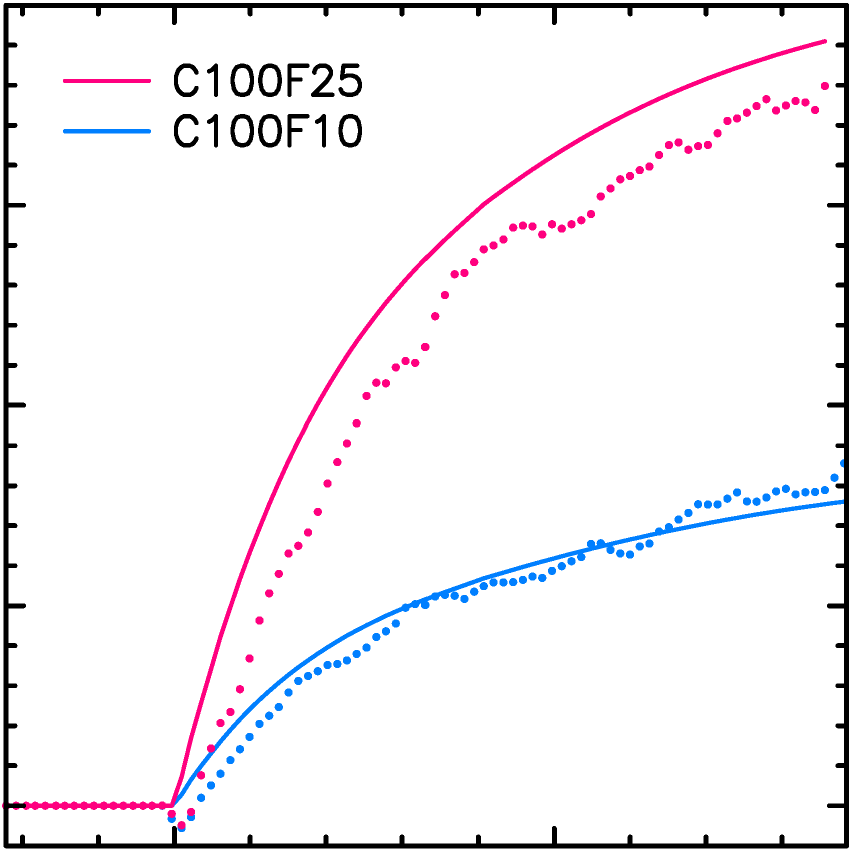}
	\linebreak
	\includegraphics[width=0.358\textwidth]{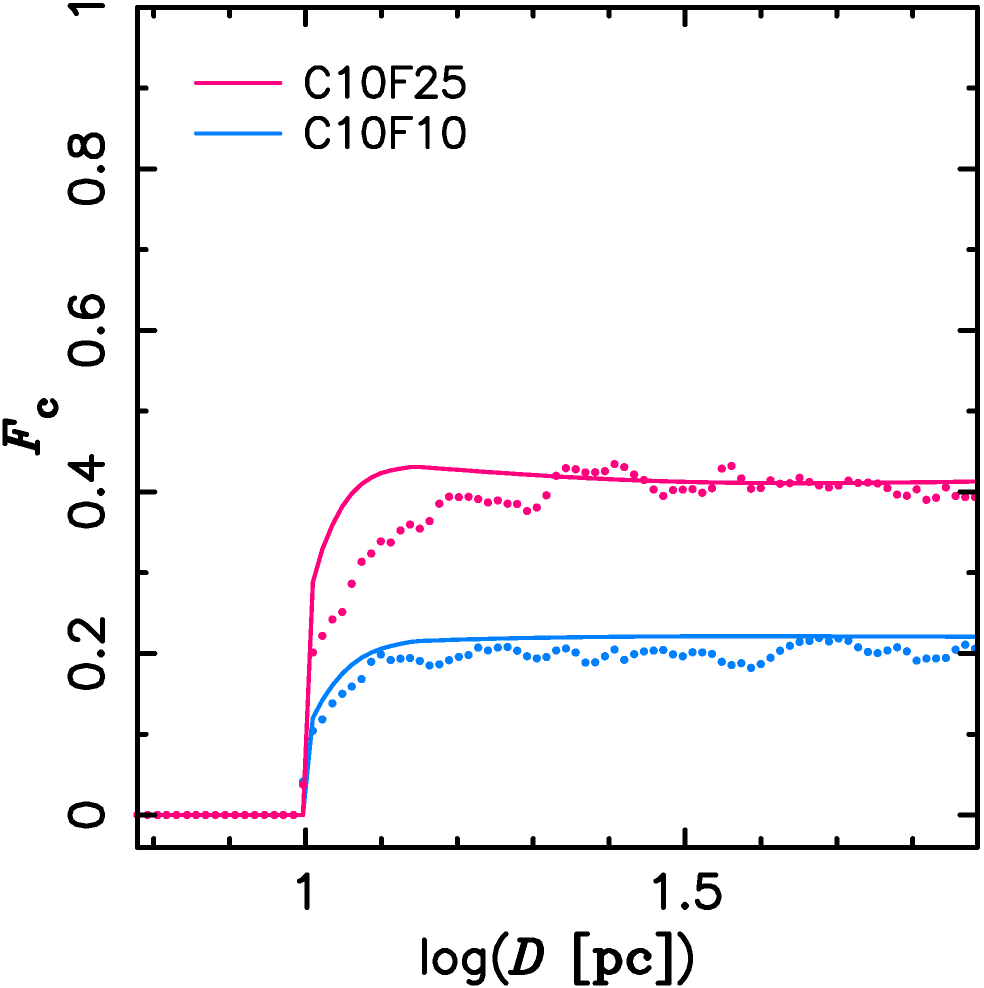}
	\includegraphics[width=0.31\textwidth]{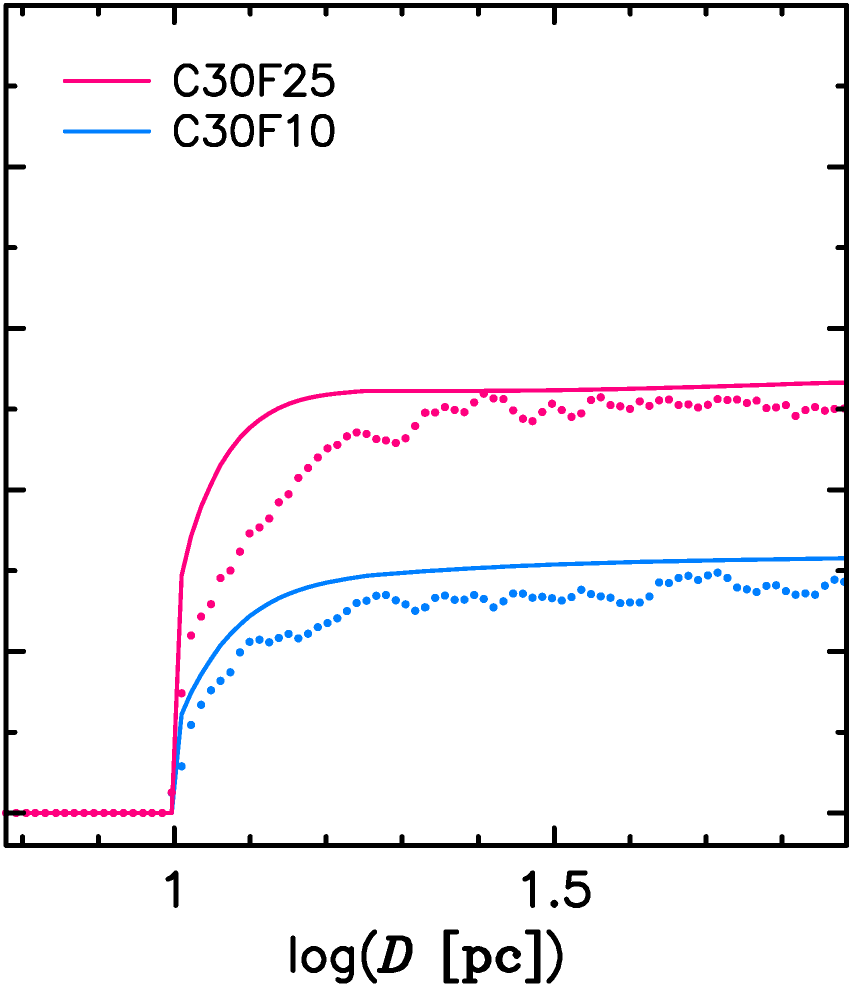}
	\includegraphics[width=0.31\textwidth]{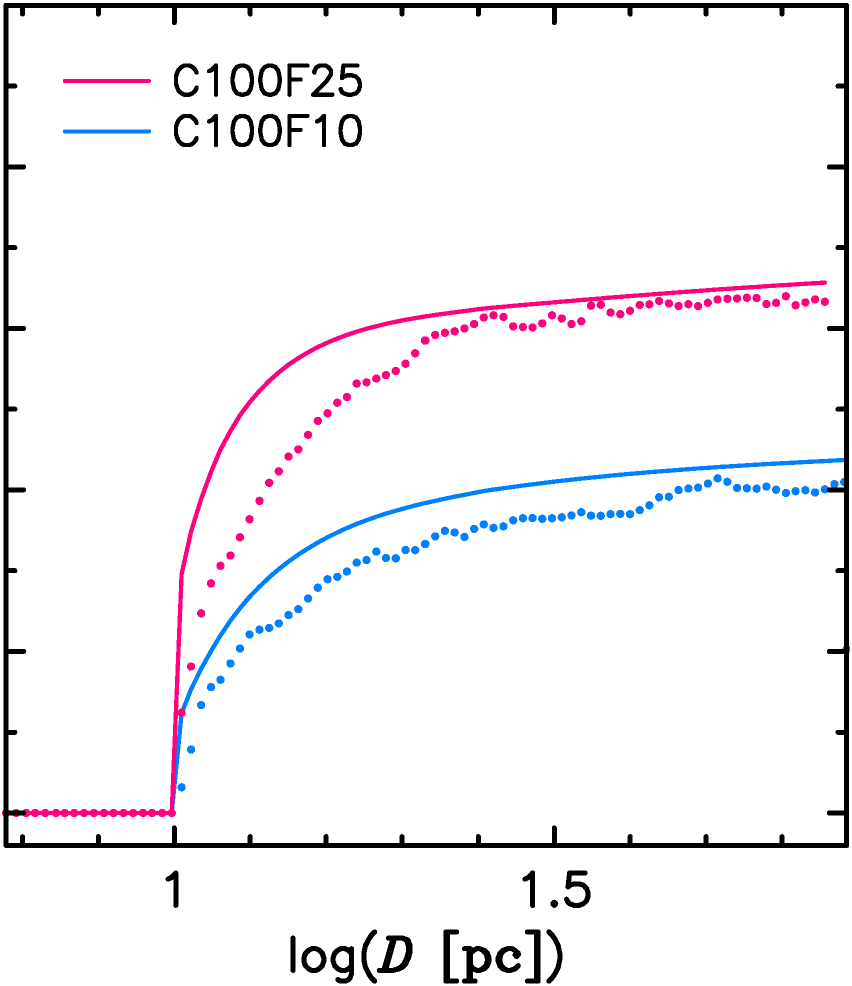}
	\linebreak
	\caption{\textit{Top}. Comparison of SNR surface area ratio (Eq.~\ref{eq:kappa}) obtained from simulations and analytical model. \textit{Bottom}. Comparison of CL shock fraction from simulations and analytic model. The dots and lines represent the simulation results and analytic model, respectively. The simulation setups are denoted in the legend.}
	\label{fig:surf_model}
\end{figure*}

\subsection{Clumpy medium - C10, C30, C100}
Figure~\ref{fig:surf_model} shows the simulations results and semi-analytical model for the shock surface ($\kappa$) and CL-shock fraction ($F_{\text{c}}$). The model agrees to the data well, although there is a systematic overestimation in both variables. Nevertheless, the surface model succeeds in estimations of $\kappa$ and $F_{\text{c}}$ within two orders of magnitude in $\rho_{\text{j}}$ (from C3 to C300), which is a good indication of its validity.

\begin{figure*}
	\centering
	\includegraphics[width=0.33\textwidth]{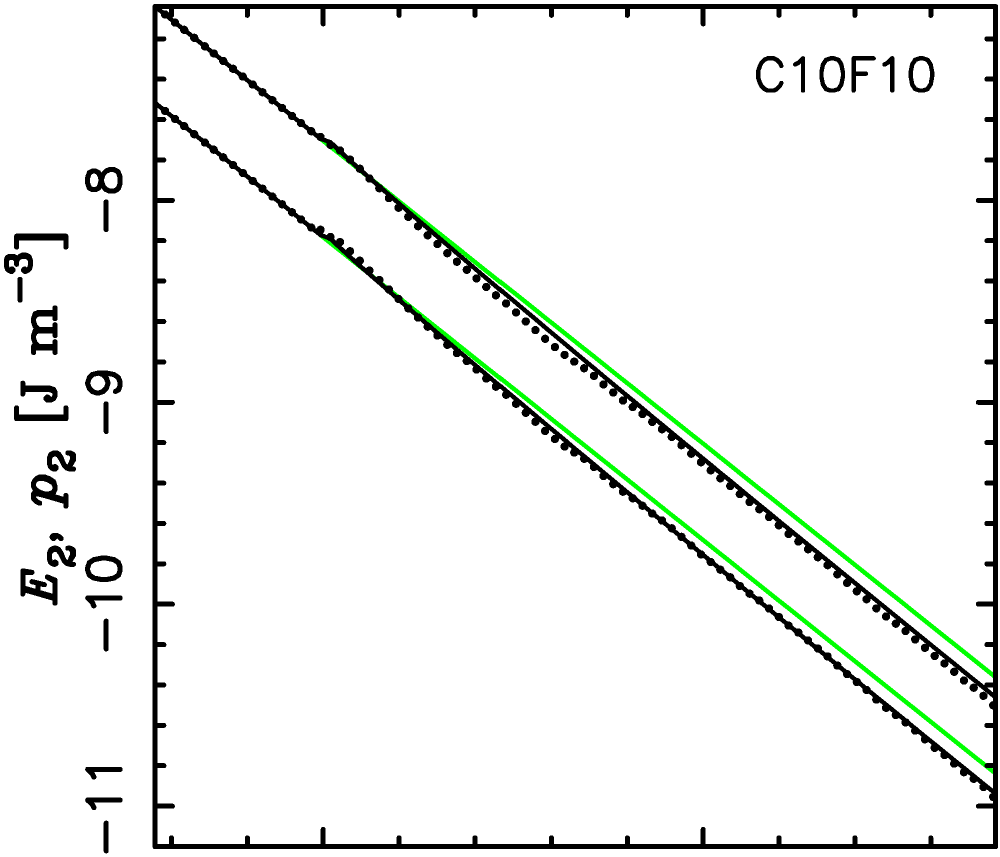}
	\includegraphics[width=0.2805\textwidth]{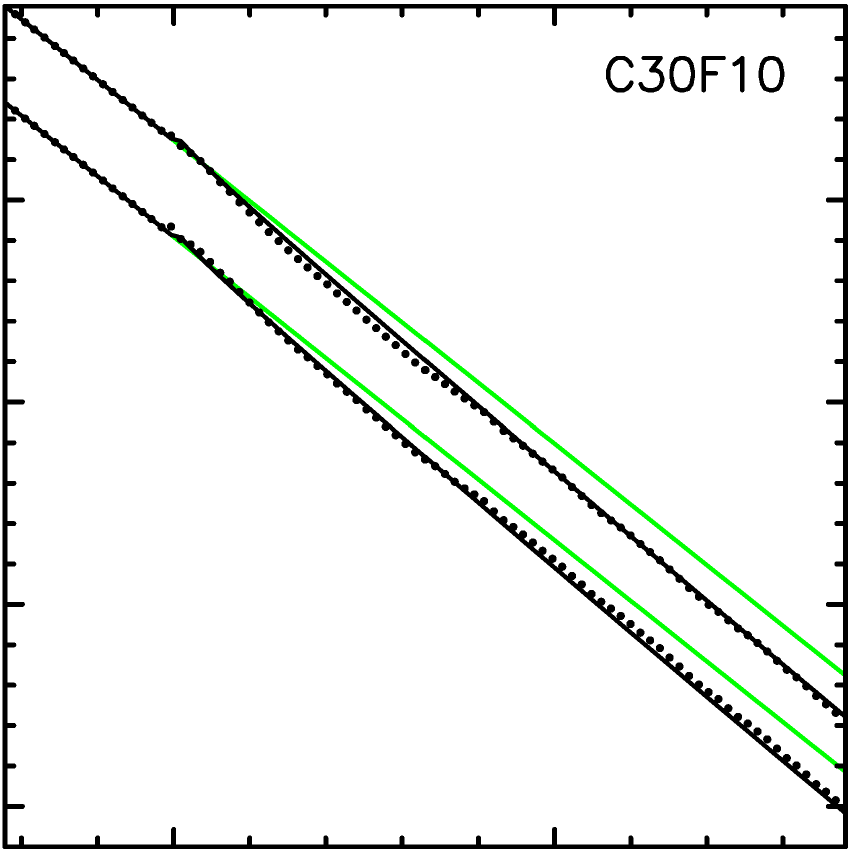}
	\includegraphics[width=0.2805\textwidth]{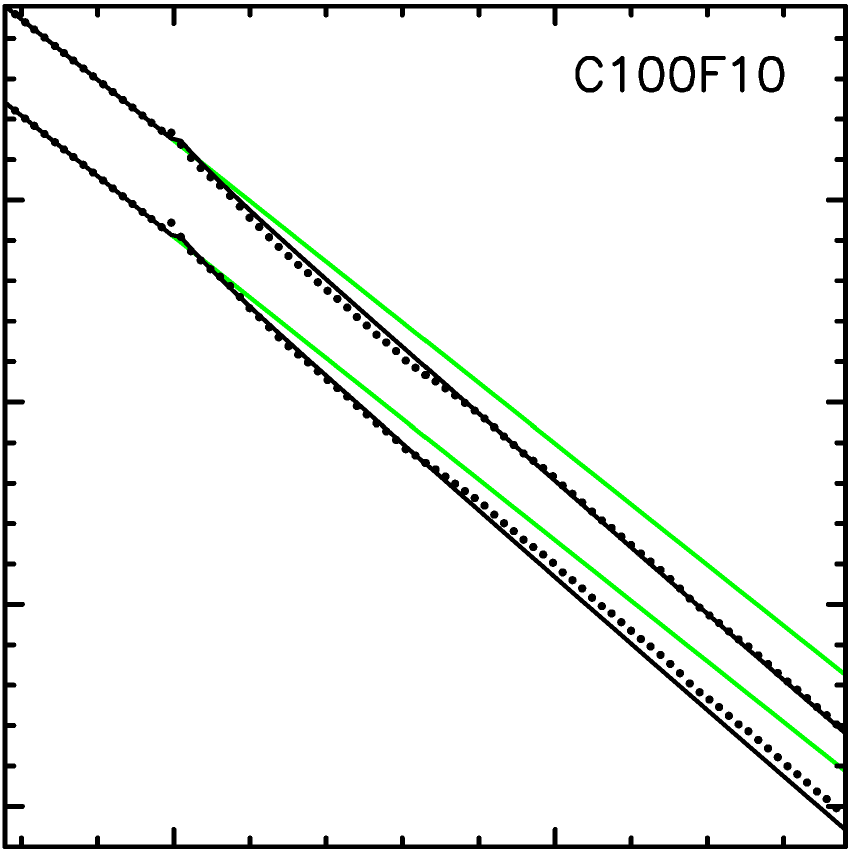}
	\linebreak
	\includegraphics[width=0.33\textwidth]{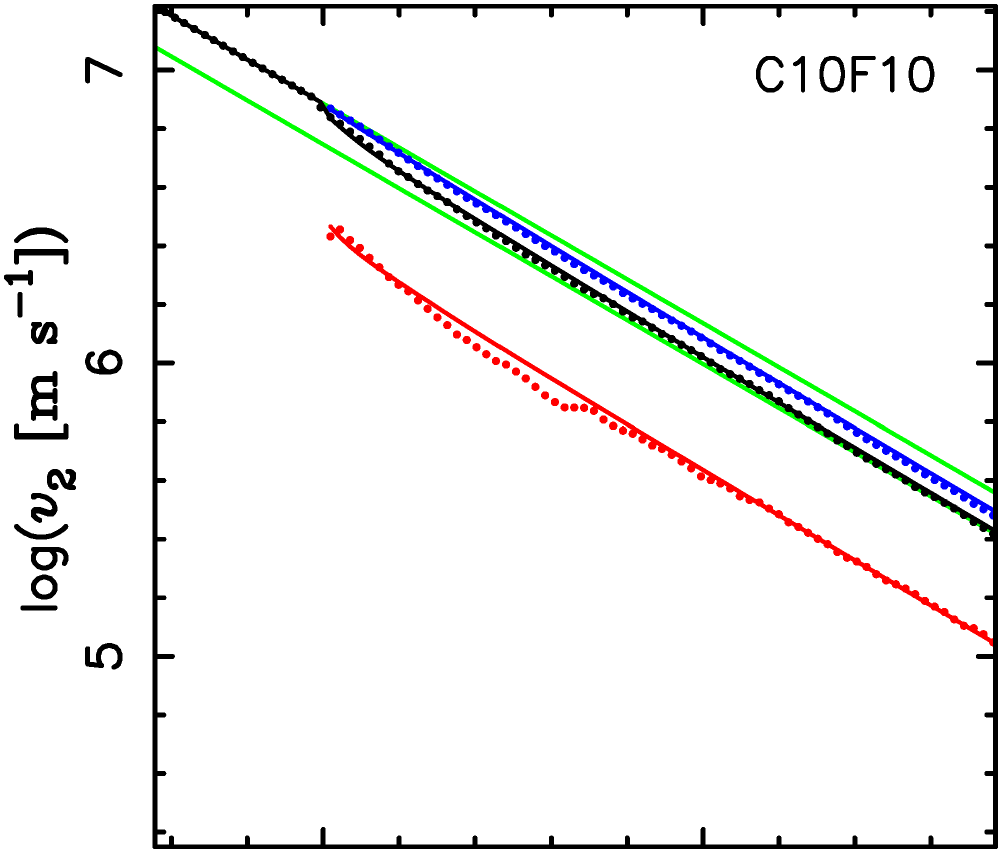}
	\includegraphics[width=0.2805\textwidth]{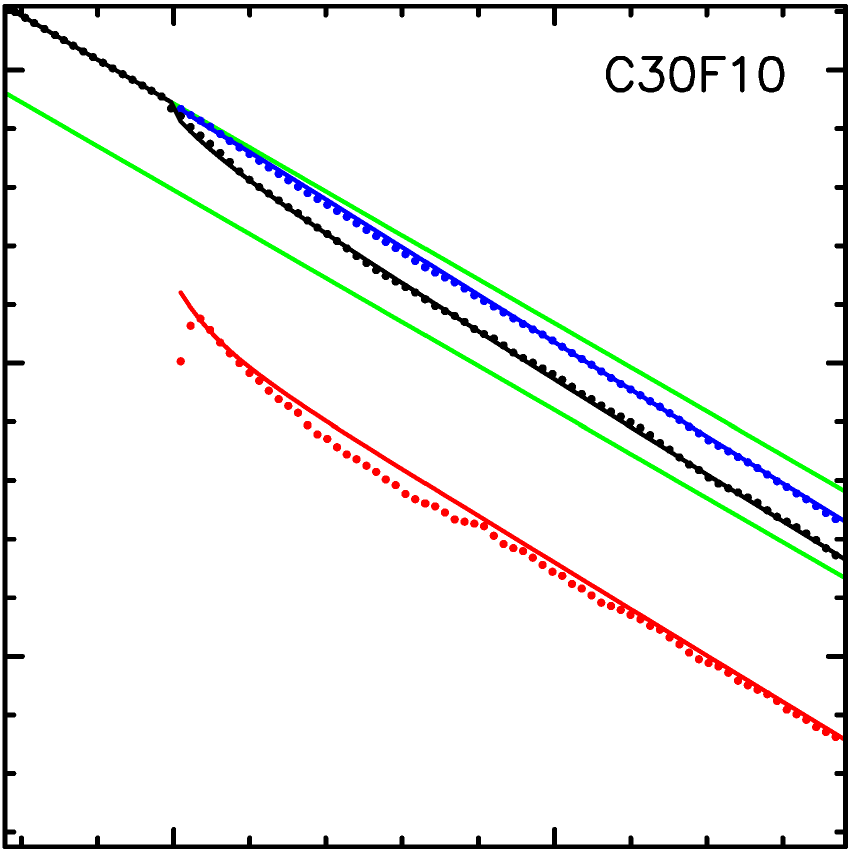}
	\includegraphics[width=0.2805\textwidth]{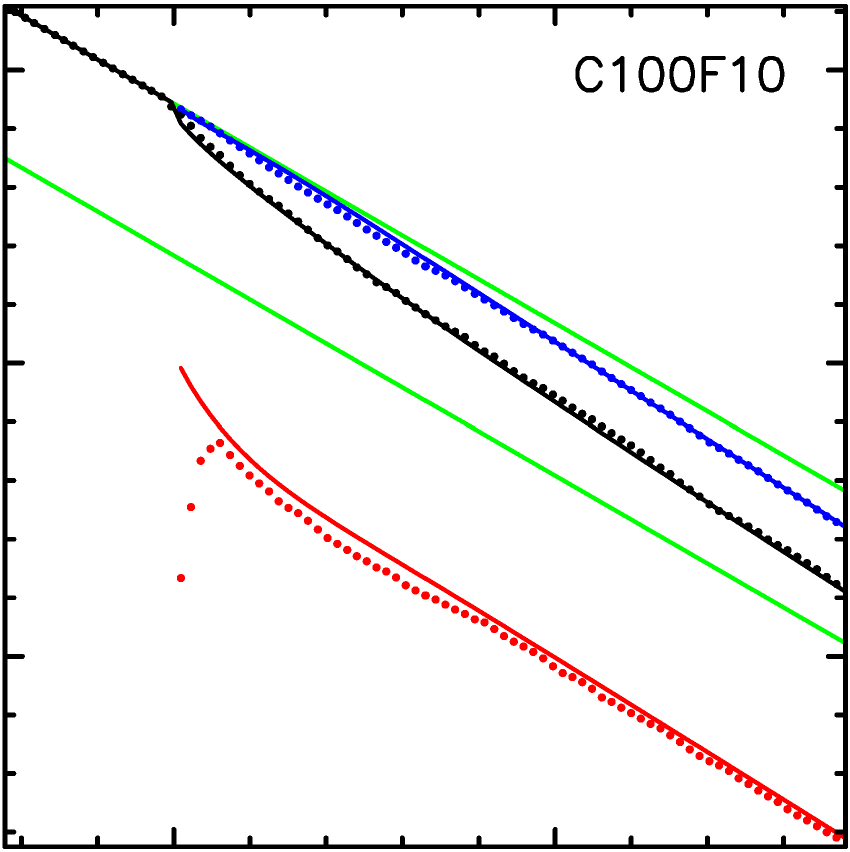}
	\linebreak
	\includegraphics[width=0.33\textwidth]{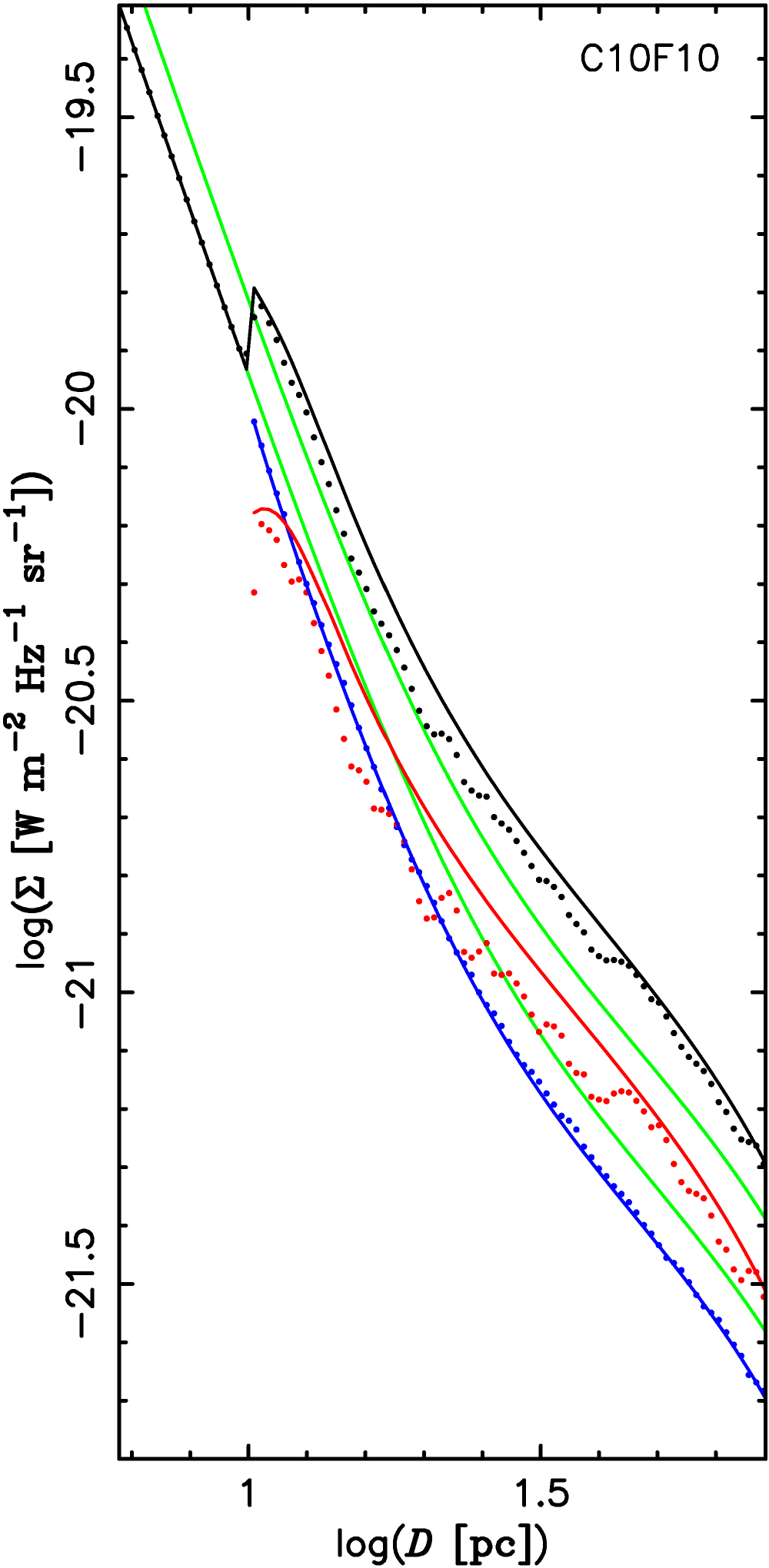}
	\includegraphics[width=0.2805\textwidth]{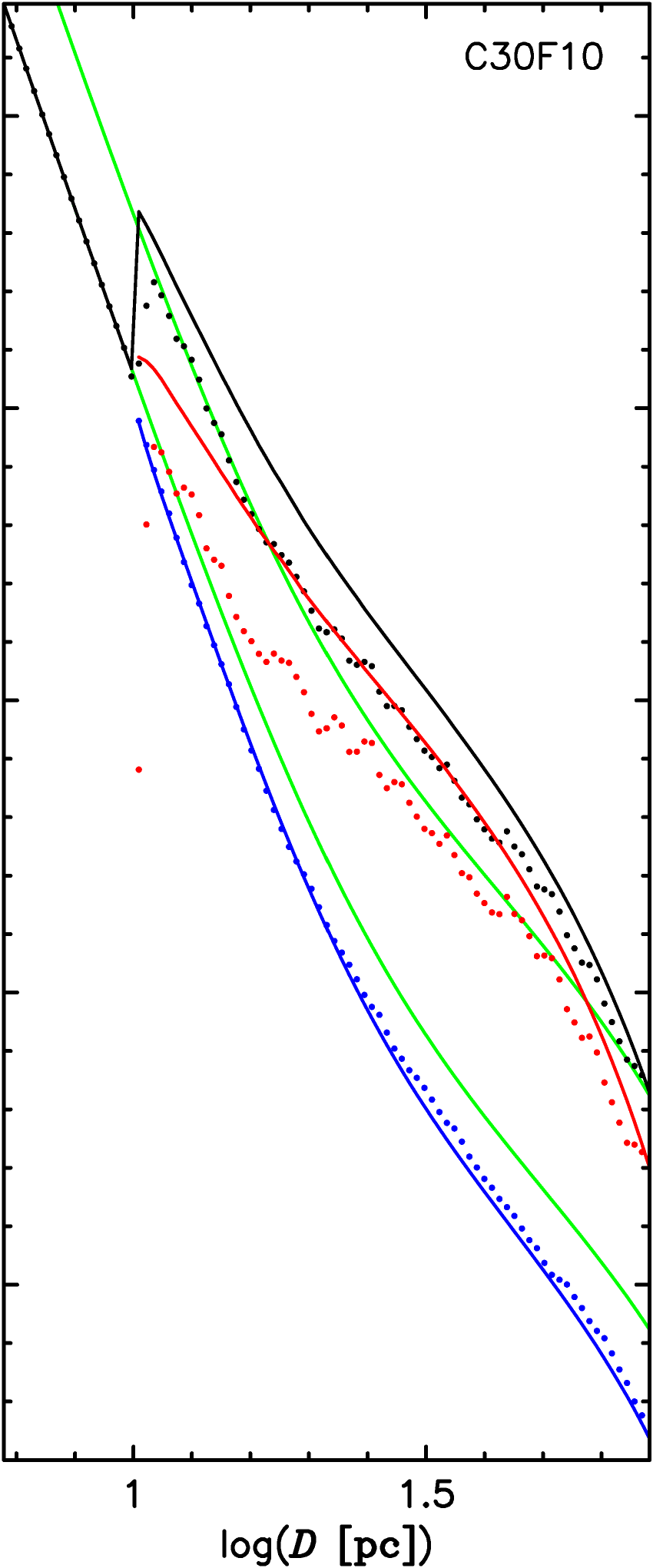}
	\includegraphics[width=0.2805\textwidth]{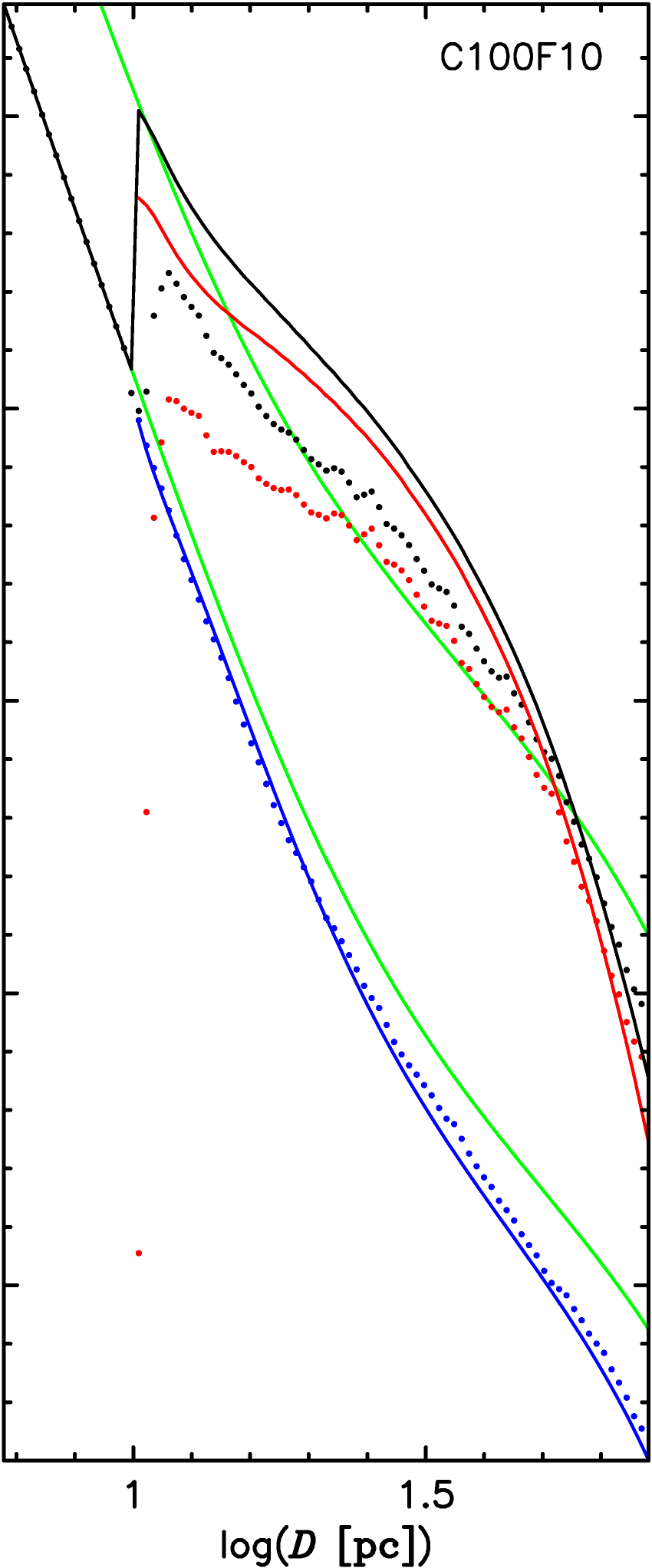}
	\linebreak
	\caption{Results for C10F10, C30F10, and C100F10. The points and lines represent the simulation results and semi-analytical model, respectively. Blue and red color stand for ICM and CL average values, respectively. Green lines show the uniform evolution in $\rho_{\text{icm}}$ and $\bar{\rho}$ ambient densities. \textit{Top}. Black color represents the mean energy (higher curves) and pressure (lower curves) of the forward shock. \textit{Middle}. Velocity graphs: Black color represents the total average. \textit{Bottom}. $\Sigma$--$D$ graphs: Black color represents the total value.}
	\label{fig:CL_F10}
\end{figure*}

\begin{figure*}
	\centering
	\includegraphics[width=0.33\textwidth]{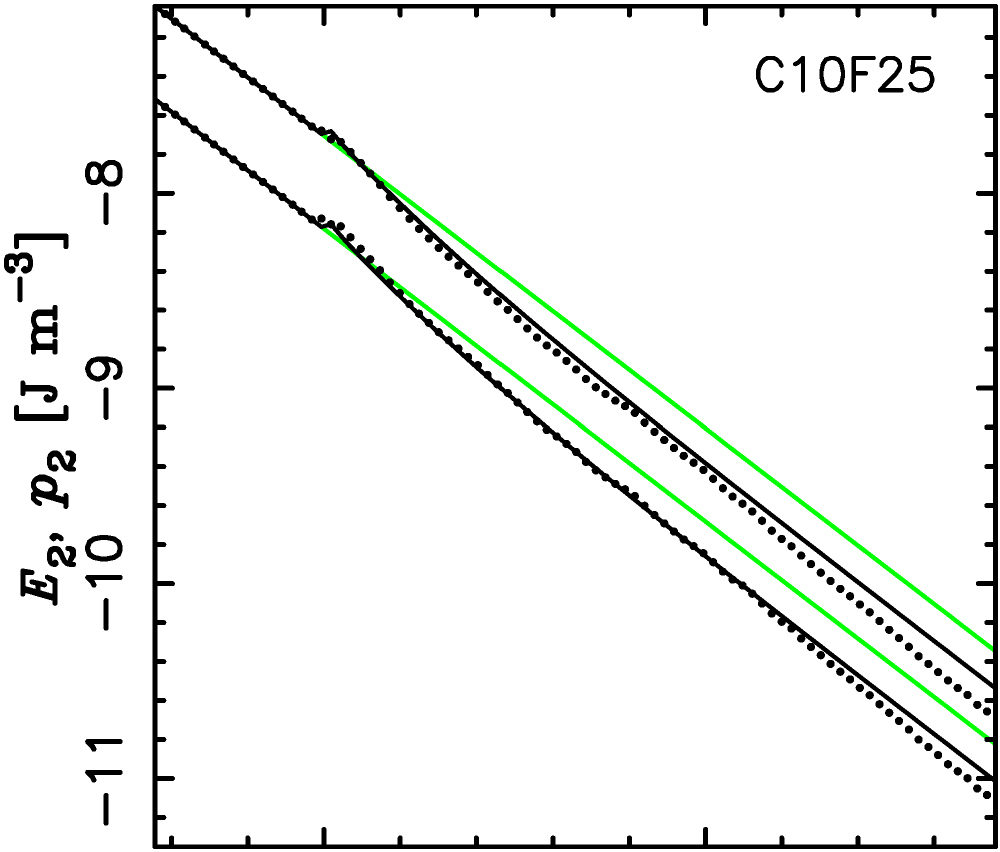}
	\includegraphics[width=0.2805\textwidth]{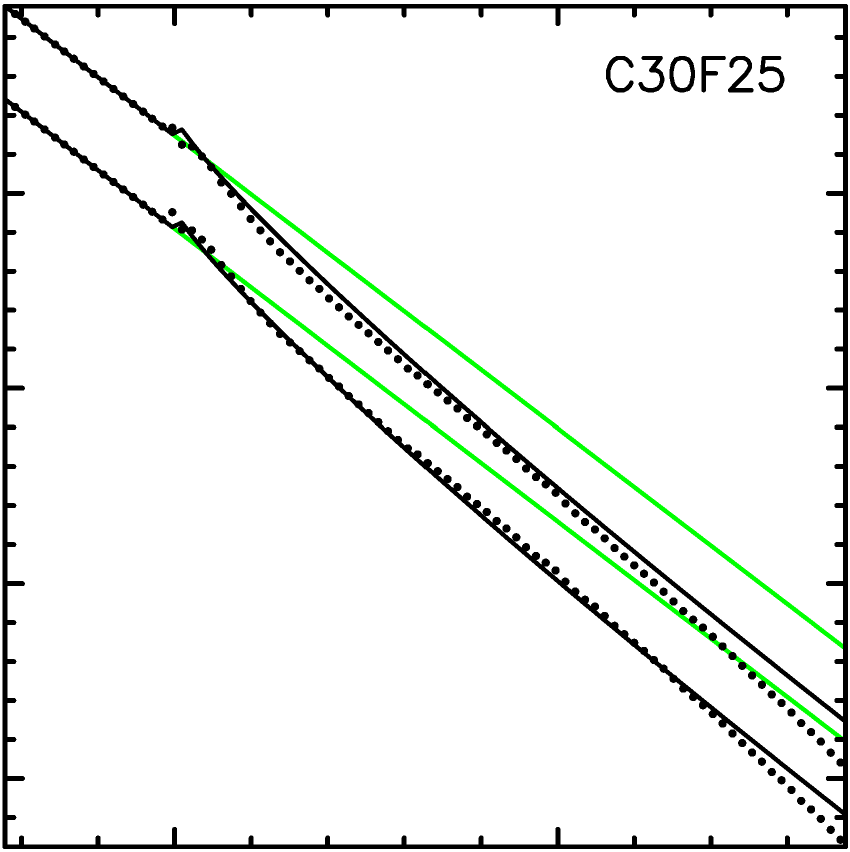}
	\includegraphics[width=0.2805\textwidth]{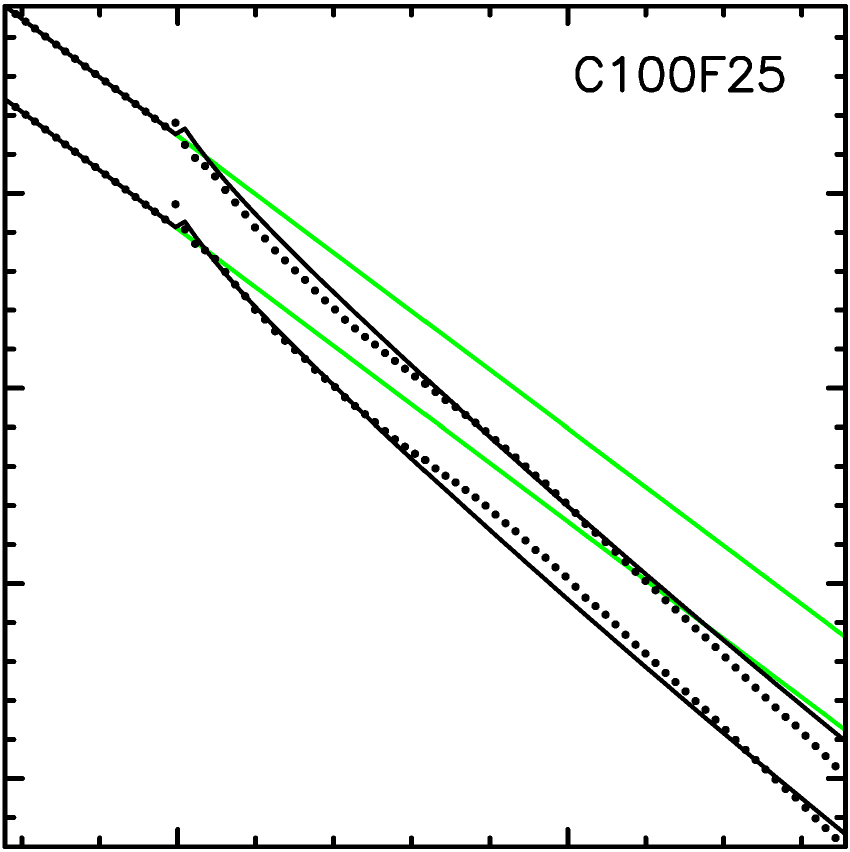}
	\linebreak
	\includegraphics[width=0.33\textwidth]{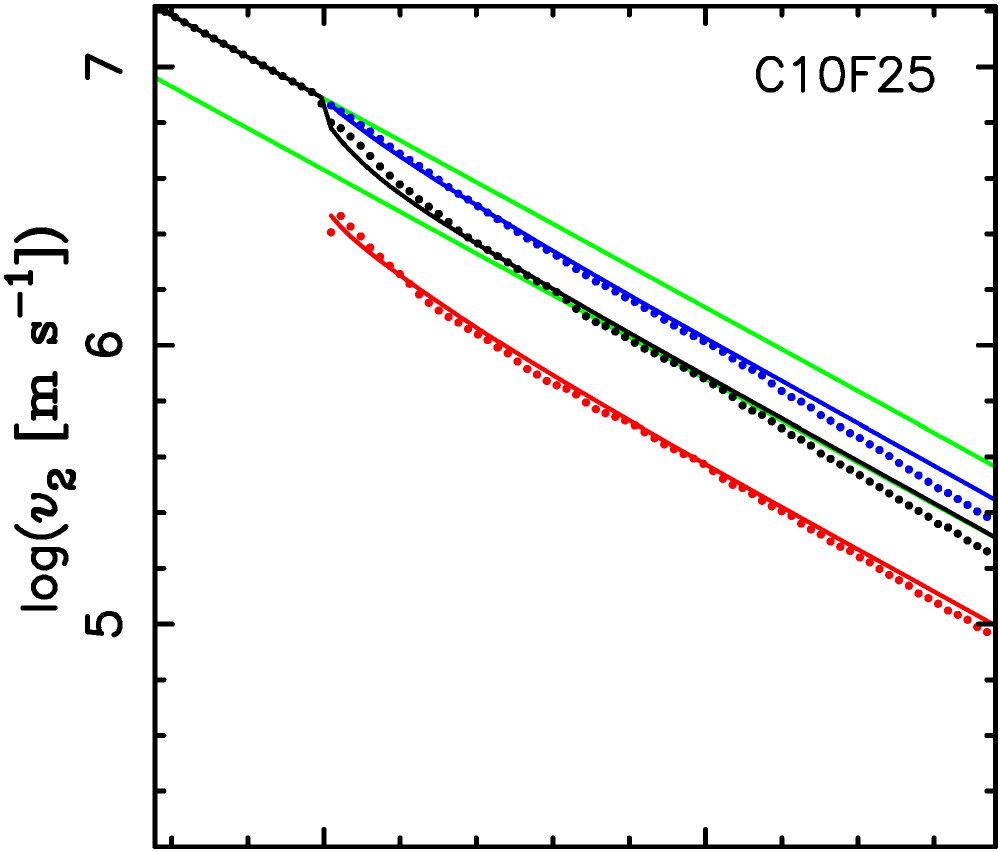}
	\includegraphics[width=0.2805\textwidth]{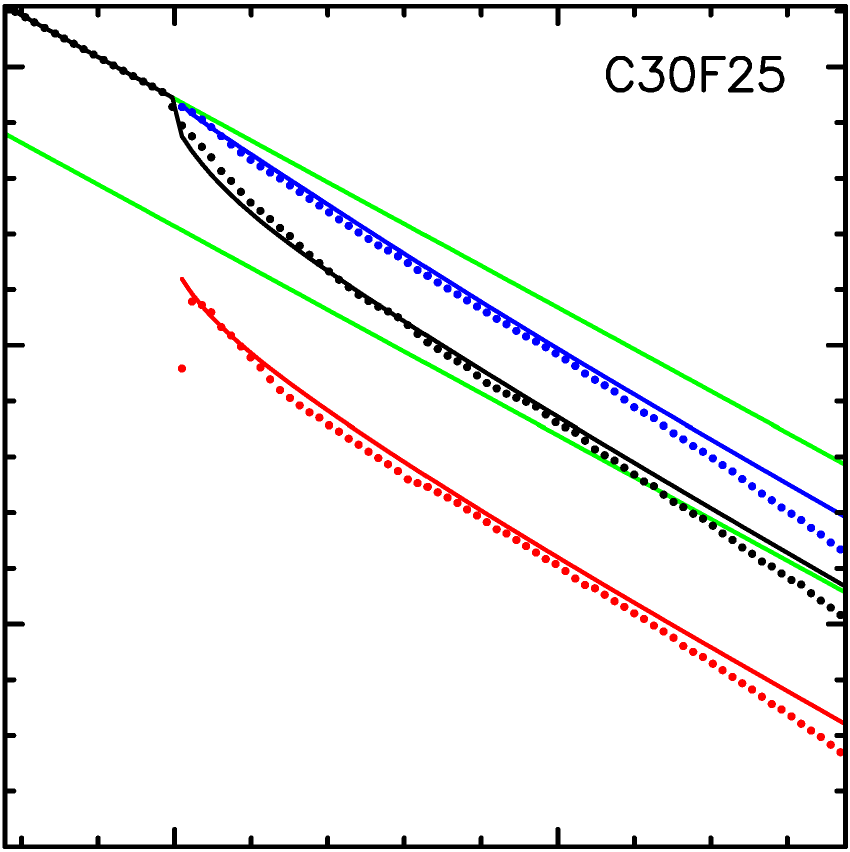}
	\includegraphics[width=0.2805\textwidth]{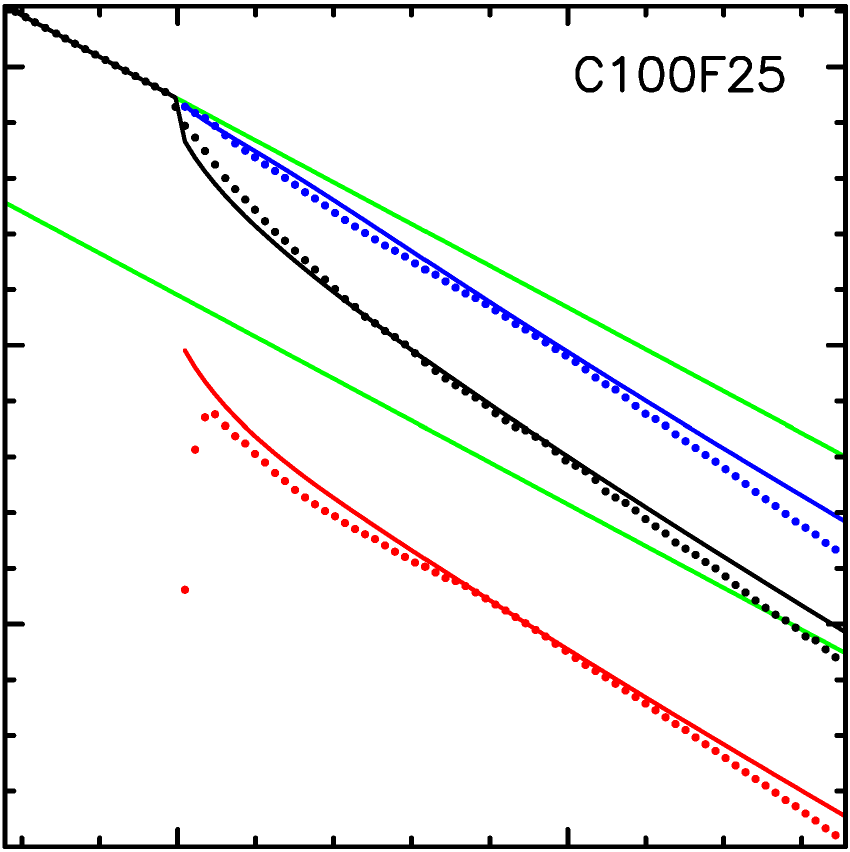}
	\linebreak
	\includegraphics[width=0.33\textwidth]{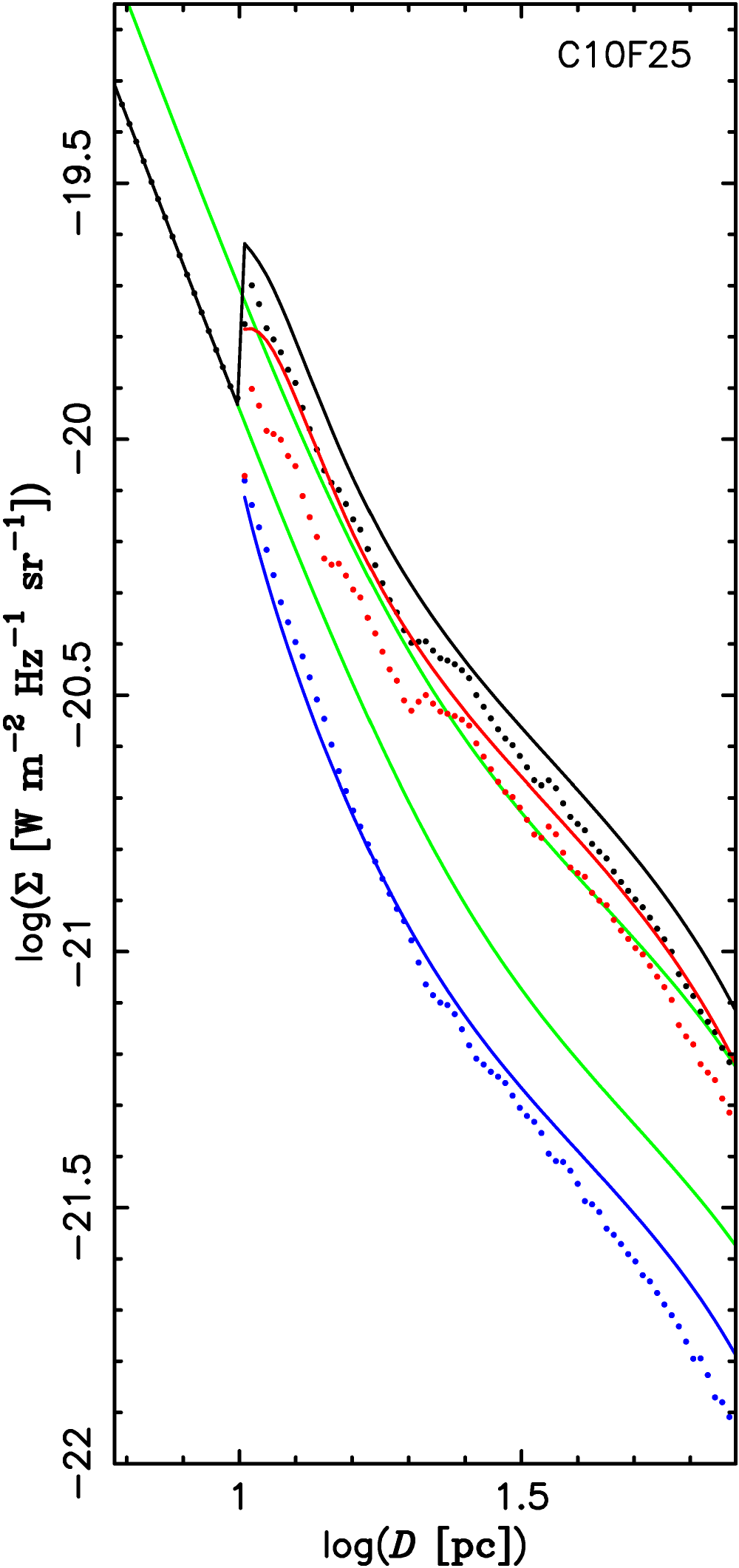}
	\includegraphics[width=0.2805\textwidth]{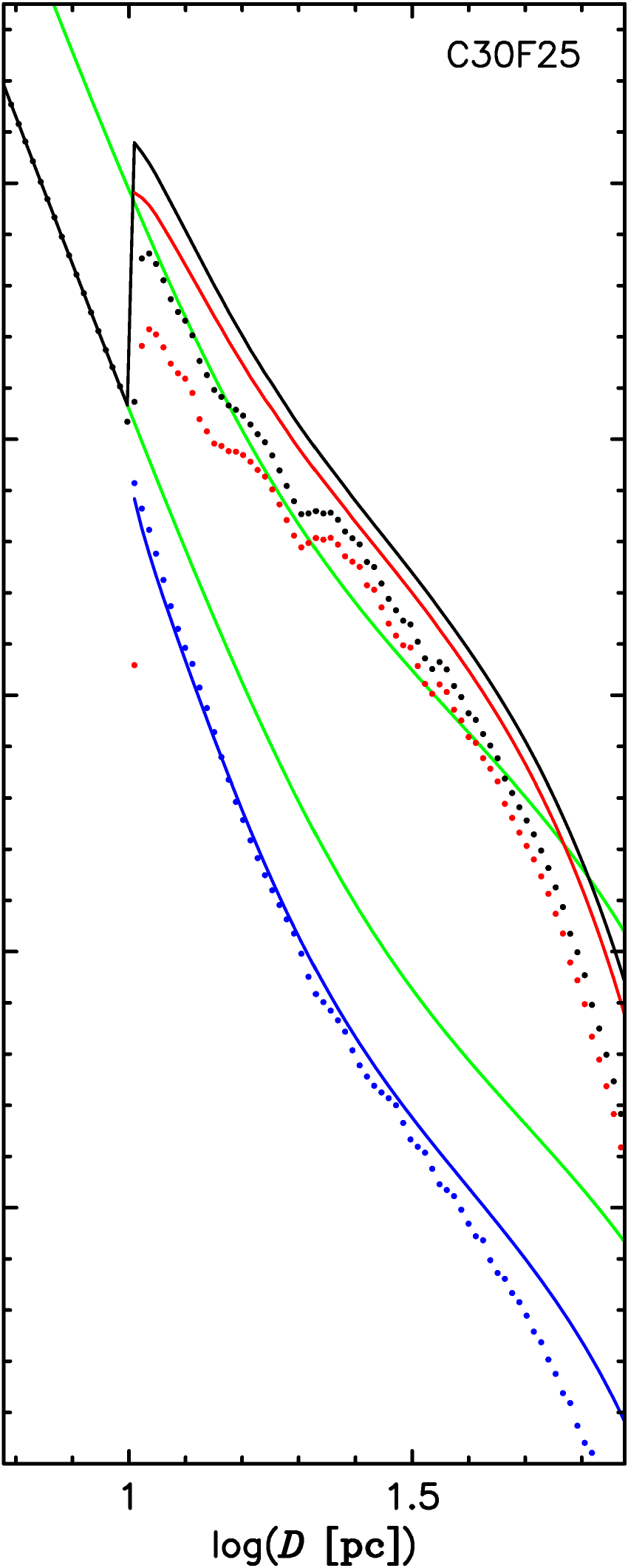}
	\includegraphics[width=0.2805\textwidth]{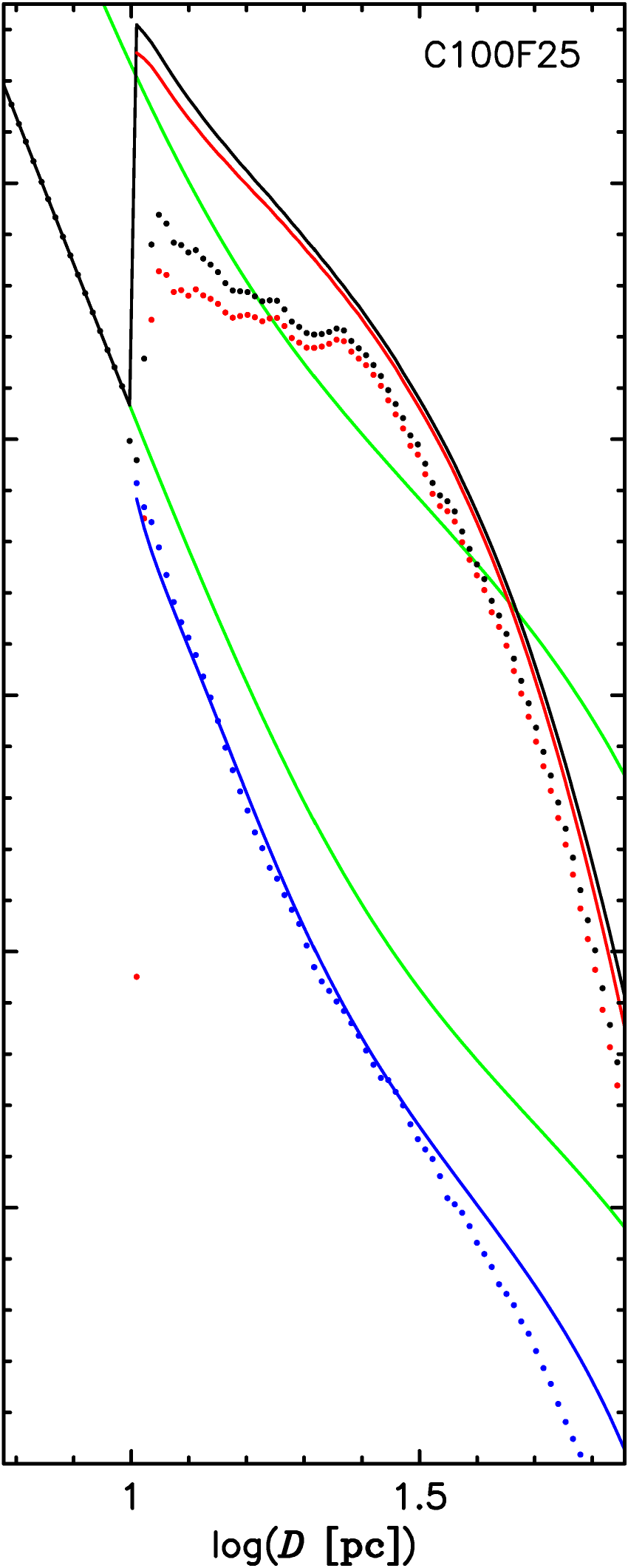}
	\linebreak	
	\caption{Results for C10F25, C30F25, and C100F25. The graphs properties are the same as in Figure~\ref{fig:CL_F10}.}
	\label{fig:CL_F25}
\end{figure*}

The results considering shock dynamics and emission, are presented in Figs.~\ref{fig:CL_F10} and \ref{fig:CL_F25}, for $f_{\text{c}}=0.1$ and 0.25, respectively. The graphs show the evolution of mean energy/pressure at the shock, mean velocity behind the shock, and surface brightness ($\Sigma$--$D$ relation). The mean velocities and surface brightnesses of the ICM and CL shocks are shown separately on the same graphs. The semi-analytical model follows the simulation results, accounting its systematic overestimation of $F_{\text{c}}$ and $\kappa$, that are responsible for the most of the surface-brightness overestimation. The reduction in energy and velocity at radii 5--10~pc (due to reflected shocks) can be noticed in simulations, but is not modeled in semi-analytical model because of complexity. Since the emission depends steeply on shock velocity, this reduction is especially noticeable in $\Sigma$--$D$ graphs.

It is obvious in setups with higher density contrasts, that the clump velocity starts at a very low level (at discontinuity) and quickly rises to the expected level, which is not expected hydrodynamic behavior. This appears also in LDB10 setup, although to a lesser extent. The most probable explanation is that the cells of the highest velocity and momentum do not match (the algorithm searches for highest momentum, see Section~\ref{sec:shock_detection}) during the first several time steps of shock transition from ICM to a clump. The clump's border cells, having a much higher density, may have more momentum even if the velocity is lower than in bordering ICM cells. In this way, the algorithm may fail in finding the cells with maximum velocity. Later on, when the shock fully enters a large number of clumps, this effect fades out and the mean shock velocity approach the expected value. Hence, we will ignore this numerical artifact (that also translates to the surface brightness) and conclude that the semi-analytical model offers a more correct result immediately after the discontinuity. 

\section{Discussion}
\label{sec:discussion}

\subsection{Interstellar and circumstellar medium}
The surveys of interstellar medium strongly indicate that it has a fractal structure and is organized in clumps of matter surrounded by lower density medium. These structures look similar at wide span of spatial scales from $\sim$0.1 to $\sim$100~pc \citep{BazellDesert1988, Vogelaaretal1991, ElmegreenFalgarone1996, Elmegreen1997, Federrathetal2009}. So, in general the ISM is not homogeneous and it is justified to assume that a large part of SNRs will interact with the clumps during their evolution. Indeed, some $\sim$30--50$\%$ of observed Milky Way (MW) SNRs are in contact with molecular clouds \citep{Vukoticetal2019, Jiangetal2010}.

However, the massive stars alter the surrounding medium by stellar winds and ionizing radiation, forming the circumstellar medium around them. These are mainly O and B stars with masses $\sim$8--35 $M_{\sun}$, most of which will end their life in red supergiant (RSG) stage. Some of the more massive stars ($>$35 $M_{\sun}$), which are much less frequent, may during the course of evolution lose outer hydrogen envelope and explode in Wolf-Rayet stage. The circumstellar medium (CSM) around the pre-SN RSG star consists of a low-density wind-blown bubble, carved by fast main sequence (MS) wind and enveloped by dense shell of swept material, with centrally distributed slow power-law RSG wind extending to a radius of $\sim$1--5~pc \citep[see e.g.][]{Dwarkadas2005}. In WR star case, the very fast WR wind pushes the RSG material forward, forming the central low-density bubble with dense shell that eventually can merge to a MS bubble shell \citep{Dwarkadas2007}. Although in theory these structures are often seen as stationary and homogenized, the studies showed that wind, even in spherically symmetric medium, develops clumps with density contrast up to $\sim$2 orders of magnitude \citep{Surlanetal2012, Surlanetal2013}. Even higher density contrast develops in the region of thin dense shell, although its density structure is more like net of filaments \citep[see][for 3D simulations of WR winds]{vanMarleKeppens2012}. Moreover, the ISM around the stars is rarely uniform, so the shell also bends around the ISM clumps, leading to further clumping.

Considering these implications, we try to reconstruct the Galactic SNR sample by applying simple models of possible types of CSM and ISM in the Milky Way. The models are not intended to give the resulting values of CSM/ISM parameters, but to demonstrate the magnitude of impact that clumping and ISM inhomogeneity could have on SNR radio-synchrotron evolution.

\subsection{Impact of the clumpy medium on $\Sigma$--$D$ curve}
\begin{figure*}
	\centering
	\includegraphics[height=0.59\textwidth]{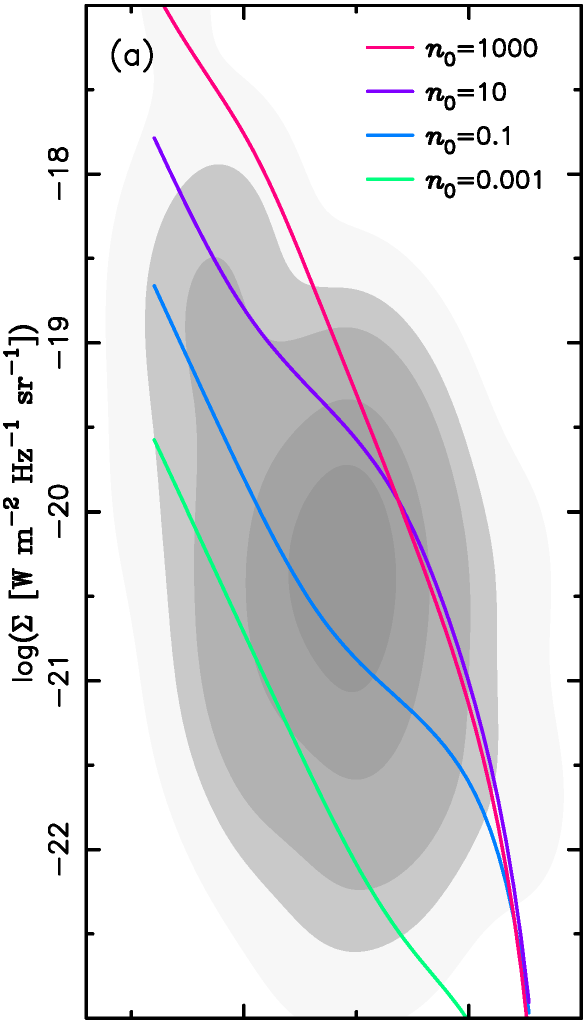}
	\includegraphics[height=0.59\textwidth]{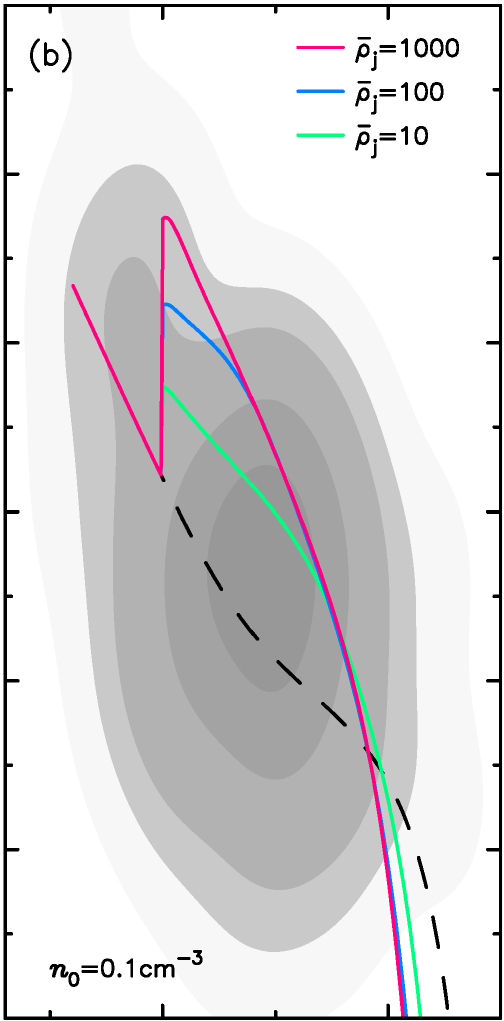}
	\includegraphics[height=0.59\textwidth]{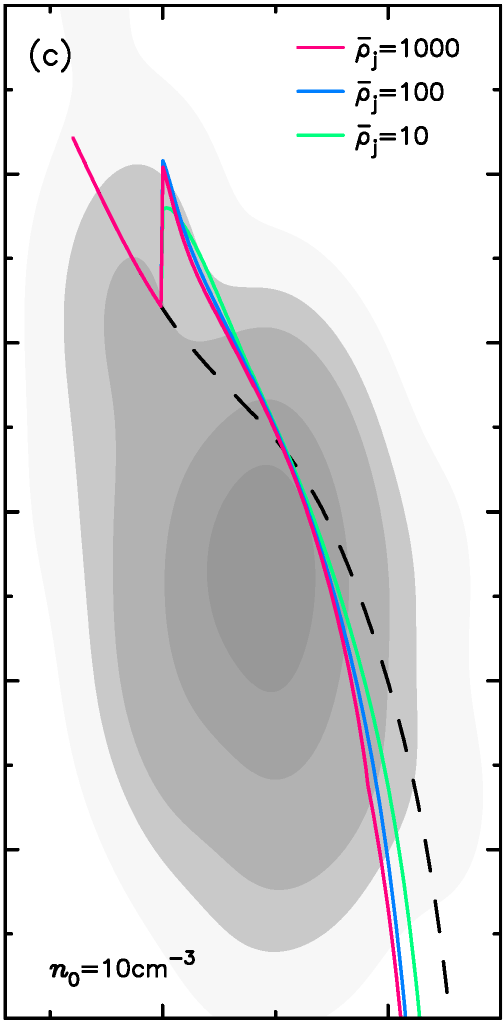}
	\linebreak
	\includegraphics[height=0.63\textwidth]{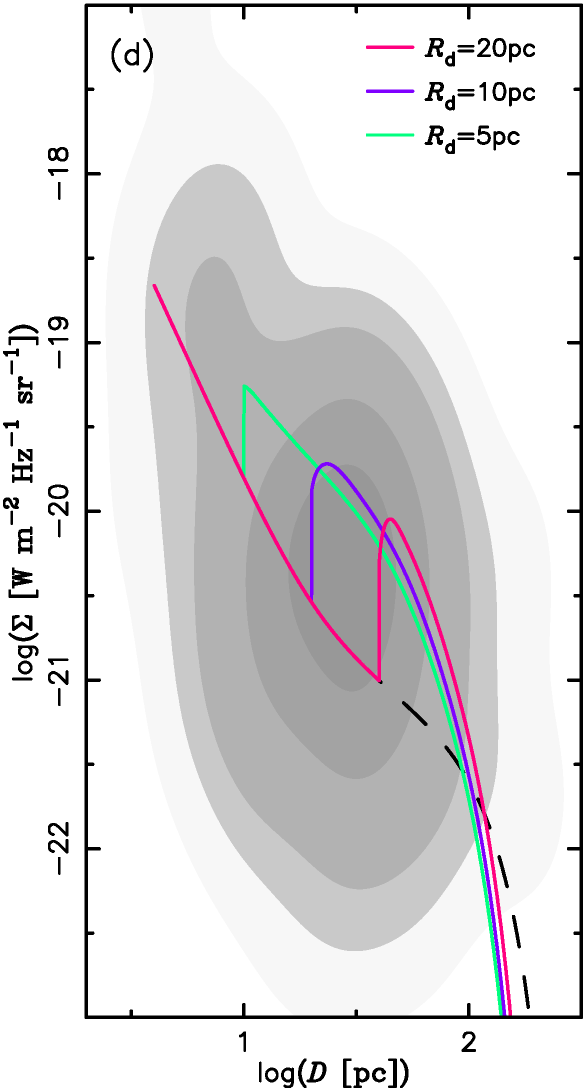}
	\includegraphics[height=0.63\textwidth]{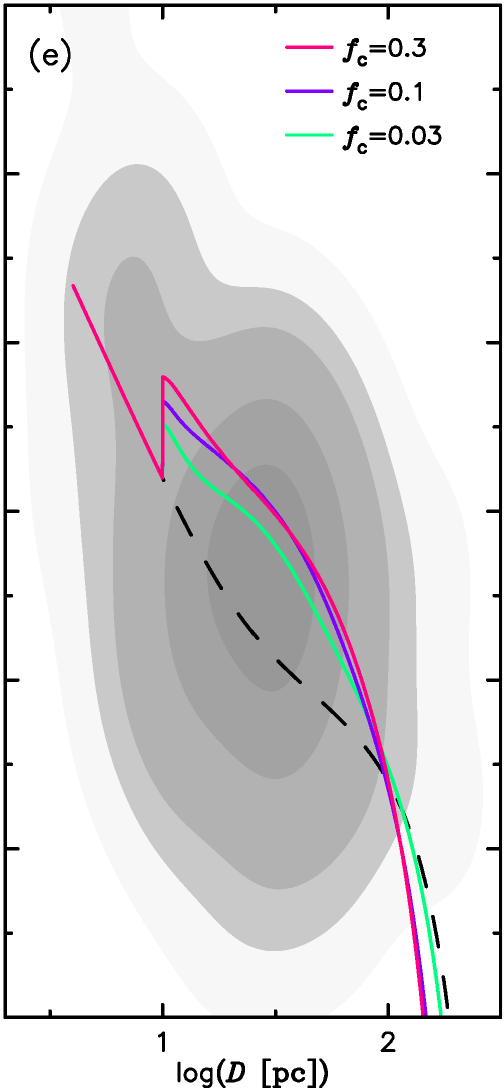}
	\includegraphics[height=0.63\textwidth]{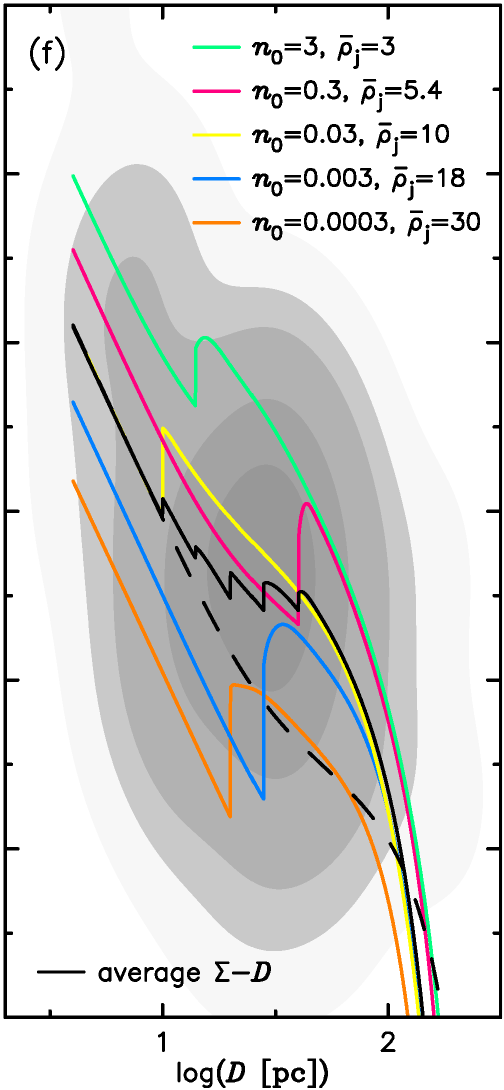}
	\linebreak
	\caption{Impact of clumpy medium properties on $\Sigma$--$D$ curve (except for the panel (a) that shows different uniform densities). The plot legend shows the variables of interest ($n_0$ is in cm$^{-3}$). The values of all other variables that are not explicitly written on graph are: $n_0=0.1~\text{cm}^{-3}$, $\bar{\rho}_{\text{j}}=10$, $f_{\text{c}}=0.2$, $R_{\text{d}}=5$~pc. The black dashed line represents the uniform case for $n_0=0.1$, for panels (a) to (e), and $n_0=0.03$, for panel (f). The clump radius $r_{\text{c}}=1$~pc is the same for all graphs.}
	\label{fig:impacts}
\end{figure*}

Now when the models of emission and HD evolution are developed (in Sects.~\ref{sec:model_emission} and \ref{sec:semi-analytical_model}), we can investigate how the clumpy medium and its properties affect the $\Sigma$--$D$ curve of the SNR evolution. The main properties of clumpy medium that have significant impact on the SNR emission are: density of the uniform/ICM medium (or ``ground'' density, $\rho_0$), density contrast between clumps and ICM ($\rho_{\text{j}}$), volume-filling factor of the clumps ($f_{\text{c}}$), and the radius of the uniform--clumpy medium border ($R_{\text{d}}$). The impact of every of these properties alone is shown in Figure~\ref{fig:impacts}. The background of the graphs is filled with probability density distribution (PDD) of Milky Way SNR sample taken from \citet{Vukoticetal2019}. PDD on graphs, obtained by the method from \citet{Duin1976}, serves for easier comparison of the $\Sigma$--$D$ curves to each other and to the MW sample. The panel (a) shows $\Sigma$--$D$ evolutions in wide range of different uniform densities, from which is obvious that in denser medium the $\Sigma$--$D$ slope tends to be steeper. This is caused by less CR acceleration efficiency at lower shock velocities. Consequently, in late stages of evolution when shock velocity weakens, the surface brightness of high density SNRs falls to a level of the ones evolving in lower densities, even lower. The slope $\beta$ is not constant but is characterized mainly by slight flattening through the evolution followed by steepening in final stages, due to the switching between non-resonant and resonant magnetic field amplification regime. The same kind of behavior is observed in simulations of \citet{Pavlovicetal2018}. The panel (b) shows how the $\Sigma$--$D$ evolution changes with different mean density jump, $\bar{\rho}_{\text{j}}=f_{\text{icm}}+f_{\text{c}}\rho_{\text{j}}$. The same is in the panel (c) but for higher ground density.
The difference between (b) and (c) is in height of the emission jump and slope of the consequent evolution, because for higher ground densities the shock is weaker, so is the particle acceleration (as shown in right panel of Figure~\ref{fig:jumps_eff}). The panel (d) shows the expected impact of changing $R_{\text{d}}$: the last jump is highest because of less density in previous evolution. The panel (e) shows a relatively small dependence on $f_{\text{c}}$. As the number of clumps increases, the CL-shock fraction rises and contributes to the brightness, but the shock is more slowed-down by the clumps (lowering the brightness), so this dependence is rather weak.

The most interesting is the panel (f) where five exemplary $\Sigma$--$D$ curves and their average are plotted. The important statistical assumption introduced here is that the mean density jump $\bar{\rho}_{\text{j}}$ decreases with increasing ground density. The idea behind this is simple: as the radiation pushes the surrounding medium away, its mass accumulates beyond some radius. So, the more density is cleared within a bubble, the more density is accumulated around. Moreover, one can assume that far away from star, even beyond the shell of swept medium, the matter on average has some universal mean density. Following this assumption, we will link the mean density jump to the ground density in the model of Galactic SNR sample. The panel (f) also shows a property that is coming from this linkage, which is useful for modeling MW sample: the average $\Sigma$--$D$ curve significantly flattens at the domain of uniform-to-clumpy medium transition. It doesn't matter if SNRs in high $\rho_0$ arrive earlier at clumps than SNRs in low $\rho_0$, or vice versa. Even in (intentionally) random distribution of $R_{\text{d}}$ among the five SNR evolutions, their average $\Sigma$ evolution steadily and gradually flattens throughout the transition domain, whereupon it continues to fall more steeply. The level of flattening can be perceived by comparing with the black dashed line representing the average uniform evolution.

From all the graphs in Figure~\ref{fig:impacts} showing the brightness jumps at clumpy medium, it is clear that after the sudden bump in the emission eventually comes the ``exhaustion'' of the shock energy due to high swept mass, so the surface brightness later falls even below the level of the uniform evolution (without any clumps). This happens to be the main property of the $\Sigma$--$D$ evolution through the clumpy medium in our model: immediate increase in brightness, followed by decline that gets steeper with decrease of shock velocity. In other words, the clumps shorten the lifetime of the SNR synchrotron evolution. The consequence on the Galactic ensemble of the SNRs is that, after the initial $\Sigma$--$D$ flattening, the average slope also gets steeper than in the uniform medium case (e.g. in Figure~\ref{fig:impacts}, after discontinuity the magenta line is globally significantly steeper than the black dashed line). This can partly explain the large scatter in Galactic $\Sigma$--$D$ samples.

\subsection{Galactic SNR sample analysis}
\label{sec:mw_models}
Based on the impact of clumpy medium properties on $\Sigma$--$D$ evolution, explored in previous subsection, the three models were made in an attempt to reconstruct the Galactic SNR sample (Model A, B, and C), using the semi-analytic model from Section~\ref{sec:semi-analytical_model}. The input parameters are: $R$, the SNR radius (predefined distribution); $\rho_0\equiv\rho_{\text{icm}}$, the uniform and ICM density (logarithmic distribution); $\bar{\rho}_{\text{j}}=f_{\text{icm}}+f_{\text{c}}\rho_{\text{j}}$, the mean density jump from uniform to clumpy medium (log. dist.); $f_{\text{c}}$, the volume-filling factor of clumps; and $R_{\text{d}}$, the inner radius of clumpy medium (log. dist.). The sample consists of 1000 points, with random inputs for listed variables (in certain ranges, with noted distributions). The comparison of the modeled with Galactic sample were done using the normalized PDDs of the points in $\Sigma$--$D$ plane (same as in Figure~\ref{fig:impacts} for MW sample). The Galactic sample was taken from \citet{Vukoticetal2019}, without the three outliers (one at $\log \Sigma >-17$ and two at $\log D > 2.15$).

The radius R is picked randomly between 2.3 and 50~pc, keeping the smoothed diameter distribution of MW sample. The range of $\Sigma$ values in MW sample allows $\rho_0$ to have a span of 3--4 orders of magnitude. With the adopted thickness of the emitting shell of $0.01R$, the matching values of ambient densities fall approximately between $5 \cdot 10^{-4}$ and $8~\text{cm}^{-3}$. The typical WBB densities are very low in this spectrum \citep[$\sim$10$^{-4}$--$10^{-3}~\text{cm}^{-3}$, see][]{Vink2020}, but the stars also explode in denser environments. The SNR shell thickness is usually assumed to be 10\% of the radius, corresponding to a compressed shell at $X=4$. However, due to a steep fall of density behind the shock, and the fact that we use $n_2$ from the shock peak, it is more accurate to use a value significantly less than 10\%, to avoid overestimation of total number of emitting electrons.

\begin{figure*}
	\centering
	\includegraphics[height=0.69\textwidth]{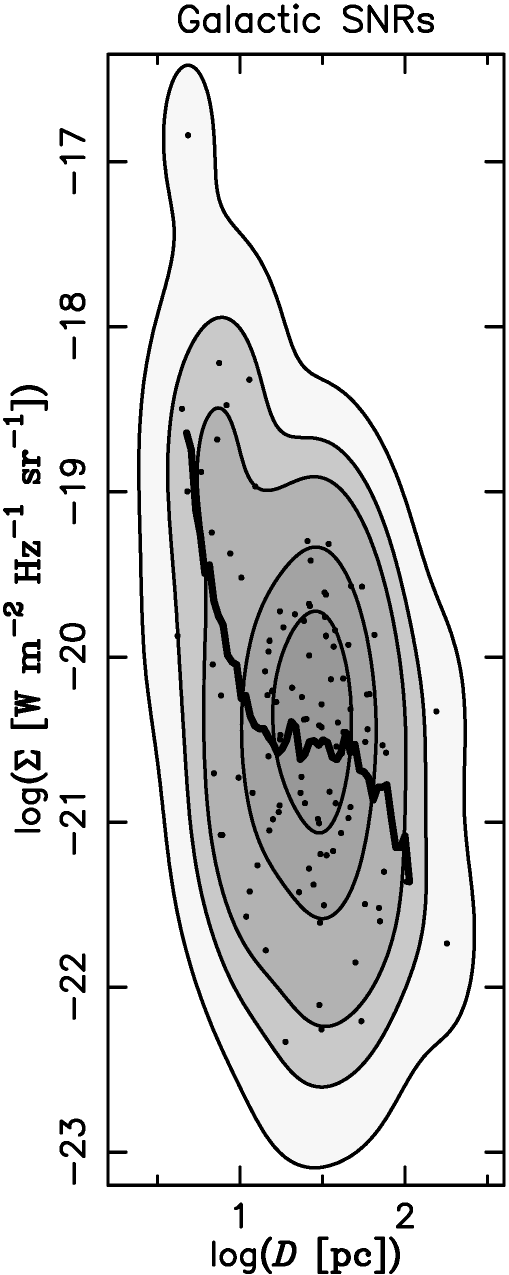}
	\includegraphics[height=0.69\textwidth]{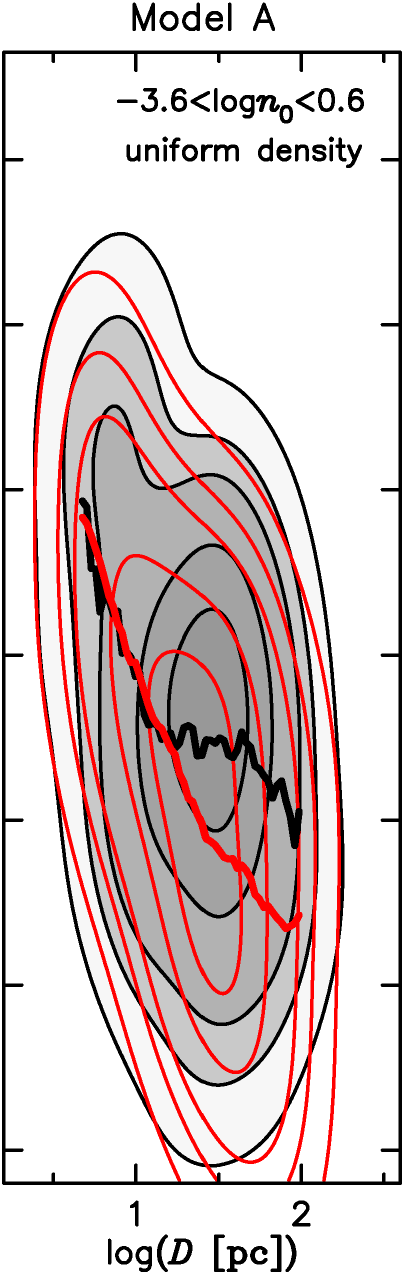}
	\includegraphics[height=0.69\textwidth]{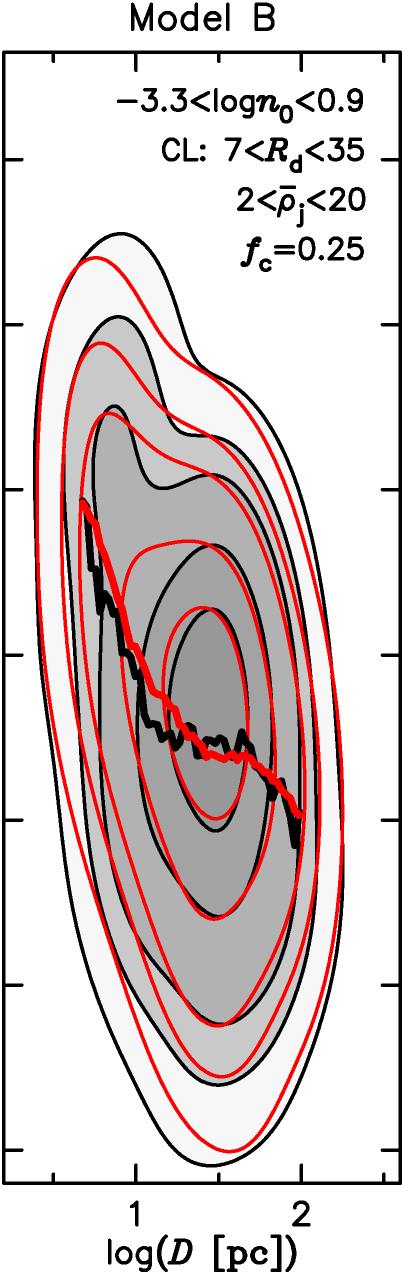}
	\includegraphics[height=0.69\textwidth]{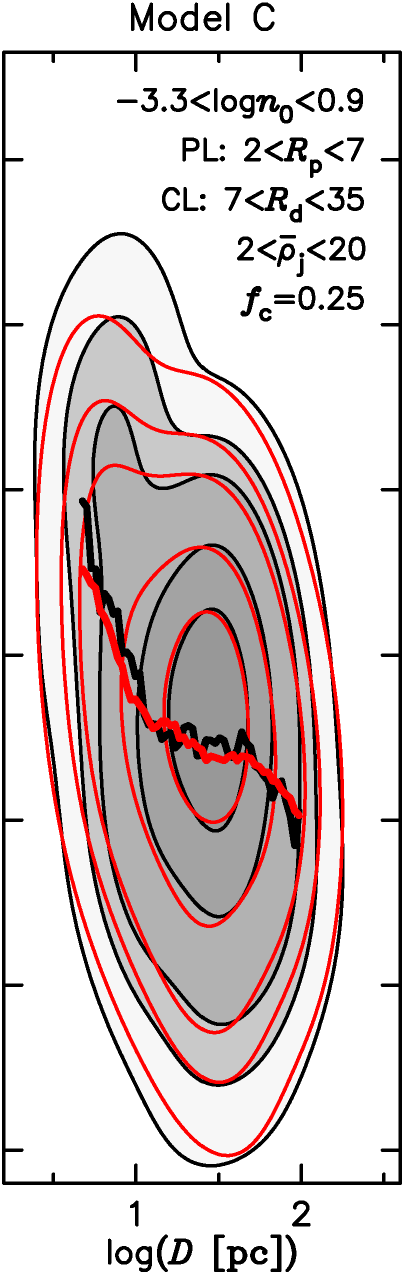}
	\caption{Panels show the Galactic SNR sample from \citet{Vukoticetal2019} and its comparison to Models A, B, and C. The contours represent the normalized probability density distributions at levels 0.025, 0.1, 0.2, 0.5, and 0.8. The probability density distribution is made using the method from \citet{Duin1976} \citep[same as in][]{Vukoticetal2019}. The thick line represents the bin-averaged value of $\Sigma$ depending on $D$ (the line connects the average $\Sigma$ in 50 bins, the width of a bin is 0.26). The black color refers to MW sample, and the red refers to a modeled sample. The model parameters are listed on graphs; PL -- power-law density distribution, CL -- clumpy medium, $n_0$ is in cm$^{-3}$, $R_{\text{p}}$ and $R_{\text{d}}$ are in pc. The three outliers that are present in MW sample are removed from the three graphs comparing models to the MW sample.}
	\label{fig:MWsample}
\end{figure*}

Looking at the distribution of MW SNRs on a $\Sigma$--$D$ plot in Figure~\ref{fig:MWsample}, it seems that the average $\Sigma$ evolution has a break at $D\approx13$~pc. This is especially indicative when comparing with Model A, the trivial artificial sample of SNRs evolving in the span of uniform densities, as shown in Figure~\ref{fig:MWsample}. It is clear that the black curve starts to flatten more than is theoretically expected (having a slope $\beta\approx0$ at one point), pretty much like the average curve in the panel (f) of Figure~\ref{fig:impacts}. The reasons for this can be numerous, but we hypothesize that it is due to a significant deceleration of SNR expansion by either molecular clouds or WBB shells, resulting in a rise of emission. The fact that the SNR distribution in diameter peaks at $\approx$31~pc and falls steeply toward higher $D$ values supports this viewpoint. However, since the surface brightness does not show dramatic fall after this peak, which would certainly follow the interaction of the shock with $\sim$10$^3$ times denser homogeneous and spherical shell \citep[due to rapid radiative cooling, see][]{Haidetal2016}, we conclude that this compressed shell must be clumpy and radially extended. Apart from the irregular structure of the local ISM, this could be caused by the asymmetric WBB shape with regard to the center of explosion, and gradual shocking of its shell.

It is shown that, when remnants arrive at higher density clumps the radio surface brightness rises, most significantly at the very beginning. So, assuming the minimum radius of WBBs or cloud-free environment, beyond which the remnants reach clumpy medium, it would be a turning point where the average surface brightness flattens. This is short description of the Model B, which is shown in Figure~\ref{fig:MWsample}. Here, the discontinuity radius $R_{\text{d}}$ is randomly chosen between 7 and 35~pc (in logarithmic scale) and the ambient concentration is in range $-3.3<\log(n_0)<0.9$. As discussed in previous subsection, the mean density jump $\bar{\rho}_{\text{j}}$ is inversely proportional to the ambient density, so that $\bar{\rho}_{\text{j}}=20$ at minimum $\rho_0$, and $\bar{\rho}_{\text{j}}=2$ at maximum $\rho_0$. The assumption is that, looking at the population average, surrounding ISM density is relatively higher for less dense WBBs. The limits were chosen so the peaks of the PDDs and their contours coincide. Model B looks significantly better than Model A, especially for SNRs at $D>20$~pc.

The Model C upgrades the Model B with a power-law density profile, $\rho=\rho_0(R/R_{\text{p}})^{-2}$, $R<R_{\text{p}}$, as a simple model of RSG wind. The maximum radius of the profile $R_{\text{p}}$ is in the range of 2--7~pc. The result is lowering the surface brightness at $R<R_{\text{p}}$ due to slower shock through higher density; velocity through power-law density is calculated using Eq.~(\ref{eq:v_mean}). Although the Model C seems to better succeeds in covering the lower-left part of MW PDD, this power-law model is very trivial and, considering that remnants at these radii are probably in free-expansion phase (due to a non-negligible ejecta mass), velocity calculation may not be correct. Moreover, the wind in the bubble is traveling with some radial velocity of the order of 10--100~km s$^{-1}$, which is not negligible, and the upstream velocity in the rest frame of the shock can be appreciably lower. This can result in significantly lower emission. Perhaps, it can cause even more $\Sigma$--$D$ flattening between the wind and the region beyond it.

The upper-left part of MW PDD ($\log \Sigma >-19.5$ and $\log D<1$), which looks like separate branch of the young remnants with slope $\beta \approx 6$, is hard or impossible to model with this paper's methods, since there is a high probability that they have not left the free-expansion phase completely. Our model is based on Sedov-Taylor expansion, which presumes enough swept-up mass so the velocity depends on it. In the case of SNRs with massive stellar ejecta and low-density medium, free-expansion phase can last several parsecs in diameter. Thus, the comparison of these models and observations in this region should be approached with necessary caution.

However, at larger diameters where the ST expansion law is valid, the Models B and C successfully reconstruct Galactic sample, which is seen from matching the average $\Sigma$ curves as well as PDD contours. We can conclude that higher density stellar surroundings, probably having the clumpy density distribution, might be a good explanation for flattening of empirical $\Sigma$--$D$ sample from $\approx$13 to $\approx$50~pc, and large sample scattering.

\subsection{Test-particle case vs. shock modification} \label{sec:TPvsMSH}

The emission model is developed using the TP approximation, with shock compression from Rankine-Hugoniot jump conditions, and calculating the MF. Here, we make a brief comparison of our homogeneous $\Sigma$--$D$ model to the one from \citet[][hereafter P18]{Pavlovicetal2018} who made a thorough approach to the time-dependent acceleration and back-reaction of the cosmic rays in modified SNR shock. They treat the shock modification using the effective adiabatic index $\gamma_{\text{eff}}$ that produces non-linear total compression (of order $X\sim7$--$10$, relevant only in early high Mach regime). The MF is calculated as in this paper. During the free expansion ($D<3$--$4$~pc), their Type Ia SNR model shows a radio flux ``brightening'' phase due to a growing number of accelerated CRs \citep[see][and references therein]{Pavlovic2017}, which is a priori not the case in this paper because of the ST character of our model.

Their inclusion of non-linear effects gives the CR pressure fraction $\sigma_{\text{CR}}$ around 0.4--0.8 during the first $\sim$10--30~pc of SNR radius, then permanently decreasing, being $\approx$0.2 at 100~pc (Figure~1 in P18). In this paper, however, during the early phase of SNR evolution $\sigma_{\text{CR}}\propto v_{\text{s}}^{-1} I(x)$, increasing with diameter approximately as $\propto D^{1.4}$ until the shock compression falls below 4. Then, it quickly starts to fall as $\propto D^{-2}$ or even steeper.

\begin{figure*}
	\centering
	\includegraphics[width=0.4\textwidth]{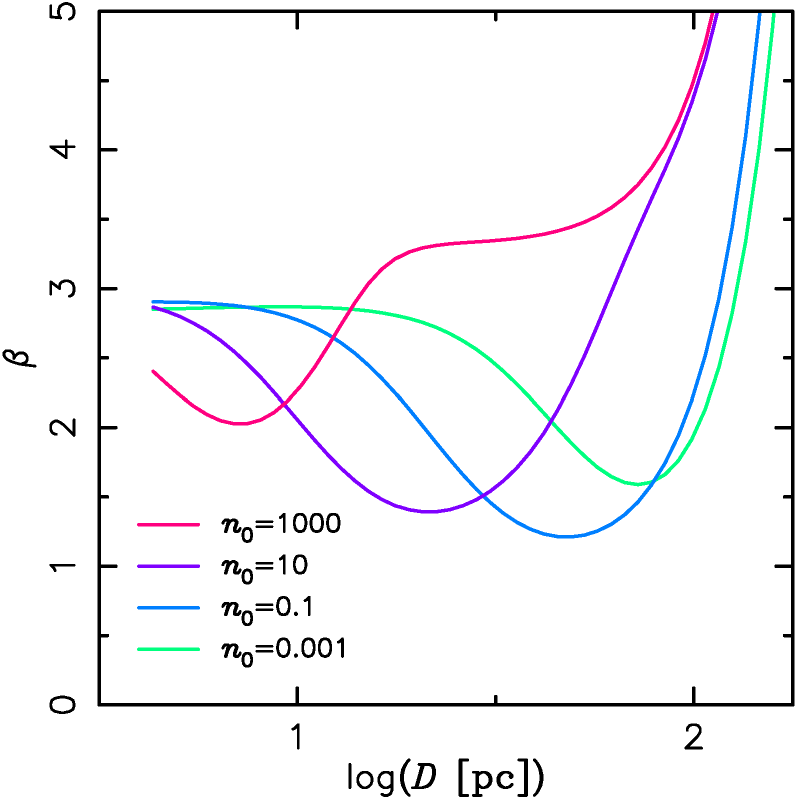}~~~~~~~
	\includegraphics[width=0.4\textwidth]{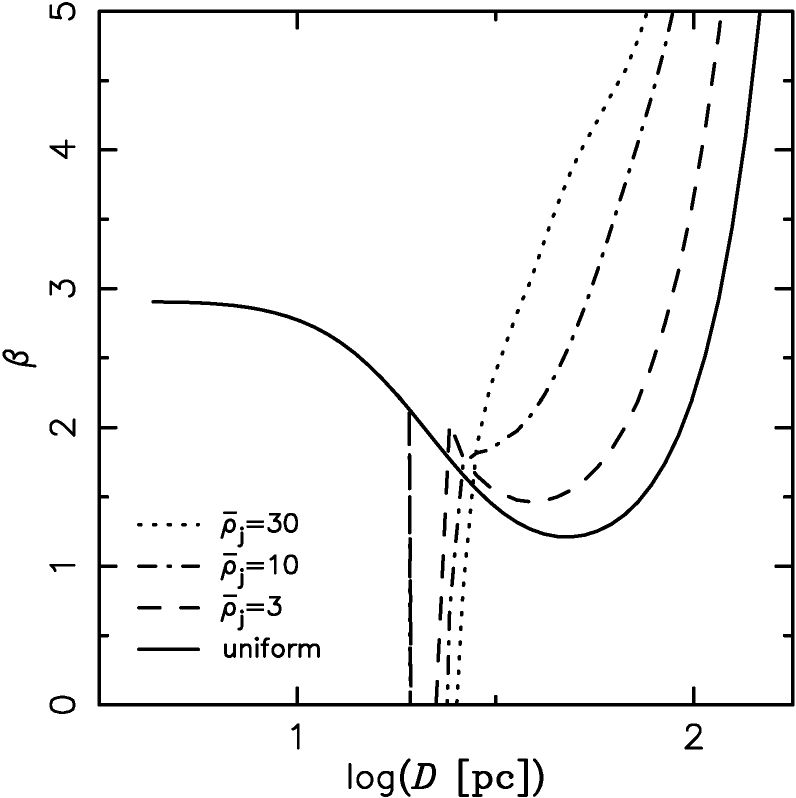}
	\caption{\textit{Left}. Slopes $\beta$ of uniform medium $\Sigma$--$D$ relations in different concentrations (in cm$^{-3}$), depending on SNR diameter. \textit{Right}. Slopes $\beta$ for $n_0=0.1$~cm$^{-3}$ in clumpy medium of different mean density jump ($R_{\text{d}}=10$~pc, $f_{\text{c}}=0.2$). At first interaction with clumps, the emission rise causes a negative slope. After the peak in $\Sigma$, further decrease is significantly steeper than in uniform medium, while the steepness depends on the mean density jump (i.e. the density contrast).}
	\label{fig:slope}
\end{figure*}

Consequently, the approach of P18 gives somewhat steeper $\Sigma$--$D$ relations in ST phase ($\beta\approx5$--6, see Figure~5 in their work) than in this paper. The left panel of Figure~\ref{fig:slope} shows the $\Sigma$--$D$ slopes of our model for different ambient densities. In early phase all the curves start at $\beta\approx3$, followed by intermediate flattening (with a limit $\beta>1$) and then rising to higher values ever after. The slopes for the cases of CC SNRs in P18 show similar behavior, although with overall higher values. The Type Ia SNRs are not really comparable to our model due to their early brightening phase, but at late stages it is shown that lower ambient density results in generally flatter relations. 

The right panel of Figure~\ref{fig:slope} shows the clumpy medium impact on the slope of the subsequent $\Sigma$--$D$ evolution. It is shown that the clumpy medium is responsible for steepening of the $\Sigma$--$D$ curve. The steepening effect is stronger with higher mean density of the clumpy medium.

\subsection{Selection effects}
The observational selection effects, which are not included in model of the artificial samples, can play a significant role in shaping the empirical $\Sigma$--$D$ slopes. The most important selection effects in detection of Galactic SNRs are \citep{Green1991,CaseBhattacharya1998,Urosevicetal2005, Pavlovicetal2013}:

\begin{itemize}
	\item[(i)] Surface brightness (or sensitivity) limit, 
	\item[(ii)] Angular-size (or resolution) limit,
	\item[(iii)] non-uniform coverage of the sky, due to variable background brightness,
	\item[(iv)] Malmquist bias, i.e. favoring of intrinsically brighter objects due to flux density limit, 
	\item[(v)] confusion for HII regions.
\end{itemize}

The Galactic SNR sample has numerous problems, because surveys have different sensitivity limits and resolutions, while sky-coverage effect is always present. It is shown that Malmquist bias steepens the $\Sigma$--$D$ slope \citep{Urosevicetal2005, Urosevicetal2009}. On the other hand, cutting off of large low-surface-brightness remnants could lead to false slope flattening. All these effects are mutually connected and so very difficult to model.

If we assume that the real SNR distribution in surface-brightness (for a fixed $D$) is symmetric, we expect that cut-off of the lower tail of that distribution would result in an asymmetry, such that the peak is positioned toward the lower $\Sigma$ values. However, from Figure~\ref{fig:MWsample} (contour plot) it is evident that the distribution peak is positioned toward higher $\Sigma$ values. This implies that, under the assumption of real distribution symmetry, the sensitivity selection effect seems not to be significant in shaping of the distribution. Nevertheless, our model shows that at higher densities evolutionary paths tend to merge together (see Figure~\ref{fig:impacts}), pulling the weight to the higher $\Sigma$ values, which undermines the assumption of the distribution symmetry. This indicates that even if the peak is positioned at higher $\Sigma$, the selection effects could still have an impact on the slope flattening of the $\Sigma$--$D$ distribution. The merging of the evolutionary paths for different densities at higher diameters also appears in a theoretical work of \citet{BerezhkoVolk2004}. On the other hand, the flattening (or even a break) around $D\approx10$~pc at high $\Sigma$ values cannot be attributed to the sensitivity selection effect. Hence, it is appropriate to seek for other causes of the break some of which are presented in this paper. We conclude that the SNR-type-dependent analysis is needed for this problem to be better diagnosed, since our preliminary analysis shows that the change in $\Sigma$--$D$ slope for a non-MC-associated SNR sample is larger than for MC-associated SNR sample (which contains brighter remnants; MC stands for ``molecular clouds''). Such an analysis should give more answers on selection effects.

\section{Summary and Conclusions}
\label{sec:conclusions}
We presented the model of radio synchrotron emission, used for calculating SNR luminosity in HD simulations, by using own \citep{Kostic2019} code. The model relies on a test-particle approximation with implied power-law distribution of emitting electrons. Based on the simulations results, we derived the semi-analytical 3D spherically-symmetric model for the HD evolution of the SNRs in an ISM evenly filled with spherical homogeneous clumps. The model links the SNR evolution to its radius and relies on a volume-filling factor, size, and density of the clumps, complex geometry of the shock and jumps in the HD states at the clumps.

We showed that the SNR brightness through the clumpy medium has a different evolution path than in the case of uniform medium of the same mean density. The surface brightness through the clumpy medium usually surpasses the uniform case initially, until the dense clumps start to damp the shock velocity and Mach number, affecting the shock dynamics. Thus, at late stages this environment leads to faster decline in the CR acceleration and emissivity, especially at clumps. Moreover, if the radiative losses would have been taken into account, the decline would be more pronounced, even at earlier stages. Hence, the phase of enhanced brightness lasts only for a limited time span during the evolution, probably up to about $\sim10$~pc after entering the clumpy medium.

The semi-analytical HD model vastly shortens the time for obtaining the radio $\Sigma$--$D$ evolutionary paths. Hence, we were able to simulate the artificial samples of Galactic SNRs and compare them with the observed one. The three simple models were made, and the results suggest that a significant part of SNR shocks must collide with and evolve through the higher density clumpy medium to reproduce the flattening in the MW sample. The significant impact of the clouds starts at diameters of $\approx14$~pc, up to $\sim70$~pc, with the average density jump at clumpy medium of $\sim2$--20 times, roughly depending on the low CSM density.

By comparing the uniform medium $\Sigma$--$D$ slopes from P18 and this paper, we conclude that in ST phase slope has to be in range from $\beta\approx1.2$, the lowest value from our model, to $\beta\approx6$, the upper limit value from P18. Since the average $\Sigma$--$D$ curve of the Galactic SNR population at $D\approx30$~pc is not nearly steep as either of these two results (being around $\beta\approx0$), we propose its flattening mechanism originating in sporadic emission jumps of individual SNR evolutionary paths in a limited diameter interval (predominantly at $D\ge14$~pc), as shown in Section~\ref{sec:mw_models}, regardless of the emission model.

\begin{acknowledgements}
	During the work on this paper the authors were financed by the Ministry of Science, Technological Development and Innovation of the Republic of Serbia (MSTDIRS) through contract no. 451-03-66/2024-03/200002 made with Astronomical Observatory of Belgrade (PK and BV), and contract no. 451-03-66/2024-03/200104 (BA and DU). BA additionally acknowledges the funding provided by the Science Fund of the Republic of Serbia through project \#7337 ``Modeling Binary Systems That End in Stellar Mergers and Give Rise to
	Gravitational Waves'' (MOBY). The authors thank the referee for helpful comments and valuable suggestions.
\end{acknowledgements}

\bibliography{mybib}{}
\bibliographystyle{aasjournal}

\end{document}